\documentclass[sigplan,10pt]{acmart}

\usepackage{booktabs}
\usepackage{amsmath}
\usepackage{xcolor}
\usepackage{graphicx}
\usepackage{caption}
\usepackage{listings}
\usepackage{url}
\usepackage[linesnumbered,lined,sright,ruled,titlenotnumbered]{algorithm2e}
\usepackage{algpseudocode}
\usepackage{enumerate}
\usepackage[shortlabels]{enumitem}
\usepackage{fancyvrb}
\usepackage{frameit}
\usepackage{graphics}
\usepackage{latexsym}
\usepackage{multirow}
\usepackage{proof}
\usepackage{sidecap}
\usepackage{url}
\usepackage{wrapfig}
\usepackage{flushend}
\usepackage{ebproof}
\usepackage{hyperref}
\usepackage{verbatim}
\usepackage{listings}
\usepackage[utf8x]{inputenc}
\usepackage{verbatim}
\usepackage{listings}
\usepackage{subfig}
\usepackage{euscript}

\usepackage[T1]{fontenc}

\DeclareSymbolFont{stmry}{U}{stmry}{m}{n}
\DeclareMathSymbol\mergemap\mathbin{stmry}{"21}

\DeclareSymbolFont{ugrf@m}{U}{eur}{m}{n}
\SetSymbolFont{ugrf@m}{bold}{U}{eur}{b}{n}
\DeclareMathSymbol{\Upupsilon}{\mathord}{ugrf@m}{"07}

\setcopyright{none}
\acmDOI{}

\acmISBN{}

\acmConference[]{}{}{}
\acmYear{}
\copyrightyear{}

\acmPrice{}

\renewcommand\footnotetextcopyrightpermission[1]{} 

\begin{document}
\captionsetup[figure]{labelfont=bf,textfont=normalfont,singlelinecheck=on}
\setlist[itemize]{leftmargin=*}
\setlist[enumerate]{leftmargin=*}

\title{Automated Modular Verification for\\
Race-Free Channels   with Implicit and Explicit Synchronization}

\author{Andreea Costea}
\affiliation{School of Computing, National University of Singapore}
\author{Chin Wei-Ngan}
\affiliation{School of Computing, National University of Singapore}
\author{Florin Craciun}
\affiliation{Faculty of Mathematics and Computer Science, Babes-Bolyai University}
\author{Shengchao Qin}
\affiliation{Computer Science, Teesside University}

\definecolor{darkgreen}{HTML}{008000}

\newcommand{\cb}{{\textit{communicates-before}}}
\newcommand{\chan}{\mathcal{C}}
\newcommand{\lchan}{\tilde{\mathcal{C}}}
\newcommand\commonspec[1]{\code{Common(#1)}}
\newcommand\partyspec[2]{\code{Party(#1,#2)}}
\newcommand\changlob[2]{\overline{\chan}(#1,#2)}
\newcommand\gloabspec[3]{\code{Glob(#1,#2,#3)}}
\newcommand{\perchanprj}[3]{\code{{#1}{\prj}{#2}{\prj}{#3}}}
\newcommand{\perpartyprj}[2]{\code{{#1}{\prj}{#2}}}
\newcommand{\All}{\code{All}}
\newcommand{\codeS}[1]{\ensuremath{\it #1}}
\newcommand{\prj}{\mytt{\#}}
\newcommand{\chpred}[2]{{\cal C}(#1,#2)}
\newcommand{\wnsay}[1]{}
\newcommand{\wnnay}[1]{}
\newcommand{\asay}[1]{}
\newcommand{\anay}[1]{}
\newcommand{\scsay}[1]{}
\newcommand{\scnay}[1]{}
\newcommand{\doubt}[1]{{\color{red}{#1}}}
\newcommand{\betw}[1]{\ensuremath{{\cdots}#1{\cdots}}}
\def\defeq{\ensuremath{\,\triangleq}}
\def\cd{\,{\cdot}\,}
\def\compl{{\sim}}
\def\pst{\omega}
\def\st{\ensuremath{\sigma}}
\def\St{\ensuremath{\mathrm{\Delta}}}
\def\Sta{\ensuremath{\mathrm{\Delta'}}}

\newcommand{\Pre}{\ensuremath{\mathsf{Pre}}}
\newcommand{\Post}{\ensuremath{\mathsf{Post}}}
\newcommand{\Curr}{\ensuremath{\mathsf{Curr}}}

\def\svar{\ensuremath{\mathsf{svars}}}
\def\rec{\ensuremath{\mathsf{rec}}}
\def\Pairs{\ensuremath{\mathsf{Pairs}}}
\def\Spec{\ensuremath{\mathsf{Spec}}}
\def\mspec{\ensuremath{\myit{mspec}}}
\newcommand{\fv}[1]{\ensuremath{\mathsf{fv}(#1)}}
\newcommand{\lv}[1]{\ensuremath{\mathsf{lv}(#1)}}
\def\Refine{\ensuremath{\mathsf{Refine}}}
\def\ReachVar{\ensuremath{\mathsf{ReachVar}}}

\newcommand{\mtt}[1]{\ensuremath{\mathtt{#1}}}
\newcommand{\mbf}[1]{\ensuremath{\mathbf{#1}}}
\newcommand{\msf}[1]{\ensuremath{\mathsf{#1}}}
\newcommand{\sevent}[2]{\ensuremath{#1^{#2}_S}}
\newcommand{\revent}[2]{\ensuremath{#1^{#2}_R}}
\newcommand{\event}[2]{\ensuremath{#1^{(#2)}}}
\def\ev{\ensuremath{E}}
\newcommand{\eventf}[3]{\ensuremath{#1^{#2}{{\at}#3}}}
\newcommand{\transf}[3]{\ensuremath{#1{{\at}#3}}}

\def\Var{\ensuremath{\mathsf{Var}}}
\def\LVar{\ensuremath{\mathsf{LVar}}}
\def\SVar{\ensuremath{\mathsf{SVar}}}

\def\nul{\texttt{null}}

\newcommand{\denoleft}{\ensuremath{|\hspace{-.06cm}[}}
\newcommand{\denoright}{\ensuremath{]\hspace{-.06cm}|}}
\newcommand{\semantic}[3]{{\denoleft}#1{\denoright}_{#2}#3}
\newcommand{\semantica}[3]{{\denoleft}#1{\denoright}_{#2}^\msf{A}#3}
\newcommand{\semantici}[3]{{\denoleft}#1{\denoright}_{#2}#3}
\newcommand{\triple}[3]{\{ #1 \}~ {\color{blue}{#2}} ~\{ #3 \}}

\def\vl{\ensuremath{{val}}}
\def\rf{\ensuremath{\mathsf{ref}}}
\def\dispose{\ensuremath{\mathsf{dispose}}}
\def\new{\ensuremath{\mathsf{new}}}
\def\cskip{\ensuremath{{skip}}}
\def\ccskip{skip}
\newcommand{\cif}[3]{\ensuremath{\mathsf{if}~#1~\mathsf{then}~#2~\mathsf{else}~#3~\mathsf{fi}}}
\newcommand{\cwhile}[2]{\ensuremath{\mathsf{while}~#1~\mathsf{do}~#2~\mathsf{od}}}
\def\known{\ensuremath{\mathsf{known}}}
\def\unknown{\ensuremath{\myit{unk}}}
\def\local{\ensuremath{\mathsf{local}}}
\def\return{\ensuremath{\mathsf{return}}}

\def\D{\Delta}
\def\Pst{\Upsilon}
\def\S{\mathsf{S}}
\def\H{\mathsf{H}}
\def\Re{\mathsf{R}}
\def\subst{\rho}
\newcommand{\hyphen}[0]{\ensuremath{\mbox{-}}}
\newcommand{\pair}[2]{\ensuremath{(#1, #2)}}
\newcommand{\unk}[2]{\ensuremath{\myit{unk}_{#1}{(#2)}}}

\def\TODO{\textsf{TODO}~}


\def\State{\ensuremath{\varsigma}}
\def\mem{\code{heap}}
\def\stk{\code{stk}}
\def\so{\textit{sizeof}}
\def\sso{\textit{ssizeof}}
\newcommand{\hp}{\ensuremath{\mu}}
\newcommand{\sk}{\ensuremath{\sigma}}
\newcommand{\err}{\ensuremath{\myit{er}}}
\newcommand{\errdead}{\ensuremath{\bot_{\it deadlock}}}
\newcommand{\memerr}{\ensuremath{\myit{er}_1}}
\newcommand{\othererr}{\ensuremath{\myit{er}_2}}

\newcommand{\hide}[1]{}
\newcommand{\nil}{\btt{null}}
\newcommand{\res}{\btt{res}}
\newcommand{\ra}{\,\rightarrow\,}
\newcommand{\emp}{\btt{emp}}
\newcommand{\veq}{\ensuremath{\equiv}}
\newcommand{\self}{\btt{root}}
\newcommand{\wwith}{\btt{inv}}
\newcommand{\equivimp}{\ensuremath{\Longleftrightarrow}}
\newcommand{\revimp}{\ensuremath{\Longleftarrow}}
\newcommand{\ndet}{\ensuremath{\wedge}}
\newcommand{\puret}{\ensuremath{\pi}}
\newcommand{\simp}{{\bf ;}}
\newcommand{\minv}{\ensuremath{\iota}}
\newcommand{\ptr}{\ensuremath{\gamma}}
\newcommand{\nm}{\ensuremath{\phi}}
\newcommand{\heap}{\ensuremath{\kappa}}
\newcommand{\constr}{\ensuremath{\mathrm{\Phi}}}
\newcommand{\pconstr}{\ensuremath{\phi}}
\newcommand{\sat}[1]{\myit{SAT}(#1)}
\newcommand{\entail}[2]{\ensuremath{#1 \vdash #2}}
\newcommand{\entailA}[2]{\ensuremath{#1 \triangleright #2}}
\newcommand{\nentail}[2]{\ensuremath{#1 \nvdash #2}}
\newcommand{\entailD}[3]{\ensuremath{#1{\vdash}#2\,{\sep}\,#3}}
\newcommand{\entailH}[3]{\ensuremath{#1{\vdash}#2\,{\sep}\,#3}}
\newcommand{\entailE}[4]{\ensuremath{#2{\vdash}_{#1}#3\,{\sep}\,#4}}
\newcommand{\entailV}[3]{\entailE{V}{#1}{#2}{#3}}
\newcommand{\entailK}[5]{\ensuremath{#3{\vdash}^{#1}_{#2}#4\,{\sep}\,#5}}
\newcommand{\entailVV}[3]{\entailK{\heap}{V}{#1}{#2}{#3}}
\newcommand{\entailI}[3]{\ensuremath{#2 \vdash_{#1} #3}}
\newcommand{\myit}[1]{\textit{#1}}
\newcommand{\mytt}[1]{\texttt{#1}}
\newcommand{\R}{\ensuremath{{\btt{R}}}}
\newcommand{\Le}{\ensuremath{{\btt{L}}}}
\newcommand{\Sh}{\ensuremath{{\btt{S}}}}
\newcommand{\A}{{\myit{A}}}
\newcommand{\at}{\mytt{@}}
\newcommand{\isview}{\myit{IsView}}
\newcommand{\isdata}{\myit{IsData}}
\newcommand{\hc}[2]{\ensuremath{{#1}{::}#2}}
\newcommand{\hx}[1]{\hc{\code{x}}{#1}}
\newcommand{\mtype}[2]{\ensuremath{{#1}{ \ :}#2}}
\def\entword{ENT}
\newcommand{\rulen}[1]{\ensuremath{{\bf \scriptstyle #1}}}
\newcommand{\entrule}[1]{\ensuremath{\scriptstyle \entword-#1}}
\newcommand{\entrulen}[1]{[\underline{{\bf \scriptstyle \entword-}\rulen{#1}}]}
\newcommand{\oprulen}[1]{[\underline{{\bf \scriptstyle \opword-}\rulen{#1}}]}
\def\opword{OP}
\def\iopword{iOP}
\newcommand{\ioprulen}[1]{[\underline{{\bf \scriptstyle \iopword-}\rulen{#1}}]}
\newcommand{\lemrulen}[1]{[{{\bf \scriptstyle LEM-}\rulen{#1}}]}
\newcommand{\lemnormrulen}[1]{\lemrulen{NORM-#1}}
\newcommand{\ebpentail}[2]{
	\begin{prooftree}
		\Hypo{\entrulen{#1}}
		\Infer[no rule]1{{\begin{prooftree}[rule style=simple] #2 \end{prooftree}}}
	\end{prooftree}
}
\newcommand{\ebpsem}[2]{
	\begin{prooftree}
		\Hypo{\oprulen{#1}}
		\Infer[no rule]1{{\begin{prooftree}[rule style=simple] #2 \end{prooftree}}}
	\end{prooftree}
}
\newcommand{\ebpisem}[2]{
	\begin{prooftree}
          \Hypo{\ioprulen{#1}}
          \Infer[no rule]1{{\begin{prooftree}[rule style=simple] #2 \end{prooftree}}}
	\end{prooftree}
}

\newcommand{\entwlbl}[2]{{\begin{array}{c} { \entrulen{#1} } \\ {#2} \end{array} }}
\newcommand{\verirulen}[1]{[\underline{{\bf \scriptstyle FV-}\rulen{#1}}]}
\newcommand{\reasonrulen}[1]{[\underline{{\bf \scriptstyle FR-}\rulen{#1}}]}
\newcommand{\markred}[1]{{{\color{red}#1}}}
\newcommand{\markblue}[1]{{{\color{blue}#1}}}
\newcommand{\ebpverif}[2]{
  \ifthenelse{\equal{#1}{} }{
    \begin{prooftree}[rule style=simple] #2 \end{prooftree}
  }{  
    \begin{prooftree}
      \Hypo{[\underline{{\bf \scriptstyle }\rulen{#1}}]}
      \Infer[no rule]1{{\begin{prooftree}[rule style=simple] #2 \end{prooftree}}}
    \end{prooftree}
  }}

\newcommand{\hformn}[3]{\ensuremath{{\btt{#1}}{\mapsto}{\btt{#2}}{\langle}{\btt{#3}}{\rangle}}}
\newcommand{\view}[2]{\ensuremath{#1{(}{#2}{)}}}
\newcommand{\pview}[2]{\ensuremath{\overline{#1}{(}{#2}{)}}}
\newcommand{\sspec}[3]{\ensuremath{\view{\mathcal{#1}}{#2,#3}}}
\newcommand{\hform}[3]{\ensuremath{{#2}({#1,#3})}}
\newcommand{\uhform}[3]{\ensuremath{\code{H}({#2})({#1,#3})}}
\newcommand{\mapview}[3]{\ensuremath{{#1}{\mapsto}{#2}{(}{#3}{)}}}
\newcommand{\transit}{\rightarrow}
\newcommand{\arrimp}{\ensremath{\,\btt{*}\!\!\!\rightarrow}\,}
\def\postst{\myit{Post}}

\def\rewrite{\leadsto}
\def\map{\myit{X{\!}Pure}}
\def\mappair{\myit{X{\!}Pure}_P}

\def\D{\Delta}
\def\myloop{\ensuremath{\alpha}}
\def\arity{\myit{arity}}
\def\max{\myit{max}}
\def\fresh{\myit{fresh}}
\def\where{\btt{where}}
\def\bag{\ensuremath{\code{B}}}
\def\bagunion{\sqcup}
\def\bagintersect{\sqcap}
\def\bagsubsume {\sqsubset}
\def\bagsubtract{-}
\def\zerobag{\{{\bf 0}\}}
\def\zerobslash{\emp}
\newcommand{\load}[1]{\mytt{load}{\langle}#1{\rangle}}
\newcommand{\store}[1]{\mytt{store}{\langle}#1{\rangle}}
\newcommand{\const}[1]{\mytt{const}{\langle}#1{\rangle}}
\def\invoke{\mytt{invoke}}
\def\invokeTail{\mytt{invoke}_{Tail}}
\def\new{\mytt{new}}
\def\dispose{\mytt{dispose}}
\def\bool{\btt{bool}}
\def\int{\btt{int}}
\def\longint{\mytt{long}}
\def\float{\mytt{float}}
\def\double{\mytt{double}}
\def\fixpt{\myit{fixpt}}
\def\rec{\mytt{Rec}}
\def\void{\btt{void}}
\def\type{\myit{type}}
\def\reference{\mytt{ref}}
\def\object{\mytt{object}}
\def\objectset{{\rm\bf ObjectType}}
\def\mn{\myit{m}}
\def\F{{\cal F}}
\def\phiinv{\phi_{\myit{inv}}}
\def\phirec{\phi_{\myit{rec}}}
\def\pre{\constr_{\myit{pr}}}
\def\post{\constr_{\myit{po}}}
\def\T{\Gamma}
\def\eqt{\myit{eq}_\tau}
\def\FS{\Pi}  
\def\true{\btt{true}}
\def\false{\btt{false}}
\def\a{s}
\def\restrict{\myit{enrich}}
\newcommand{\parg}[1]{\ensuremath{{\langle}#1{\rangle}}}
\newcommand{\num}[1]{|#1|}
\newcommand{\ifEq}[2]{\mytt{if}~#1~#2}
\newcommand{\whileEq}[1]{\mytt{while}~#1}
\newcommand{\myif}[3]{\btt{if}~(#1)~#2~\btt{else}~#3}
\newcommand{\mywhile}[3]{\btt{while}~#1~\btt{where}~#3~\btt{do}~#2}
\newcommand{\mybind}[3]{\mytt{bind}~#1~\mytt{to}~#2~\mytt{in}~#3}
\def\skipcmd{\mytt{nop}}
\newcommand{\defs}{\ensuremath{=_{df}}}
\newcommand{\snd}[1]{\myit{snd}(#1)}
\newcommand{\guarding}{{\ensuremath{\rightarrow}}}
\newcommand{\nonneg}[1]{\myit{notneg}(#1)}
\newcommand{\code}[1]{{\ensuremath{\tt #1}}}
\newcommand{\sm}[1]{\mbox{$#1$}}
\newcommand{\btt}[1]{{\tt #1}}
\newcommand{\todo}[1]{{\bf(* #1 *)}}
\newcommand{\passref}{\btt{ref}}
\newcommand{\ensures}{\btt{ensures}}
\newcommand{\requires}{\btt{requires}}

\newcommand{\llin}[1]{\texttt{#1}}

\newtheorem{thm}{Theorem}
\newtheorem{pty}{Property}
\newtheorem{defn}{Definition}
\newtheorem{idfn}{Informal Definition}
\def\proof{\vspace{-0.2cm}\noindent{\bf Proof}\hspace*{0.5cm}}
\newcommand{\tcv}[3]{#1\,\vdash\,#2,\,#3}
\newcommand{\tcvT}[2]{\tcv{}{#1}{#2}}

\newcommand{\fsr}[5]{#1 \vdash_{\small\myit{F}} #2 \rightsquigarrow #3, #4, #5}
\newcommand{\fsinf}[3]{\ensuremath{\frac{\begin{array}{c}[\underline{\rulen{FS-#1}}]\\[0.5ex]
#2\end{array}}{#3}}}

\newcommand{\asr}[4]{#1 \vdash_{\small\myit{A}} #2 \rightsquigarrow #3, #4}
\newcommand{\asinf}[3]{\ensuremath{\frac{\begin{array}{c}[\underline{\rulen{AS-#1}}]\\[0.5ex]
#2\end{array}}{#3}}}

\newcommand{\ssrA}[2]{\ssr{\myit{a}}{#1}{#2}}
\newcommand{\ssr}[3]{#1 \vdash_{\small\myit{S}} #2 \rightsquigarrow #3}
\newcommand{\ssinf}[3]{\ensuremath{\frac{\begin{array}{c}[\underline{\rulen{SS-#1}}]\\[0.5ex]
#2\end{array}}{#3}}}

\newcommand{\hsrX}[5]{#5,#1 \vdash_{\myit{H}} #2 \rightsquigarrow #3,#4}
\newcommand{\hsr}[4]{\hsrX{#1}{#2}{#3}{#4}{\myit{a}}}
\newcommand{\hsinf}[3]{\ensuremath{\frac{\begin{array}{c}[\underline{\rulen{HS-#1}}]\\[0.5ex]
#2\end{array}}{#3}}}

\newcommand{\oprerr}[5]{\ensuremath{\config{#1}{#2}{#3}{#4}\hookrightarrow #5}}
\newcommand{\config}[4]{\ensuremath{{\langle}{#1},{#2},{#3},{#4}{\rangle}}}
\newcommand{\opr}[8]{\ensuremath{\config{#1}{#2}{#3}{#4} \hookrightarrow \config{#5}{#6}{#7}{#8}}}
\newcommand{\oprd}[4]{\opr{#1:\FS}{\O}{#2}{\ss,\mem}{#3:\FS}{\O}{#4}{\ss,\mem}}
\newcommand{\oprds}[6]{\opr{#1:\FS}{\O}{#2}{\ss,\mem}{#3:\FS}{#4}{#6}{\ss,#5}}
\newcommand{\oprc}[5]{\config{#1}{#2}{#3}{#4}\hookrightarrow^* {#5}}
\newcommand{\oprule}[3]{\ensuremath{\frac{\begin{array}{c}[\underline{\rulen{OP-#1}}]\\[0.5ex]
#2\end{array}}{#3}}}

\def\EI{\epsilon}
\def\Y{\Upsilon}
\def\L{\Lambda}
\def\O{\omega}
\def\i{\iota}
\def\MemErr{\mytt{MemInadequate}}
\def\OF{\mytt{FrameErr}}
\def\cf{\myit{f}}
\def\ret{\mytt{ret}}
\def\size{\myit{size}}
\def\cxt{\mytt{cxt}}
\newcommand\reqs[4]{\ensuremath{#1 \vdash #2 \rightsquigarrow #3,#4 }}

\def\G{\Gamma}
\def\nochange{\myit{nochange}}
\def\FSV{\myit{V}}
\newcommand{\sep}{\ensuremath{*}}
\newcommand{\sepc}{\ensuremath{*}}
\newcommand{\rview}[2]{\code{\ensuremath{#1{(}{\self,#2}{)}}}}
\newcommand{\sepimpl}{\ensuremath{\code{-\!\!\!\!-\!\!\!*}}}
\newcommand\eq[3]{\ensuremath{\myit{eq}_{#1}(#2,#3)}}
\newcommand{\tj}[5]{#1,#2\,{\vdash}\,#3~{::}~#4, #5}
\newcommand{\tr}[3]{\ensuremath{\frac{\begin{array}{c}[\underline{\rulen{#1}}]\\[0.5ex]
#2\end{array}}{#3}}}
\newcommand{\hlr}[3]{\ensuremath{\frac{\begin{array}{c}\reasonrulen{#1}\\[0.5ex]
#2\end{array}}{#3}}}
\newcommand{\hlrs}[3]{\ensuremath{\frac{\begin{array}{c}
#2\end{array}}{#3}}}
\newcommand{\htriple}[3]{\ensuremath{\vdash \{#1\}\,#2\,\{#3\}}}
\newcommand{\iview}[2]{\ensuremath{#1{\langle}#2{\rangle}}} 
\newcommand{\ihc}[2]{\ensuremath{#1{::}#2}}                 
\newcommand{\subtr}[3]{\entailH{#1}{#2}{#3}}
\newcommand{\reasonrulenm}[1]{[\underline{{\bf \scriptstyle }\rulen{#1}}]}
\newcommand{\hlrm}[3]{\ensuremath{\frac{\begin{array}{c}\reasonrulenm{#1}\\[0.5ex]
#2\end{array}}{#3}}}

\newcommand{\raws}[9]{{#1} ; {#2} ; {#3} ;{#4} \vdash {#5} :: {#6} , {#7},{#8} ,{#9}}
\newcommand{\rawss}[5]{\T ; \D; \L; \Y  \vdash {#1} :: {#2} , {#3}, {#4}, {#5}}

\newcommand{\mymodel}{\models}
\def\Flds{\myit{Fields}}
\def\Store{\myit{Heaps}}
\def\Stack{\myit{Stacks}}
\def\Locations{\myit{Loc}}
\def\Val{\myit{Val}}
\def\Var{\myit{Var}}
\newcommand{\mo}[3]{#1,#2 \mymodel #3}
\newcommand{\notmo}[3]{#1,#2 \not\mymodel #3}
\newcommand{\mos}[1]{\ensuremath{\mo{s}{h}{#1}}}
\newcommand{\mosp}[1]{\ensuremath{{s}\mymodel{#1}}}
\def\iffs{\mbox{\tt iff~}}
\def\dom{\myit{dom}}
\def\loc{\iota}
\newcommand{\s}[1]{[\![#1]\!]s}
\def\inv{\myit{inv}}
\def\getP{\myit{getProperty}}
\newcommand{\menv}[2]{\frac{\begin{array}{c}#1\end{array}}{#2}}
\def\fresh{\myit{fresh}}
\def\length{\myit{length}}

\newcommand{\mysf}[1]{\textsf{\bf\small #1}}
\newcommand{\presize}{\phi_{\myit{pr}}}
\newcommand{\postsize}{\phi_{\myit{po}}}
\newcommand{\sizev}{{\cal V}}
\newcommand{\pure}{\ensuremath{\pi}}
\newcommand{\sarg}[1]{{\la}#1{\ra}}
\def\la{\langle}
\def\ra{\rangle}
\def\bagconstr{\varphi}

\newcommand{\pnt}[2]{\ensuremath{#1 {\mapsto} #2}}
\newcommand{\lseg}[2]{\ensuremath{\mathsf{ls}(#1, #2)}}
\newcommand{\lsegs}[3]{\ensuremath{\mathsf{ls}(#1, #2, #3)}}

\newcommand{\hipsleek}{\textsc{Hip/Sleek}~}
\newcommand{\hip}{\textsc{Hip}~}
\newcommand{\sleek}{\textsc{Sleek}~}
\newcommand{\M}{\mathsf{M}}


\def\hip{\textsc{Hip}~}
\def\sleek{\textsc{Sleek}~}
\def\Prog{\ensuremath{\mathcal{P}}}
\def\shps{\ensuremath{\mathcal{S}}}
\def\fail{\textsf{fail}}
\def\FAIL{\textsf{FAIL}}

\lstset{ language=[ANSI]C,
  basicstyle=\ttfamily\small,
  keywordstyle=\color{black}\ttfamily,
  commentstyle=\small\color{red},
  framexleftmargin=0mm,
  xleftmargin=0mm,
  xrightmargin=0mm,
  captionpos=b, 
  showspaces=false,
  showstringspaces=false,   
  showtabs=false, 
  tabsize=4,
  breaklines=true,
  extendedchars=false,
  escapeinside= {(*@} {@*)}
}



\newcommand{\dist}{0em}
\newcommand{\distv}{0em}
\newcommand{\disth}{0em}
\newcommand{\distt}{0em}

\newcommand{\distf}{-1.6ex}



\newcommand{\sstate}[1]{\textcolor{MidnightBlue}{\code{#1}}}
\newcommand{\obar}[1]{\code{\bar{#1}}}

\newcommand{\viewbody}{\L}
\newcommand{\prot}{{G}}
\newcommand{\cproj}{{Z}}
\newcommand{\cprojtail}{{Z^t}}
\newcommand{\cprojres}{{Z^\prime}}
\newcommand{\prottail}{{\prot^t}}
\newcommand{\protres}{{\prot'}}
\newcommand{\gp}{H}
\newcommand{\prott}{\ensuremath{\prot^\mathcal{T}}}
\newcommand{\proj}{{L}}
\newcommand{\projp}{\Upupsilon}
\newcommand{\projc}{\proj}
\newcommand{\unkbox}{\square}
\newcommand{\sessend}{end}

\newcommand{\send}[2]{
  \ifthenelse{\equal{#1}{} }{
     !{#2}
   }{  !{#1}{\cd}{#2}}
 }
\newcommand{\recv}[2]{
   \ifthenelse{\equal{#1}{} }{
     ?{#2}
   }{  ?{#1}{\cd}{#2}}
 }
\newcommand{\sendc}[3]{
  \ifthenelse{\equal{#2}{} }{
     #1!{#3}
  }{ #1!{#2}{\cd}{#3} }
}
\newcommand{\recvc}[3]{
  \ifthenelse{\equal{#2}{} }{
     #1?{#3}
  }{ #1?{#2}{\cd}{#3} }
}
  
\newcommand{\trans}[2]{\unkbox {#1}{\cd}{#2}}

\newcommand{\seq}{;} 
\newcommand{\useq}{*}
\newcommand{\disj}{\vee}
\newcommand{\chani}[3]{\ensuremath{\view{\chan}{#1,#2,#3}}}
\newcommand{\lchani}[3]{\ensuremath{\view{\lchan}{#1,#2,#3}}}
\newcommand{\hov}{V}
\newcommand{\hovars}{\hov}
\newcommand{\hovarh}{\hov}
\newcommand{\hosets}{\mathcal{I}}
\newcommand{\hoseth}{\mathcal{I}}
\newcommand{\vdashs}{\vdash_{s}}
\newcommand{\vdashe}{\vdash_{e}}
\newcommand{\entailS}[3]{\ensuremath{#1 \vdash #2 ~\leadsto ~#3}}

\newcommand{\projectsymbol}{\downharpoonright}
\newcommand{\project}[2]{(#2){\projectsymbol}_{#1}}

\newcommand{\bind}[2]{#1~$\leftarrow$~#2}
\newcommand{\mapschan}[2]{\ensuremath{#1\vDash#2}}

\def\pequiv{\dashv\vdash}
\def\wrt{w.r.t.~}
\definecolor{projcolor}{RGB}{0, 176, 80} 
\definecolor{protcolor}{RGB}{222, 71, 142} 
\definecolor{fencecolor}{RGB}{0, 176, 240} 

\newcommand{\setfont}[2]{{\fontfamily{#1}\selectfont #2}}

\newcommand{\savespace}{\vspace{-2mm}}
\def\dual{\ensuremath{\sim}}
\def\resid{S}
\def\chanvar{c}
\def\lchanvar{c}
\def\echanvar{\tilde{c}}
\def\rolevar{P}
\def\msgvar{\myit{m}}
\def\employ{\in}
\def\notemploy{\notin}
\def\eqdef{\mathrel{\stackrel{\makebox[0pt]{\mbox{\normalfont\tiny
          def}}}{=}}}

\newcommand{\wait}[2]{\code{join~{#1}~{#2}}}
\newcommand{\joinfunction}[2]{\code{join~{#1}~{#2}}}
\def\SAFE{\btt{SAFE}}
\def\FAIL{\btt{FAIL}}
\def\HOLE{\btt{HOLE}}
\def\ANS{\btt{ANS}}
\def\BLOCK{\btt{BLOCK}}
\def\YES{\btt{YES}}
\def\NO{\btt{NO}}
\def\safesend{\btt{S_{safe}}}
\def\saferecv{\btt{R_{safe}}}
\def\fid{\btt{FID}}
\def\safeempty{\btt{EMPTY}}
\def\error{error}
\def\abort{abort}
\newcommand{\notiferr}[1]{{\color{red}#1}}
\def\RACE{{\btt{RACE\_ERR}}}
\def\UNEXPPROT{{\btt{PROT{\_}ERR}}}
\def\RESERR{{\btt{RES{\_}ERR}}}
\def\LEAKERR{{\btt{LEAK{\_}ERR}}}

\def\idt{idt}
\def\idp{idp}
\def\idc{idc}

\newcommand\figref[1]{Fig. \ref{#1}}
\newcommand\secref[1]{Sec. \ref{#1}}
\newcommand\defref[1]{Def. \ref{#1}}
\newcommand\lemref[1]{Lemma \ref{#1}}
\newcommand\propref[1]{Prop. \ref{#1}}

\def\sessionlogic{Mercurius}

\newcommand{\explicitseq}[1]{
  \ifthenelse{\equal{RR}{#1} }{
    {\downdownharpoons} 
  }{
    \ifthenelse{\equal{SS}{#1} }{
      \upupharpoons 
    }{
      \ifthenelse{\equal{RS}{#1} }{
        \downupharpoons 
      }{
        \ifthenelse{\equal{SR}{#1} }{
          \updownharpoons
        }{

        }         
      }   
    }
  }
}

\newcommand{\gseq}[1]{\ensuremath{\seq^{\explicitseq{#1}}}}
\newcommand{\gseqA}[1]{\ensuremath{\seq}}
\newcommand{\guard}[1]{\ensuremath{{\ominus}(#1)}}
\newcommand{\assume}[1]{\ensuremath{{\oplus}(#1)}}
\newcommand{\assumeL}[2]{\ensuremath{\oplus(#1)}}

\newcommand{\transmitbase}[5]{
  \ifthenelse{\equal{#3}{} \AND \equal{#4}{}}{
    #1{#5}#2
  }{
    \ifthenelse{\NOT \equal{#3}{} \AND \equal{#4}{}}{
      #1{#5}#2:#3
    }{
      \ifthenelse{\NOT \equal{#4}{} \AND \equal{#3}{}}{
        #1{#5}#2:#4
      }{
        #1{#5}#2:#4{\langle}#3{\rangle}
      }
    }
  }
}

\newcommand{\transmit}[4]{\transmitbase{#1}{#2}{#3}{#4}{\rightarrow} }
\newcommand{\bringson}[2]{{#1}{\rightharpoonup}{#2} }
\newcommand{\delegate}[3]{\transmitbase{#1}{#2}{#3}{}{\xrightarrow{d}}}
\newcommand{\atransmit}[5]{\transmitbase{#1}{#2}{#3}{#4}{\xrightarrow{#5}}}
\newcommand{\msg}[2]{#1\,{\cdot}\,#2}
\newcommand{\omsg}[2]{#2}


\newcommand{\gdown}[3]{#1:#2 {\langle} DOWN(#3) {\rangle}}
\newcommand{\gawait}[3]{#1:#2 {\langle} AWAIT(#3) {\rangle}}
\newcommand{\pdown}[1]{down(#1)}
\newcommand{\pawait}[1]{await(#1)}

\newcommand{\gprot}[4]{
  \ifthenelse{\equal{#4}{}}{
    #1(#2,#3) } {
    #1(#2,#3) \leadsto @O(#4)}
}

\def\atid{\textit i}
\def\thdid{\textit t}
\def\atloc{\textit l}
\def\atzero{\bot}

\def\implies{\Rightarrow}

\newcommand{\CB}{CB}
\newcommand{\HB}{HB}
\newcommand{\SEQ}{SEQ}
\newcommand{\precx}[1]{\ensuremath{\prec_{#1}}}
\newcommand{\lt}[2]{#1\,{\prec}\,#2}
\newcommand{\ltset}[2]{merge(#1,\,#2)}
\newcommand{\ltcb}[2]{#1\!{\prec_{\CB}}#2}
\newcommand{\lthb}[2]{#1\!{\prec_{\HB}}#2}
\newcommand{\seqhb}[2]{#1\!{\prec_{\SEQ}}#2}
\newcommand{\lthbw}[2]{#1{\preceq_{\HB}}\,#2}
\newcommand{\lthbmay}[2]{#1{\prec^{maybe}_{HB}}\,#2}
\newcommand{\rl}[2]{\event{#1}{#2}}
\newcommand{\rrl}[3]{\atransmit{#1}{#2}{}{}{#3}}
\newcommand{\rk}[3]{\rl{#1}{#3}{:}{#2}}
\def\roles{\mathcal{R}}
\newcommand{\graph}[1]{
  \ifthenelse{\equal{#1}{}}{\mathcal{G}}{\mathcal{G}(#1)}}

\def\oplusseq{\otimes}
\def\oplusstar{\circledast}
\def\opluschoice{\oplus}

\def\roletyp{\mathcal{R}{{ole}}}
\def\eventtyp{\mathcal{E}{{vents}}}
\def\labeltyp{{{Nat}}}
\def\idtyp{nat}
\def\lchantyp{\mathcal{C}{{han}}}
\def\chantyp{\mathcal{{L}}{{chan}}}
\def\echantyp{\mathcal{{E}}{{ndpt}}}
\def\operationtyp{\mathcal{O}}

\def\ordering{\pure}
\newcommand{\chanmap}{\ensuremath{\mathrm{\Gamma}}}
\def\rolemap{K}
\def\locof{\mathcal{L}}
\def\loccompstar{\useq}
\def\loccompseq{\seq}
\def\loccompvee{\vee}

\def\composeseqop{h_1}
\def\composestarop{h_2}
\def\composeveeop{h_3}
\newcommand{\composeseq}[2]{{\composeseqop}(#1,#2)}
\newcommand{\composestar}[2]{{\composestarop}(#1,#2)}
\newcommand{\composevee}[2]{{\composeveeop}(#1,#2)}

\def\bordermergebase{\sqcup}
\newcommand{\bordermerge}[1]{\lfloor{#1}\rfloor}
\newcommand{\mergeseq}{\bordermerge{\loccompseq}}
\newcommand{\mergestar}{\bordermerge{\loccompstar}}
\newcommand{\mergevee}{\bordermerge{\loccompvee}}

\def\mergeset{merge\_adjacent}
\newcommand{\mergeseqtier}[2]{\mergeset(#1,#2)}
\newcommand{\mergestartier}[2]{merge\_star(#1,#2)}

\def\mfront{F}
\def\mback{B}
\newcommand{\morders}{\ensuremath{\mathrm{\Pi}}}
\def\mguards{A^{\ominus}}
\def\massum{A^{\oplus}}
\def\terminal{T}
\newcommand{\notop}[1]{\neg(#1)}
\def\orderas{\vartheta}
\def\racefreeas{\ensuremath{\mathrm{\Psi}}}
\def\os{\mathcal{S}}
\newcommand{\osbck}[1]{{#1}^{bt}}
\newcommand{\osfrt}[1]{{#1}^{ft}}
\newcommand{\osasm}[1]{{#1}^{\oplus}}
\newcommand{\osgrd}[1]{{#1}^{\ominus}}

\newcommand{\bordrmap}[1]{{#1}.\rolemap}
\newcommand{\bordcmap}[1]{{#1}.\chanmap}

\newcommand{\racefree}[2]{\ensuremath{#1 \,{\vDash_{RF}}\,  #2}}
\newcommand{\racefreerel}[2]{{\ifthenelse{\equal{#2}{}}{RF(#1)}{RF(#1,#2)}}}
\def\cbformula{BForm}
\def\cformula{EForm}
\def\eborder{\beta}
\def\tborder{\eborder^T}
\def\oborder{\eborder^E}

\definecolor{navyblue}{HTML}{336699}
\definecolor{denimblue}{HTML}{151B8D}
\newcommand\blueord[1]{\textcolor{denimblue}{#1}}
\newcommand\grayord[1]{\textcolor{gray}{#1}}

\def\fmsg{\St}
\newcommand\chanspec[3]{\chani{#1}{#2}{#3}}

\lstdefinestyle{informal}{
  belowcaptionskip=1\baselineskip,
  breaklines=true,
  showstringspaces=false,
  basicstyle=\footnotesize\ttfamily\color{blue},
  keywordstyle=\bfseries\color{blue},
  commentstyle=\itshape\color{blue},
  identifierstyle=\color{blue},
  stringstyle=\color{blue},
}

\newcommand{\openall}[3]{openall(#1,#2,#3)}
\newcommand{\opened}[3]{opened(#1,#2,#3)}
\newcommand{\closed}[1]{closed(#1)}
\newcommand{\emptyc}[2]{empty(#1,#2)}
\def\simpag{\textbackslash{\sep}}
\def\init{init}
\def\initall{initall}

\newcommand{\peer}[1]{Peer(#1)}
\newcommand{\bindv}[2]{bind(#1,#2)}


\newcommand{\EV}[1]{EV(#1)}
\newcommand{\TR}[1]{TR(#1)}
\newcommand{\first}[1]{TR^{fst}(#1)}
\newcommand{\sub}[1]{sub(#1)}

\newcommand{\safesnd}[4]{\safesend(#2, #4, #3, #1)}
\newcommand{\safercv}[2]{\saferecv(#2, #1)}
\newcommand{\safefid}[3]{\fid(#1, #2, #3)}
\newcommand{\safeemp}[2]{\safeempty(#1, #2)}

\def\sendfunction{send}
\def\recvfunction{recv}
\def\notifyAllfunction{notifyAll}
\def\waitfunction{wait}
\newcommand{\tsend}[1]{\sendfunction(#1)}
\newcommand{\notifyAll}[1]{\notifyAllfunction(#1)}
\newcommand{\mwait}[1]{\waitfunction(#1)}
\newcommand{\msend}[2]{\sendfunction(#1, #2)}
\newcommand{\trecv}[1]{\recvfunction(#1)}
\newcommand{\esend}[1]{{\sendfunction_e}(#1)}
\newcommand{\erecv}[1]{{\recvfunction_e}(#1)}
\newcommand{\tlbl}[1]{lbl(#1)}
\newcommand{\tchan}[1]{chan(#1)}
\newcommand{\tmsg}[1]{msg(#1)}
\newcommand{\mdisj}[2]{#1 \# #2}
\newcommand{\evpeer}[1]{#1.peer}
\newcommand{\evlbl}[1]{#1.id}
\newcommand{\cnil}{\btt{NULL}}
\newcommand{\cpar}{\|} 

\newcommand{\RF}{RF~}
\newcommand{\SO}{SO}
\newcommand{\ltseq}[2]{#1\!{\prec}#2}
\newcommand{\ord}[1]{{\ifthenelse{\equal{#1}{}}{seq}{seq(#1)}}}
\newcommand{\adj}[2]{Adj(#1,#2)}
\newcommand{\link}[2]{Adj^+(#1,#2)}
\def\proof{\vspace{-0.2cm}\noindent{\it Proof}\hspace*{0.5cm}}
\renewenvironment{proof}{\noindent{\it Proof:}}{\qedsymbol}
\newcommand{\satisf}[3]{\ensuremath{(#1,#2) \vDash #3}}
\newcommand{\Satisf}[2]{\ensuremath{#1 \vDash #2}}
\newcommand{\ISatisf}[3]{\ensuremath{#1,#2 \vDash #3}}
\newcommand{\Valid}[1]{\ensuremath{ \vDash #1}}

\def\smstack{\myit{s}}
\def\smheap{\myit{h}}
\newcommand{\node}[3]{\hformn{#1}{#2}{#3}}
\def\disjoint{\perp}
\def\mord{\myit{o}}
\newcommand{\eeval}[1]{[[#1]]}
\newcommand{\beval}[1]{[[#1]]}
\def\expr{e}
\def\pstate{\btt{PState}}
\def\PS{\btt{PS}}
\def\tconf{TConfig}
\def\TC{TC}
\def\sstore{SS}
\def\css{CSS}
\def\TS{\btt{IState}}
\def\LS{\btt{TState}}
\def\SS{\btt{State}}
\def\CH{CH}
\def\channel{chan}
\def\CC{\btt{CC}}
\def\cconf{\btt{CConfig}}
\def\cstate{CS}
\def\cstore{CStore}
\def\CS{\btt{CS}}
\def\Prot{\btt{Prot}}
\def\SStore{\btt{SStore}}
\def\tst{\sigma}
\def\ltst{\overline{\sigma}_L}
\def\itst{\overline{\sigma}_i}
\newcommand{\ttrans}{\ensuremath{\hookrightarrow}}
\newcommand{\mstate}[2]{{\langle}#1,#2{\rangle}}
\newcommand{\tstate}[3]{{\langle}#1,#2,#3{\rangle}}
\newcommand{\ltstate}[2]{{\langle}#1,#2{\rangle}}
\newcommand{\tconfig}[2]{{\langle}#1,#2{\rangle}}

\newcommand{\comp}[2]{comp(#1,#2)}

\newcommand{\appsem}[2]{ \ifthenelse{\equal{#1}{appendix}}{{#2}}{} }
\newbool{appendix}
\newbool{paper}

\begin{abstract}


Ensuring the correctness of software for communication-centric programs
is important but challenging. Previous approaches, based 
on session types, have been intensively investigated over the
past decade. They provide a concise way to express protocol specifications
and a lightweight approach for checking their implementation.
Current solutions are based on only implicit synchronization, and
are based on the less precise types rather than logical formulae.
In this paper, we propose a more expressive {\em session logic}
to capture multi-party protocols. By using two kinds of ordering constraints,
namely {\em ``happens-before''} \code{\precx{HB}} and 
{\em ``communicates-before''} \code{\precx{CB}},
we show how to ensure from {\em first principle} race-freedom over common
channels. Our approach refines each specification with
both assumptions and proof obligations to ensure compliance to some global
protocol. Each specification is then projected for each party and 
then each channel, to allow {\em cooperative proving} through localized
automated verification. 
Our primary goal in
automated verification is to ensure
race-freedom and communication-safety, but the approach is 
extensible for deadlock-freedom as well.
We shall also describe how modular protocols can be
captured and handled by our approach.

\end{abstract}

\maketitle

\lstdefinestyle{MCStyle}{
  language=[LaTeX]Tex,
  keywordstyle=\bfseries\color[rgb]{0,0,0},
  texcsstyle=*\color{black}
  captionpos=b,
  numbers=left,
  basicstyle=\scriptsize\normalfont,
  stepnumber=1,
  numbersep=10pt,
  tabsize=2,
  showspaces=false,
  showstringspaces=false,
  numbersep=0pt,
}

\lstdefinelanguage{MC}
{
	morekeywords={data,int},
	sensitive=false,
	morecomment=[l]{@},
	morecomment=[s]{/*}{*/},
	morestring=[b]"
}

\newcounter{ExampleNo}
\setcounter{ExampleNo}{0}
\newcommand{\setExampleNo}[1]{\refstepcounter{ExampleNo}\arabic{ExampleNo}\label{#1}}

\section{Introduction}
\hide{
  Plan:
  complex -> high level of concurrency
  communication

  relax the  Lamport’s “happens before” relation

  interlace of communicating models

  dynamic sync needed because
  (1) need to give an explicit semantic to ; which is not context-dependent - most often it
  imposes sequencing only on the actions of the same participant thus leading to a causal ordering
  which flows naturally from Lamport's “happens before” relation. The assumption of
  these works is the ... of asynchronous communication is solely govern by correct
  ordering of message passing (2) This assumption however is broken somehow in
  the current systems which require high level of concurrency (eg such as Android), performance
  and reliability. The concurrency in these systems 
  is obtained by using a combination of synchronization mechanism, which often
  do not exclude each other but are orthogonal to each other. For example, the Android
  systems recognized the performance benefits from highly using message passing,
  but
  are not shy in combining this style with
  other synchronization mechanisms such as CountDownLatch
  where needs be.

  cannot trust correctnes ==> need logic to capture values

  fix fonts in code either all italics or no italics at all

}


The notable performance and scalability of
our computing platform
can be attributed to the virtue of distributed computation.
In turn, distributed computation has seen a rise in the level
of concurrency with the adoption of the message passing model,
where concurrent programs communicate with each other via
communication channels.
Given the somewhat complex interaction schemes used,
ensuring the safety and
correctness 
of such system is challenging.
Some research progress 
towards specification-based development
for the message passing paradigm 
have mostly revolved around the session-based
concurrency. In these approaches, the communication patterns
are abstracted as session types for
variants of \code{\pi}-calculus
\cite{Honda:2008:MAS:1328897.1328472,Denielou:2011:DMS:1926385.1926435}
which are either statically checked against given specifications
or are being used to guide the generation of
deadlock-free code \cite{Carbone:2013:DMA:2480359.2429101}.

While providing a foundational approach towards communication
correctness, most of these works make the assumption
that the underlying system communicates exclusively via message passing
using unlimited channels. That is, extra channels are created
when necessary to resolve any ambiguity in the communication
protocol.
This assumption is broken 
in 
current systems which may 
rely on more tightly synchronized concurrency. Concurrency 
is often managed by using a combination of synchronization mechanisms
which are designed to work together 
for better performance over shared channels. For instance,
the Android systems recognize the benefits of
using message passing,
but are not hesitant in combining this style with
other synchronization mechanisms such as CountDownLatch.
Moreover, while sequential and disjunctive combinations of
communication channels are quite well understood, the same cannot
be said for concurrent compositions of transmissions over shared channels.

In summary, most of the current approaches 
which guarantee communication safety,
conveniently limit their results to implicit synchronization
of protocols. Should race-related
conflict arises, extra transmission channels 
are arbitrary added to ensure unambiguous communication.
In this paper, we consider how to abstractly extend
communication protocols to those which require 
{\em explicit synchronization}.
Our thesis is that specification in concurrency should be
abstract but with sufficient detail so 
that implementation can be safely written to meet
the intention of protocol designers. 
We aim for an
expressive logic for session-based concurrency; and 
a set of tools for automated reasoning of our enhanced
session logic. We consider how to ensure well-formedness
of session logic, and how to build summaries to
ensure race-freedom and type safety
amongst shared channels.

As an example, session types approaches only
impose sequencing 
on the actions of the same participant, 
as illustrated here:
~~~~\markred{{\code{(\transmit{A}{C}{}{})\,\seq\,(\transmit{B}{C}{}{}})}}.
This example involves three communicating parties with \code{C}
receiving data from \code{A} and \code{B}, respectively.
For simplicity, we assume asynchronous communication
with non-blocking send and blocking receive.
We further assume that messages are received in a FIFO
order in each of these communication channels.
In terms of communications, the two sends (by \code{A} and \code{B}) are
un-ordered, while the two receives by \code{C} are strictly
sequentialized (by \code{C} itself)
with the receipt of message from \code{A}
expected to occur before the message from \code{B}.
Such implicit synchronization may only 
work if two distinct 
channels
are being used to serve the transmissions of
\code{\transmit{A}{C}{}{}} and that of 
\code{\transmit{B}{C}{}{}}, respectively,
since the two sends by \code{A} and \code{B}  
would then arrive un-ambiguously,
irrespective of their arrival order over the two distinct channels,
say \code{\chanvar_1} and \code{\chanvar_2}, as shown next:
\markred{\code{(\transmit{A}{C}{}{\chanvar_1})\,\seq\,(\transmit{B}{C}{}{\chanvar_2})}}.
However, if we had to use
a common channel, say \code{\chanvar} for mailbox of party \code{C},
we would have the following
specification.

\centerline {\markred{\code{(\atransmit{A}{C}{t_1}{\chanvar}{\atid_1})\,\seq\,(\atransmit{B}{C}{t_2}{\chanvar}{\atid_2})}}}

This is perfectly legitimate depending on the
programming models used for
implementing the given protocol. For example,
given the following two code snippets,
the second implementation (b) of the above protocol
is safe, while the first one (a) contains a race potentially
leading to an unsafe communication:
\vspace{-1em}

\begin{figure}[h]
\scriptsize
\begin{subfloat}[][]    {
    \label{unsafe}
    $
\begin{array}{ l || l || l }
  \multicolumn{1}{c}{\code{A}} & \multicolumn{1}{c}{\code{B}} & \multicolumn{1}{c}{\code{C}} \\
  \code{...} & \code{...} & \code{...} \\
  \code{\msend{\chanvar}{"Lorelei"}} & \code{\msend{\chanvar}{60}} & \code{book~=~\trecv{\chanvar}} \\
                                     &  & \code{price~=~\trecv{\chanvar}} \\
                                     & &\\
\end{array}\hfill
$}\end{subfloat} \hfill
\begin{subfloat}[][]    {
    \label{safe}
    $
\begin{array}{ l || l || l }
  \multicolumn{1}{c}{\code{A}} & \multicolumn{1}{c}{\code{B}} & \multicolumn{1}{c}{\code{C}} \\
  \code{...} & \code{...} & \code{...} \\
  \code{\msend{\chanvar}{"Lorelei"}} &  & \code{book~=~\trecv{\chanvar}} \\
  \code{\textbf{\notifyAll{w}}} & \code{\textbf{\mwait{w}}} & \code{price~=~\trecv{\chanvar}} \\
  & \code{\msend{\chanvar}{60}} & \\
\end{array}
$}\end{subfloat}
\end{figure}
\vspace{-1em}

We also provide unique 
labels \code{\atid_1} and \code{\atid_2} for each transmission,
and its corresponding 
message types \code{t_1} and \code{t_2}.
We assume each party never sends
message to itself.
This allows us to use \code{\event{R}{\atid}} to unambiguously refer
to a send or receive event by party \code{R} at a
transmission labelled \code{\atid}.
We now have a situation where two senders 
(by \code{A} and \code{B}) must be
explicitly ordered to have their messages arrive in 
strict sequential order. (The two receives by the same
party \code{C} are already implicitly ordered.) It may even become a communication safety issue if
\code{t_1{\neq}t_2} and should the two messages arrive in
the wrong order. We refer to this problem as a {\em channel race}
where messages could be sent to unintended
destinations. In this case, we 
have to arrange
for the two sends (initiated by \code{A} and \code{B})
to be strictly ordered. For this to be supported, we propose
an {\em explicit synchronization}
mechanism that 
would force
the second send by \code{B} to occur after
the first send by \code{A} has been initiated.
While explicit synchronization 
can be handled by a number of 
mechanisms,
such as \code{notifyAll-wait} or 
\code{CountDownLatch}, our
specification logic shall abstractly
capture this requirement by an
 {\em ordering}, namely
\code{\lthb{\event{A}{\atid_1}\,}{\,\event{B}{\atid_2}}}, which requires
that event \code{\event{A}{\atid_1}} to happen before
event \code{\event{B}{\atid_2}}.


Such explicit orderings should 
be minimised, where possible,
but they are 
an essential component of concurrency control for the message
passing paradigm.
\hide{
As another twist to this example,
it may be that messages read by the  common receiver \code{C}
do not need to be aware of the senders, or that
the message itself has already
embedded its sender's information. Under this scenario,
there is no need for any sequentialization on the
two transmissions. We prefer to explicitly
capture this relaxation in our logic
by the following pair of
spatial transmissions:
\asay{should we use distinct channels here? since we dropped
  non-determinism}

\centerline {\code{(\atransmit{A}{C}{t_1}{\chanvar_c}{\atid_1})\,\star\,(\atransmit{B}{C}{t_2}{\chanvar_c}{\atid_2})}}

With this, the two messages, of types \code{t_1} and \code{t_2},
may now arrive in either order. We refer to this as
{\em non-deterministic} communication. To ensure message safety,
we further require a side-condition \code{t_1{=}t_2} that has
to be satisfied,
for every pair of non-deterministic communication over the same
channel. Apart from greater concurrency with less synchronization 
overheads, this side-condition ensures soundness
of messages received through such
communications
.
}
Our  
contributions are:
\begin{itemize}
\item We design \sessionlogic, an {\em expressive session logic} that is 
both precise and concise for modelling multi-party 
 protocols. 
 Our approach utilizes two fundamental ordering constraints.
\item We ensure {\em race-freedom} in communications over common channels
with both {\em implicit} and {\em explicit} synchronization.
Prior works, based on session types,
relied on only implicit synchronization.
\item We provide a {\em specification refinement} that
explicitly introduces both assumptions and proof obligations.
These may be either local or global events and proofs, but can be
used together
to ensure adherence to the global protocol,
communication safety and race-freedom. 
\item Each refined specification may be firstly projected 
for each party and then for each channel, with a set
of shared global assumptions. There
are three novelties. 
First, global proof obligations
may be projected 
to support {\em cooperative proving},
where the proof by one party can be relied
as assumption by the other concurrent parties. 
Second, we use {\em event guards}
to ensure that channel specifications are 
verified in correct sequence.
Lastly, multi-endpoints for channels are supported.
\item We show how {\em automated verification} is applied
on a per party basis with global assumptions and proof obligations.
Assumptions are released as soon as possible. Proof
obligations can be delayed and locally proven at the
appropriate time.
\item We propose an approach to express global protocols
in a modular fashion, 
where ``plug-and-play'' protocols are only designed and refined
  once,
  and re-used multiple times in different contexts. This greatly
facilitate specification re-use and forms the basis
for supporting both inductive protocol specification and
their recursive implementation.
\hide{
\item We show how to extend our approach to
{\em recursive} protocols.
\item We have implemented our proposal and show
 a set of small examples to confirm its viability.
}
\end{itemize}

{\noindent}{\bf{Limitation}:} For simplicity, we currently
restrict our protocol to well-formed disjunction
where a common sender and receiver is always
and exclusively present in each disjunction
to express communications across its branches.
This restriction provides a {\em simplification}
where every \code{\precx{HB}}
ordering discovered is a {\em must}-ordering, since transitivity
holds globally whenever it holds in one of the branches.
Lifting this restriction will require orthogonal
mechanisms to be supported, such as partial orderings 
and synchronization across
multiple parties in conditional branches.


\hide{
\begin{verbatim}
deadlock-freedom
delegation?
safety and well-formedness
explicit synchronization
  - with CDL?
  - semophore?
semantics
automated verification
experiments
produce:
    P(emptyCount)
    P(useQueue)
    putItemIntoQueue(item)
    V(useQueue)
    V(fullCount)
The consumer does the following repeatedly
consume:
    P(fullCount)
    P(useQueue)
    item ← getItemFromQueue()
    V(useQueue)
    V(emptyCount)

\end{verbatim}
}




\hide{
  Prof. Chin:
 - existing protocol are based on implicit synch
    for example   A->C; B->C involve 3 parties
   with C receiving two sends from A and B, respectively.
  In terms of communications, the two sends (by A and B) are
  un-ordered, while the two receives by C are strictly
  sequentialized (by C itself) with the receipt of message from A
  occurring before the message of B.
  Such implicit synchronization may work if two bi-directional channels
  are being used by both A->C and B->C, respectively,
  since the two sends by A and B  can arrive un-ambiguously,
  in any 
  order, in two distinct channels. 
  However, if we had assumed the use of 
  a mailbox channel (asynchronous buffered channel)
  for C instead that supports multiple senders but a single
  receiver, we may wish for the two sends (by A and B)
  to be strictly ordered; so as
  For this to be supported, we propose the following
  specification A->C ;_{SS} B->C using an operator
  with explicit synchronization, ;_{SS}, which ensures that
  the send in predecessor communication, always occur
  before the send of the successor communication.

  More generally, to support such explicit synchronization,
  we say propose the use of a new operator, ;_M,
  with a set of explicit synchroziations M whose
  members are either:
   (i) SS - for ordered sends
   (ii) RR - for ordered receives
   (iii) RS -for receive of predecessor to happen before the send of successor communication),
  (iv) SR - for sends of predecessor communicattio to happen before the receive of successor communication.

for the send of predewith strict may have multiple which  un-ordered
 A, Bthe two comm are independent of each other.
    as a result, session type allows these pair of 
    communication to be concurrently scheduled.
   this begs the question what sequentializatio
   is being enforced by seq operaotr ":"
 - in IECCS15, we introduce an explicit spatial
   conjunction operator to denote independence
     A->B * C-> D
   where the two pairs of communication are
   explicitly specified to be mutually independent of each other.

 - in the case of comm: A->B; C->D
   Asynchronous comm of blocking receive currently guarantees that
   the send of A occurs before the receive of B; and
   similarly the send of C occurs before the receive of D.
 - there is currently no strict relation between the two sends,
   A and C,
   neither is there an explicit relation between the two receipts
   at B and D
   
}


\renewcommand\labelitemi{\tiny$\bullet$}
\section{Overview}
\label{sec.overview}
To capture a broad range of 
communication patterns, we propose an
expressive logic for specification of global protocols
in \figref{fig.global_syntax_prot}. 
This protocol uses
\code{\prot_1\,{\useq}\,\prot_2} for
the concurrency of \code{\prot_1} and \code{\prot_2}, and
\code{\prot_1\,{\vee}\,\prot_2} for disjunctive choice between
either \code{\prot_1} and \code{\prot_2},
and finally \code{\prot_1\,{\gseqA{}}\,\prot_2} on
the implicit sequentialization of \code{\prot_1} before \code{\prot_2}
for either the same party or same channel communication.
Moreover, we express the message type of a transmission
\code{\atransmit{S}{R}{v\,{\cdot}\,\fmsg}{c}{\atid}}
by $\code{{v\,{\cdot}\,\fmsg}}$
which expects a message \code{v} on some resource
expressed in logical form \code{\fmsg} (defined in \figref{fig.global_syntax_msg})
transmitted over channel \code{c}.
As an example, \code{v\,{\cdot}\,\view{ll}{v,n}_{\frac{1}{2}}} would capture
the transmission of half fractional permission\footnote{Fractional permissions
are important for supporting concurrent programming.
Its use allows read access when a resource
is shared by multiple parties, but also
write access when a resource
later becomes exclusively owned.} of a linked list with
\code{n} elements
\code{\view{ll}{v,n}} rooted at \code{v}. 
Apart from messages in logical form, our 
session logic 
captures 
logical disjunction and 
can enforce explicit synchronization to
ensure communication safety.


We provide a specification refinement
that adds unique transmission labels \codeS{i},
guards (of the form
\code{\guard{\racefreeas}}) and assumptions (of
the form \code{\assume{\racefreeas}}), as more precise
denotation of each global protocol. Guards
capture assertions (or proof obligations), such as
explicit synchronization, for its
global protocol to ensure race-free communications;
while assumptions
support flow of information 
across multiple parties to facilitate
local verification. To sketch the main
purpose of using these guards and assumptions,
we briefly introduce some
elements of \code{\racefreeas}
later in this section.

These new specification
constructs are added automatically by our 
refinement procedure
to help ensure race-freedom in shared channels.
Our refinement 
could also 
add channel
information based on some expected communication patterns.
For example, if mailbox communication
style is preferred, each communication always directs its
message to the mailbox of its receiver. 
If bi-directional channels of communication are preferred,
a distinct channel is provided
for every pair of
parties that are in communication.
To support diversity, we 
assume channels are captured 
explicitly in our
protocol specifications.
\wnnay{change R (Role) to P (Party)?}
\begin{figure*}
\centering
\captionsetup[subfigure]{farskip=1pt,captionskip=1pt}  
\setcounter{subfigure}{0}
\subfloat[][A Logic for Global Protocol Specification]
{\footnotesize
  \label{fig.global_syntax_prot}
  $ 
  \arraycolsep=1pt
  \begin{array}{llll}
 \textit{Single ~ transmission} & \code{\terminal}& {~ ::=}&  \code{\atransmit{S}{R}{v\,{\cdot}\,\fmsg}{\chanvar}{\atid}}\hspace*{4mm}\\
  \textit{Global~protocol}& \code{\prot
                   } &{~ ::=} &\code{~~~ \terminal}\\
  \textit{Concurrency} &&& ~|~ \code{\prot\,{\useq}\,\prot} \\
  \textit{Choice} &&& ~|~ \code{\prot\,{\vee}\,\prot } \\
  \textit{Sequencing}
  &&& ~|~ \code{\prot\,{\gseq{}}\,\prot} \\
  \textit{Guard} &&& ~|~ \code{\guard{\racefreeas}} \\
  \textit{Assumption} &&& ~|~ \code{\assume{\racefreeas}} \\
  \textit{Inaction} &&&   ~|~ {\emp}
  \\ \\
  \code{\text{\emph{(Parties)}}} ~&&& \multicolumn{1}{l}{\code{S,R,\rolevar \in \roletyp}} \\
  \code{\text{\emph{(Channels)}}}~&&& \multicolumn{1}{l}{\code{\lchanvar \in \lchantyp }} \\
  \code{\text{\emph{(Labels)}}}~&&& \multicolumn{1}{l}{\code{{\atid} \in \labeltyp }} \\
  \code{\text{\emph{(Messages)}}}~&&&\multicolumn{1}{l}{\code{v\,{\cdot}\,\fmsg ~\text{~~~s.t.~}~ ~v \in \Var }}
  \end{array}
  $}
\hspace*{10mm}
\subfloat[][The (Program) Specification Language]{\footnotesize
  \label{fig.global_syntax_msg}
$\arraycolsep=1pt
\begin{array}{llll}
  \textit{Symb.~ pred.} & \myit{pred} &\!\!\!\!::=& \code{\view{\myit{p}}{\self, v^*}~{\veq}~ \constr
\mid \view{\gp}{\rolevar^*,\lchanvar^*}~{\veq}~\prot}
  \qquad

  \\
  \textit{Formula} &   \constr & \!\!\!\!::=& \code{\bigvee \St \qquad  \St  ::= \exists ~ v^*{\cdot}\heap{\wedge}\pure{\wedge}\racefreeas 
~|~ \St * \St  } 
  \\
  \textit{Separation} & \heap &\!\!\!\!::=& \code{\emp 
      ~|~ \mapview{v}{d}{v^*}
      ~|~ \view{\myit{p}}{v^*}
      ~|~ \chani{\chanvar}{\rolevar}{\proj}
      ~|~ \heap {\sep} \heap
    ~|~ \hov}
  \quad
  \\
  \textit{Pure} &\pure &\!\!\!\!::=& \code{v:t 
  \mid \myit{b} \mid \myit{a} \mid 
\pure {\wedge} \pure \mid \pure {\vee} \pure \mid \neg \pure } \\
  &&&
  \mid \exists v \cdot \pure \mid \forall v \cdot \pure \mid \ptr
  \\
  \textit{Ptr~ eq./diseq.}& \ptr &\!\!\!\!::=& \code{v {=} v ~|~ v{=}\nil ~|~ v{\neq}v ~|~ v{\neq}\nil}
  \\
  \textit{Boolean} & \myit{b} &\!\!\!\!::=& \code{~\!\true \mid \false 
  \mid \myit{b} \!=\! \myit{b} \qquad
  \myit{a} ::= \!\a {=} \a
  \mid \a {\leq} \a
}
  \\
  \textit{Presbg.~ arith.} & \a &\!\!\!\!::=&~ 
  \code{\!k^{\int} \mid v \mid k^{\int}{\times}\a
  \mid \a \!+\! \a 
\mid - \a }  \\
  &&& \\
  &\text{where} && \code{ k^{\int}} \text{~: ~integer~constant;}~\code{d} \text{~:~data~ structure}\\
  &&& \code{\hovars \text{~:~ second-order~ variable};~\rolevar} \text{~:~ session~ role}\\
  &&& \code{L} \text{~:~local~protocol~(defined~in~\figref{fig.proj_syntax})}\\
  &&& \code{\racefreeas} \text{~:~race-free~assertions~(defined~in~\figref{ALsyntax})}
\end{array}
$}

\hrule
\vspace*{-3mm}
\caption{\sessionlogic~}
\label{fig.global_syntax}
\vspace*{-3mm}
\end{figure*}

Our approach thus takes in a protocol
specified in an expressive logic, {\em refine} it,
then {\em project} it on a per party and per channel basis.
Once these are done, we can proceed to apply automated
verification on the entire program code using the
derived specs.

\hide{
 \begin{figure}[htb]
\vspace*{-3mm}
  \begin{center}
    \includegraphics[scale=0.45]{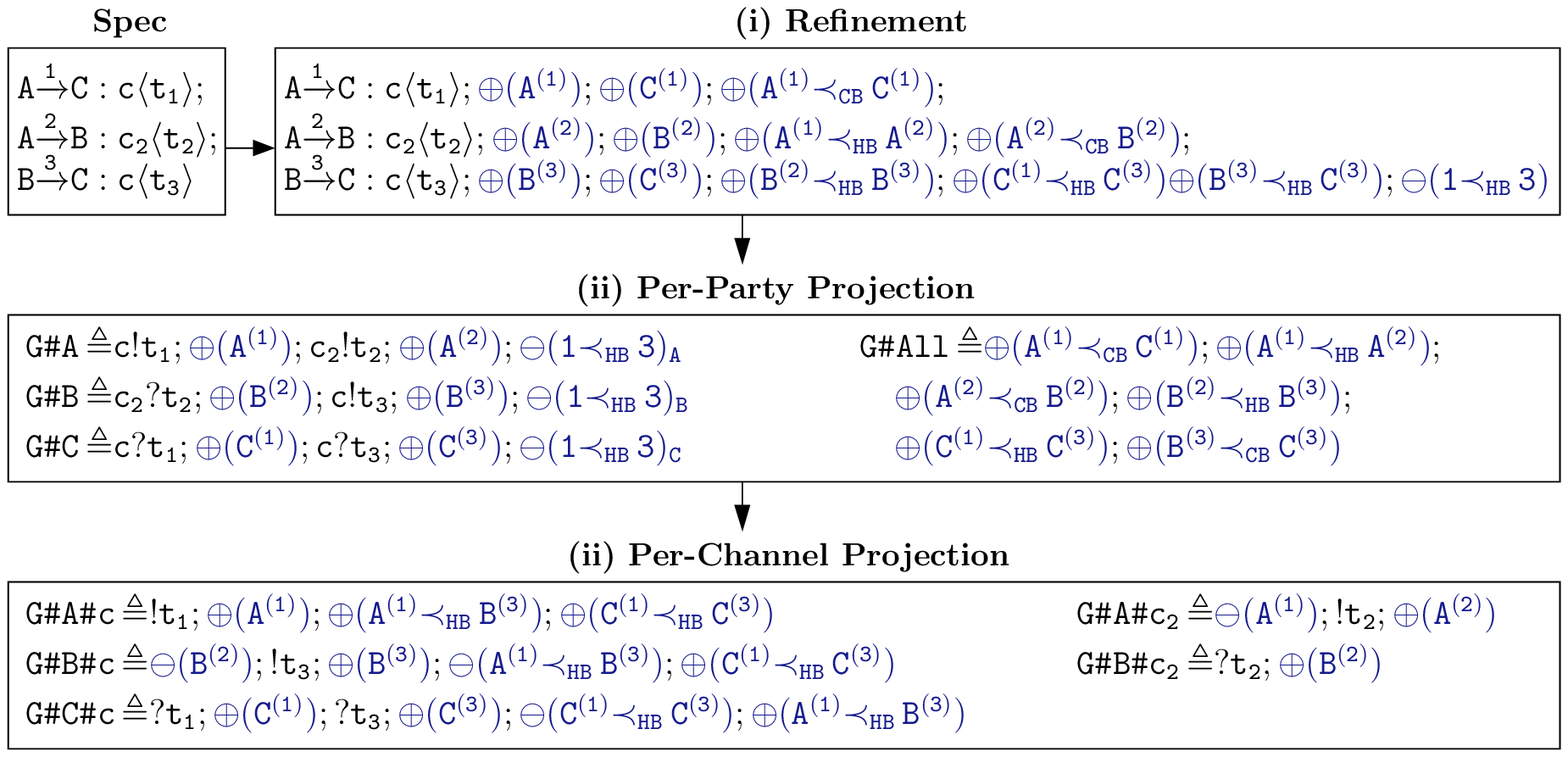}
  \end{center}
  \vspace*{-3mm}
  \caption{Overview on Spec Refinement and Projections}
\label{fig:ExampleDiagram}
\vspace*{-3mm}
\end{figure}

As a simple example, consider the following
specification over three parties \code{A, B} and \code{C}
which performed three transmissions.
Two of these transmissions
are over
a common channel \code{\chanvar} while a third one
uses another channel \code{\chanvar_2}, as shown below.

\centerline {\code{(\atransmit{A}{C}{t_1}{\chanvar}{1})\,\seq\,(\atransmit{A}{B}{t_2}{\chanvar_2}{2}) \,\seq\,(\atransmit{B}{C}{t_3}{\chanvar}{3}) }}

This example relies on only implicit 
synchronization to ensure race-freedom.
Our approach would refine the global specification with
assumptions and proof obligations, as shown in Fig~\ref{fig:ExampleDiagram}.
It would then project it on per-party and then per-channel basis,
prior to code verification. The refinement contains an
obligation to 
prove race-freedom between the \code{1^{st}} and \code{3^{rd}}
transmissions. This proof obligation \code{\guard{\lthb{1}{3}}} is broken into three components
\code{\guard{\lthb{1}{3}}_{A} {\sep} \guard{\lthb{1}{3}}_{B} {\sep} \guard{\lthb{1}{3}}_{C}}
for it to be cooperatively proven between the
three parties. In fact, we only need to prove the relevant obligations
at the sender and receiver of the  \code{3^{rd}} transmission only, 
as shown below.

\hide{
\begin{verbatim}
-(1 <_HB 3)_B = -(A^1 <_HB B_3);+(C^1 <_HB C^3)
-(1 <_HB 3)_C = -(C^1 <_HB C^3);+(A^1 <_HB B_3)
\end{verbatim}
}

{\footnotesize
  \centering
  \vspace*{-2mm}
  \arraycolsep=1pt
  \[\vspace*{-2mm}
    \begin{array}{lll}
      \code{\guard{\lthb{1}{3} }_{B}} & \code{=} &
      \code{\guard{\lthb{\event{A}{1}}{\event{B}{3}}} \seq
      \assume{\lthb{\event{C}{1}}{\event{C}{3}}} }
      \\
      \code{\guard{\lthb{1}{3} }_{C}} & \code {=} &
      \code{
      \guard{\lthb{\event{C}{1}}{\event{C}{3}}} 
      \seq
      \assume{\lthb{\event{A}{1}}{\event{B}{3}} } 
      }
    \end{array}
  \]
}

Our cooperative proof here is novel.
It allows the proof outcome(s) of one party
to be assumed by the other parties. 
As party \code{A} did not have to prove anything,
it could just take the entire proof obligation
(proven by \code{B} and \code{C} instead),
as assumption. Thus:

\hide{
\begin{verbatim}
-(1 <_HB 3)_A = +(A^1 <_HB B_3);+(C^1 <_HB C^3)
\end{verbatim}
}

{\footnotesize
  \centering
  \vspace*{-2mm}
  \arraycolsep=1pt
  \[\vspace*{-2mm}
    \begin{array}{lll}
      \code{\guard{\lthb{1}{3} }_{A}} & \code{=} &
      \code{\assume{\lthb{\event{A}{1}}{\event{B}{3}}} \seq
      \assume{\lthb{\event{C}{1}}{\event{C}{3}}} }
    \end{array}
  \]
}

Furthermore,
the proof obligations
to ensure race freedom for this example
can be entirely proven by the 
orderings that hold implicitly.
\hide{
as highlighted by
the following two implications, where the relation 
 \code{\precx{CB}}  denotes an expected {\cb} ordering between the sending and receiving events in a communication.

\hide{
\begin{verbatim}
(A^1 <_CB C^1)/\ (A^1 <_HB A^2)/\ (A^2 <_CB B^2) /\ 
  (B^2 <_HB B^3) /\ (C^1 <_HB C^3) /\ (B^3 <_CB C^3) 
  |- (A^1 <_HB B_3)
(A^1 <_CB C^1)/\ (A^1 <_HB A^2)/\ (A^2 <_CB B^2) /\ 
  (B^2 <_HB B^3) /\ (C^1 <_HB C^3) /\ (B^3 <_CB C^3) 
  |- (C^1 <_HB C^3)
\end{verbatim}
}

{\footnotesize
  \centering
  \vspace*{-2mm}
  \arraycolsep=1pt
  \[\vspace*{-2mm}
    \begin{array}{lll}
     \begin{array}{l} 
      \code{(\ltcb{\event{A}{1}}{\event{C}{1}}) {\wedge} 
       (\lthb{\event{A}{1}}{\event{A}{2}}) {\wedge}
       (\ltcb{\event{A}{2}}{\event{B}{2}}) {\wedge}
      } \\
      \code{
       (\lthb{\event{B}{2}}{\event{B}{3}}) {\wedge}
       (\lthb{\event{C}{1}}{\event{C}{3}}) {\wedge}
       (\ltcb{\event{B}{3}}{\event{C}{3}})
      } 
      \end{array}
      & \code{\implies} &
      \code{\lthb{\event{A}{1}}{\event{B}{3}}}
      \\
    \begin{array}{l} 
      \code{(\ltcb{\event{A}{1}}{\event{C}{1}}) {\wedge} 
       (\lthb{\event{A}{1}}{\event{A}{2}}) {\wedge}
       (\ltcb{\event{A}{2}}{\event{B}{2}}) {\wedge}
      } \\
      \code{
       (\lthb{\event{B}{2}}{\event{B}{3}}) {\wedge}
       (\lthb{\event{C}{1}}{\event{C}{3}}) {\wedge}
       (\ltcb{\event{B}{3}}{\event{C}{3}})
      } 
      \end{array}
      & \code{\implies} &
      \code{\lthb{\event{C}{1}}{\event{C}{3}}}
     
    \end{array}
  \]
}

To prove the former implication we use the propagation lemmas (from the next
section), first the lemma \code{(2)} as:

{\footnotesize
  \centering
  \arraycolsep=1pt
  \[
    \begin{array}{lll}
      \code{\assume{\ltcb{\event{A}{2}}{\event{B}{2}}} {\wedge}
       \assume{\lthb{\event{B}{2}}{\event{B}{3}}}} & {\implies} &
      \code{
      \assume{\lthb{\event{A}{2}}{\event{B}{3}}} }
    \end{array}
  \]
}

\noindent
and then the transitivity lemma as:

{\footnotesize
  \centering
  \arraycolsep=1pt
  \[
    \begin{array}{lll}
      \code{\assume{\lthb{\event{A}{1}}{\event{A}{2}}} {\wedge}
       \assume{\lthb{\event{A}{2}}{\event{B}{3}}}} & {\implies} &
      \code{
      \assume{\lthb{\event{A}{1}}{\event{B}{3}}} }
    \end{array}
  \]
}

\hide{
Note that we used a stronger definition of the 
guard \code{\guard{\lthb{1}{3}}} to simplify the illustration.
}
}
In general, some explicit
synchronization may be needed to ensure the communication
safety and race-freedom. This need
for explicit synchronization is 
orthogonal to our approach as our
requirement is expressed abstractly
using \code{\precx{HB}} ordering constraints.

}


\wnnay{Florin : Can help typeset and improve above}
\wnnay{Florin: Can we have a diagrammatic overview of
our approach by a very simple example. Highlight (i) specification refinement (ii) spec projection (iii) automated verication}








The automated verification is conducted against a general and
expressive specification language based on Separation Logic
\cite{reynolds2002separation,Chin:2012:AVS:2221987.2222283},  as shown
in ~\figref{fig.global_syntax_msg}, 
which supports the use of (inductive) predicates,  the specification of separation and numerical properties, as well as channel predicates.

\subsection{Specification Refinement}

Let us use a well-known 
example
to illustrate the key ideas of our approach.
Consider the multi-party
protocol from \cite{Honda:2008:MAS:1328897.1328472} where we use a
distinct channel as mailbox of each of the
three parties, namely seller \code{S},
first buyer \code{B_1} and the second collaborative buyer \code{B_2}.
\hide{
\begin{verbatim}
B1->S:s(Order);
(S->B1:b(Price) * S->B2:b(Price));
(B1->B2:b(Amt));
(B2->S:s(No) \/ B2->S:s(Yes);B2->S:s(Addr))
\end{verbatim}
}



{\footnotesize
  \centering
  \vspace*{-2mm}
  \arraycolsep=1pt
  \[\vspace*{-2mm}
    \begin{array}{l}
      \code{
      \transmit{B_1}{S}{\omsg{v}{Order}}{s}~ \seq~} 
      \code{(\transmit{S}{B_1}{\omsg{v}{Price}}{b_1} \useq
      \transmit{S}{B_2}{\omsg{v}{Price}}{b_2} )~ \seq~~} 
      \code{\transmit{B_1}{B_2}{\omsg{v}{Amt}}{b_2} ~\seq~ } \\
      \code{ (\transmit{B_2}{S}{\omsg{vR}{No}}{s}
      \vee
      ( \transmit{B_2}{S}{\omsg{v}{Yes}}{s} ~\seq~
      \transmit{B_2}{S}{\omsg{v}{Addr}}{s})
      )
      }
    \end{array}
  \]
}

\noindent
Note that \code{T} is a
short-hand for \code{\msg{v}{T(v)}} where
\code{T} is a type or predicate for 
some resource, such as
\sm{\code{Order}}, \sm{\code{Price}}, \sm{\code{Amt}} or \sm{\code{Addr}}.

For the above example, our refinement would add
information to signify the timely completion
of each communication completed using
\code{\assume{\atransmit{S}{R}{}{}{\atid}}}.  
This event itself
captures two locally observable events and a
globally observable ordering, as follows:
{\footnotesize
\[
  \code{\assume{\atransmit{S}{R}{}{}{\atid}}} ~{\implies}~ \code{\assumeL{\event{S}{\atid}}{R}~{\wedge}~\assumeL{\event{R}{\atid}}{R}}~{\wedge}~\code{\assume{\ltcb{\event{S}{\atid}}{\event{R}{\atid}}}}
\]
}
\noindent
which indicates the occurrence of the send
and receive events
and the expected {\cb} 
ordering\footnote{Under race-free assumption,
\code{\precx{CB}} ordering is the same as \code{\precx{HB}}-ordering.},
denoted by \code{\precx{CB}}, for its
successfully completed communication.
The send and receive events are only observable locally.
For example, the completion of event \code{\assumeL{\event{\rolevar_1}{\atid}}{S}}
is only observable in party \code{\rolevar_1}, but not by other distinct
parties, such as \code{\rolevar_2}. Its occurrence in the global protocol
is thus location sensitive and can be used to
capture relative orderings within the same party.
However, the ordering
\code{\assume{\ltcb{\event{S}{\atid}}{\event{R}{\atid}}}} is globally observable
since it relates two parties and can
rely on our assurance
that all send/receive events are uniquely labelled.

We also generate immediate \code{\precx{HB}} orderings
between each pair of adjacent events of the same party
in the global protocol, as follows:
{\footnotesize
\[
\code{..\assumeL{\event{P}{\atid_1}}{P}.. ~\seq~
  ..\assumeL{\event{P}{\atid_2}}{P}..} \implies
\code{\assume{\lthb{\event{P}{\atid_1}}{\event{P}{\atid_2}}}} 
\]
}
%
%
\indent
When common channels are involved, we
also use proof obligations of the form
\code{\guard{\lthb{\atid_1}{\atid_2}}} to denote two transmissions,
labelled \code{\atid_1} and \code{\atid_2}, that are 
expected to be ordered to ensure race freedom.
Assuming each transmission \code{\atid} be denoted
by \code{\atransmit{S_i}{R_i}{v{\cdot}\fmsg_i}{c}{\atid}}
with \code{\chanvar} as common channel,  
these proof obligations can be
further reduced, as follows:
{\footnotesize
\[
\arraycolsep=1pt
\begin{array}{ll}
  \code{\guard{\lthb{\atid_1}{\atid_2}}}  &
  \code{{\equiv}~
  \guard{\lthb{\event{S}{\atid_1}_1}{\event{S}{\atid_2}_2}}\,{\wedge}\,
  \guard{\lthb{\event{R}{\atid_1}_1}{\event{R}{\atid_2}_2}}} \\
\end{array}
\]
}
\indent
The \code{\lthb{}{}} is overloaded for both transmissions
e.g. \code{\lthb{\atid_1}{\atid_2}} 
and events e.g. \code{\lthb{\event{P}{\atid_1}_1}{\event{P}{\atid_2}_2}}. 
Our global specification \code{G} is now refined to \code{G@Refine} as shown below.
\hide{
From
the definition of \code{\guard{\lthb{\atid_1}{\atid_2}}}, there are
two different ways of ensuring
\code{\precx{HB}} ordering over two
sequential transmissions sharing the same channel.
The former condition is the minimum one required to ensure
race-freedom. The latter condition is a stronger condition
that implies the former, but needs only a
single synchronization object (that blocks at both \code{S_2} and
\code{R_2} waiting for \code{R_1} to be completed),
as opposed to two synchronization objects
(that blocks at \code{S_2} waiting for \code{S_1}
and at \code{R_2} waiting for \code{R_1}, respectively) in the former condition.}
\hide{The definition of \code{\guard{\atid_3{\useq}\atid_4}} ensures that messages over shared
channel in non-deterministic communication are logically equivalent to each other,
up to the abstraction level used to ensure protocol correctness.}

{\footnotesize
  \centering
  \vspace*{-2mm}
  \arraycolsep=1pt
  \vspace*{-2mm}
      \[
    \begin{array}[t]{l}
     \code{\prot{\at}Refine} \code{\defeq} 
      \code{
      \atransmit{B_1}{S}{\omsg{v}{Order}}{s}{1}~ \seq~
      \blueord{\assume{\atransmit{B_1}{S}{}{}{1}}~ \seq}} \\
      \code{((\atransmit{S}{B_1}{\omsg{v}{Price}}{b_1}{2} ~\seq~
      \blueord{\assume{\atransmit{S}{B_1}{}{}{2}}
      {\seq}\assume{{\lthb{\event{S}{1}}{\event{S}{2}}}}  
       {\seq}\assume{{\lthb{\event{B_1}{1}}{\event{B_1}{2}}}}  
       })}\\
     \code{~ \useq~
      (\atransmit{S}{B_2}{\omsg{v}{Price}}{b_2}{3} ~
      \seq~ \blueord{\assume{\atransmit{S}{B_2}{}{}{3}}
      {\seq}\assume{{\lthb{\event{S}{1}}{\event{S}{3}}}}  
      })) ~\seq~} 
      \\
     \code{\atransmit{B_1}{B_2}{\omsg{v}{Amt}}{b_2}{4} ~\seq~
      \blueord{\assume{\atransmit{B_1}{B_2}{}{}{4}}
      {\seq}\assume{{\lthb{\event{B_1}{2}}{\event{B_1}{4}}}}  
      {\seq}\assume{{\lthb{\event{B_2}{3}}{\event{B_2}{4}}}}  
      {\seq} \guard{\lthb{3}{4}}}   ~\seq} \\
     \code{ ((\atransmit{B_2}{S}{\omsg{v}{No}}{s}{5} ~\seq~
      \blueord{\assume{\atransmit{B_2}{S}{}{}{5}}
      {\seq}\assume{{\lthb{\event{B_2}{4}}{\event{B_2}{5}}}}  
      {\seq}\assume{{\lthb{\event{S}{2}}{\event{S}{5}}}}  
      {\seq}\assume{{\lthb{\event{S}{3}}{\event{S}{5}}}} ~\seq~ }}
  \\
     \code{ \blueord{\qquad\qquad\qquad\qquad ~~ \guard{\lthb{1}{5}}})
     }\\
      \!\!~\vee~ \code{\;\,( \atransmit{B_2}{S}{\omsg{v}{Yes}}{s}{6} ~\seq~
      \blueord{\assume{\atransmit{B_2}{S}{}{}{6}}
      {\seq}\assume{{\lthb{\event{B_2}{4}}{\event{B_2}{6}}}}  
      {\seq}\assume{{\lthb{\event{S}{2}}{\event{S}{6}}}}  
      {\seq}\assume{{\lthb{\event{S}{3}}{\event{S}{6}}}}  ~\seq}}
  \\\code{\  \qquad\qquad\qquad\qquad ~~~~
        \blueord{~\guard{\lthb{1}{6}}~\seq~} 
      }\\
      \code{\!\qquad
      \atransmit{B_2}{S}{\omsg{v}{Addr}}{s}{7} ~\seq~ 
      \blueord{\assume{\atransmit{B_2}{S}{}{}{7}}
      {\seq}\assume{{\lthb{\event{B_2}{6}}{\event{B_2}{7}}}}  
      {\seq}\assume{{\lthb{\event{S}{6}}{\event{S}{7}}}}  
      } ~\seq~
      \blueord{\guard{\lthb{6}{7}}}
      )
      )
      }
    \end{array}
  \]
}

\subsection{Specification Projection}
\wnnay{projection \& automated verification?
Does projection require some global assume/spec that become
available gradually or immediately. For the latter,
is it sound to do so?}

Once we have a specification that has been annotated with assumptions
and obligations, we can proceed to project them for use in
code verification. The projection can be done in two phases. Firstly, on
a per-party basis, and secondly on
a per-channel basis. A per-channel within each party
specification allows us to observe all communication
and proof obligation 
activities within each channel. We propose to
undertake per-channel
projections on specifications for two
reasons. Firstly, it supports backward compatibility, as an earlier
two-party session logic verification system,
\cite{Florin:IECCS15}, was designed based on per-channel
specifications. Secondly, it is more natural
for communication primitives, such as \code{send}/\code{receive},
to be denoted by their channel's specifications.
Lastly, we also 
project assumptions on
global orderings, 
so that they may be 
utilized by each party for their respective
local verification.

Let us first discuss the specification that would
be projected for each party. For the running example,
we would project the following local
specifications for the three parties.

{\footnotesize
  \centering
  \vspace*{-2mm}
  \arraycolsep=1pt
  \[
    \begin{array}{ll}
      \code{\prot{\prj}B_1{\defeq}} & 
      \code{\sendc{s}{}{Order} \seq
      \blueord{\assume{\event{B_1}{1}}} \seq
      \recvc{b_1}{}{Price} \seq
      \blueord{\assume{\event{B_1}{2}}} \seq
      \sendc{b_2}{}{Amt} \seq                          
      \blueord{\assume{\event{B_1}{4}}} \seq
      \blueord{\guard{\lthb{3}{4}}_{B_1}}. } \\
      \code{\prot{\prj}B_2{\defeq}} & 
      \code{\recvc{b_2}{}{Price} \seq
      \blueord{\assume{\event{B_2}{3}}} \seq
      \recvc{b_2}{}{Amt} \seq
      \blueord{\assume{\event{B_2}{4}}} \seq
      \blueord{\guard{\lthb{3}{4}}_{B_2}}  \seq}\\
      &\code{
       ((
        \sendc{s}{}{No} \seq                                             
        \blueord{\assume{\event{B_2}{5}}} \seq
        \blueord{\guard{\lthb{1}{5}}_{B_2}}
        ) ~\vee}
      \code{
        \;\,(\sendc{s}{}{Yes} \seq                                             
         \blueord{\assume{\event{B_2}{6}}} \seq
         \blueord{\guard{\lthb{1}{6}}_{B_2}} \seq}\\
      & \code{
         \sendc{s}{}{Addr} \seq                                             
         \blueord{\assume{\event{B_2}{7}}} \seq
         \blueord{\guard{\lthb{6}{7}}_{B_2}}
        )
       ).
     } 
      \\
      \code{\prot{\prj}S{\defeq}} & 
      \code{\recvc{s}{}{Order} \seq
      \blueord{\assume{\event{S}{1}}} \seq
      (\sendc{b_1}{}{Price} \seq
      \blueord{\assume{\event{S}{2}}})
      \useq
      (\sendc{b_2}{}{Price} \seq
      \blueord{\assume{\event{S}{3}}}) \seq
      \blueord{\guard{\lthb{3}{4}}_S}  \seq }\\
      &\code{
       ((
        \recvc{s}{}{No} \seq                                             
        \blueord{\assume{\event{S}{5}}} \seq
        \blueord{\guard{\lthb{1}{5}}_S}
        ) ~\vee}\\
      &\code{
        \;\,(\recvc{s}{}{Yes} \seq                                             
         \blueord{\assume{\event{S}{6}}} \seq
         \blueord{\guard{\lthb{1}{6}}_S} \seq
         \recvc{s}{}{Addr} \seq                                             
         \blueord{\assume{\event{S}{7}}} \seq
         \blueord{\guard{\lthb{6}{7}}_S}
        )
       )
     }   
      
    \end{array}
  \]
}
\hide{
\begin{verbatim}
B1:
 B1!s;+(B1^1);B1?(b1);+(B1^2);
 B1!b2;+(B1^4);-(3 <_HB 4)_{B1};

B2:
 B2?b2;+(B2^3);
 B2?b2;+(B2^4);-(3 <_HB 4)_{B2};
(B2!s;+(B2^5);-(1 <_HB 5)_{B2};
 \/B2!s;+(B2^6);-(1 <_HB 6)_{B2};B2!s;+(B2^7)-(6 <_HB 7)_{B2})

S:
 S?s;+(S^1);
 (S?b1;+(S^2)*S?b2;+(S^3));-(3 <_HB 4)_S;
(S?s;+(S^5);-(1 <_HB 5)_{S};
 \/S?s;+(S^6);-(1 <_HB 6)_{S};S?s;+(S^7);-(6 <_HB 7)_{S};)
\end{verbatim}
}
Each of the local specification contains send/receive
proof obligations, such as \code{\sendc{s}{}{Order}} and
\code{\recvc{s}{}{Order}},
which capture expected 
channel and message type (or property). It also
contains assumptions on the local events, such as  \code{\assume{\event{B_1}{1}}}
and \code{\assume{\event{S}{1}}} that are immediately released once
the corresponding send/receive obligations have been locally
verified. We may also have global
proof obligation, such as \code{\guard{\lthb{3}{4}}}, 
that is meant to ensure
that the two transmissions, labelled as \code{3} and \code{4}.
are race-free since they share a common channel.
Our projection of each
global proof obligation to multiple parties
supports the concept of {\em cooperative proving}, whereby the
proving effort by one party is utilized as an assumption by
the other parties. 
Thus, \code{\guard{\lthb{3}{4}}} is translated to
the following cooperative proving and assumptions:

{\footnotesize
  \centering
  \vspace*{-2mm}
  \arraycolsep=1pt
  \[\vspace*{-2mm}
    \begin{array}{lll}
      \code{\guard{\lthb{3}{4} }_{B_1}} & {=} &
      \code{\guard{\lthb{\event{S}{3}}{\event{B_1}{4}} 
     } \seq
      \assume{\lthb{\event{B_2}{3}}{\event{B_2}{4}}} }
      \\
      \code{\guard{\lthb{3}{4} }_{B_2}} & {=} &
      \code{
      \assume{\lthb{\event{S}{3}}{\event{B_1}{4}}} 
      \seq
      \guard{\lthb{\event{B_2}{3}}{\event{B_2}{4}}
      } 
      }
      \\
      \code{\guard{\lthb{3}{4} }_{S}} & {=} &
      \code{\assume{\lthb{\event{S}{3}}{\event{B_1}{4}} 
      } \seq
      \assume{\lthb{\event{B_2}{3}}{\event{B_2}{4}}} }     
    \end{array}
  \]
}

\hide{
\begin{verbatim}
-(3 <_HB 4)_B1 = -((S^{3} <_{HB} B1^{4} \vee B2^{3} <_{HB} B1^{4}))
    ;+(B2^{3} <_{HB} B2^{4})
-(3 <_HB 4)_B2 =  +(B2^{3} <_{HB} B2^{4})
    ;-((S^{3} <_{HB} B1^{4} \vee B2^{3} <_{HB} B1^{4}));
-(3 <_HB 4)_S =  +(B2^{3} <_{HB} B2^{4})
    ;+((S^{3} <_{HB} B1^{4} \vee B2^{3} <_{HB} B1^{4}));
\end{verbatim}
}


The other global orderings from each transmission,
such as {\cb} 
ordering
like \code{\assume{\ltcb{\event{B_1}{1}}{\event{S}{1}}}}, and 
{\em happens-before} ordering between adjacent events within the same
party, like \code{\assume{\lthb{\event{S}{1}}{\event{S}{2}}}}, are placed in a shared space, as say \code{{\prot}{\prj}{\All}}, that
is visible to all parties. As these global ordering
information are modelled for unique events and are also
immutable, we propose to release all orderings in
a single step at the beginning of the protocol for simplicity.
Our proposal to release all global ordering together
as a single assumption is done for simplicity.
Though future orderings are not required to prove 
current global obligations, they never cause any
inconsistency to the current state,
but merely
add some new orderings that are available in advance.
\scnay{Should we hide the next sentence?}
\hide{The presence of disjuncts cause imprecision 
which could be rectified by tree-share mechanism
that we will
describe in Sec~\ref{}.}


\wnnay{Can we label below as \code{G@All} below?}
\wnnay{should we use ``='' rather than ``:''}

{\footnotesize
  \centering
  \vspace*{-2mm}
  \arraycolsep=1pt
  \[\vspace*{-2mm}
    \code{{\prot}{\prj}{\All}}{\defeq}~\begin{array}[t]{l}
      \code{\assume{\ltcb{\event{B_1}{1}}{\event{S}{1}}} \seq~} \\
      \code{((\assume{\lthb{\event{B_1}{1}}{\event{B_1}{2}}} \seq
      \assume{\lthb{\event{S}{1}}{\event{S}{2}}} \seq
      \assume{\ltcb{\event{S}{2}}{\event{B_1}{2}}}) ~\useq
      }\\
      \code{\;\,
      (\assume{\lthb{\event{S}{1}}{\event{S}{3}}} \seq
      \assume{\ltcb{\event{S}{3}}{\event{B_2}{3}}})) \seq } \\
      \code{\assume{\lthb{\event{B_1}{2}}{\event{B_1}{4}}} \seq
      \assume{\lthb{\event{B_2}{3}}{\event{B_2}{4}}} \seq
      \assume{\ltcb{\event{B_1}{4}}{\event{B_2}{4}}} \seq~} \\
      \code{((
      \assume{\lthb{\event{B_2}{4}}{\event{B_2}{5}}} \seq
      \assume{\ltcb{\event{S}{2}}{\event{S}{5}}} \seq
      \assume{\ltcb{\event{S}{3}}{\event{S}{5}}} \seq
      \assume{\ltcb{\event{B_2}{5}}{\event{S}{5}}}
      )}\\
      \code{\vee ~(
      \assume{\lthb{\event{B_2}{4}}{\event{B_2}{6}}} \seq
      \assume{\lthb{\event{S}{2}}{\event{S}{6}}} \seq
      \assume{\lthb{\event{S}{3}}{\event{S}{6}}} \seq
      \assume{\ltcb{\event{B_2}{6}}{\event{S}{6}}} \seq} \\
      \code{\;\,
      \assume{\lthb{\event{B_2}{6}}{\event{B_2}{7}}} \seq
      \assume{\lthb{\event{S}{6}}{\event{S}{7}}} \seq
      \assume{\ltcb{\event{B_2}{7}}{\event{S}{7}}}
      ))
      }
    \end{array}
  \]
}

\hide{
\begin{verbatim}
Global Constraints:
 +(B1^1 <_CB S^1);
 (+(B1^1 <_CB B1^2);+(S^1 <_HB S^2); +(S^2 <_CB B1^2) 
  * +(S^1 <_HB S^3); +(S^3 <_CB B2^3));
 +(B1^2 <_HB B1^4);+(B2^3 <_HB B2^4); +(B1^4 <_CB B2^4);
((-(B2^5)\/-(S^5));+(B2^4 <_HB B2^5);+(B2^5 <_CB S^5) S?s;+(S^5);
 \/(-(B2^6)\/-(S^6));+(B2^4 <_HB B2^6);+(S^2 <_HB S^6);
   +(S^3 <_HB S^6);+(B2^6 <_CB S^6);
   +(B2^6 <_HB B2^7);+(S^6 <_HB S^7);+(B2^7 <_CB S^7))
\end{verbatim}
}


Given global specification \code{G} over 
\code{n} parties, \code{P_1{\cdots}P_n},
our projection would 
transform \code{G} 
into a per-party specification, as follows:
\begin{small}
\[\vspace*{-2mm}
\begin{array}{lcl}
\code{\prot} & \implies & \code{\prot{\prj}P_1} ~{\sep}~ {\cdots} ~{\sep}~\code{\prot{\prj}P_n} ~{\sep}~\code{\prot{\prj}{\All}}
\end{array}
\]
\end{small}
\noindent
where \code{\prot{\prj}P_1~{\sep}~ {\cdots} ~{\sep}~\prot{\prj}P_n} denote the per-party specifications
and \code{\prot{\prj}{\All}} denotes the 
global orderings that are shared by all parties.
Once we have a per-party specification, say \code{\prot{\prj}P_i}
over channels \code{\chanvar_1{\cdots}\chanvar_m}, we can further project
each of these specifications into a
spatial conjunction of several per-channel specifications,
namely \sm{\code{\perchanprj{\prot}{P}{\chanvar_1}{\sep}{\cdots}~{\sep}~\perchanprj{\prot}{P}{\chanvar_m}}}, as follows:
\hide{%
\begin{verbatim}
G    ==> G@P1 * .. * G@Pn * G@All
G@P  ==> G@P<k1> * .. * G@P<kn> 
\end{verbatim}
}%
\begin{small}
\[\vspace*{-2mm}
\begin{array}{lcl}
\code{\prot{\prj}P_i} & \implies & \code{\perchanprj{\prot}{P_i}{\chanvar_1}} ~{\sep}~ {\cdots} ~{\sep}~\code{\perchanprj{\prot}{P_i}{\chanvar_m}}  
\end{array}
\]\vspace*{-2mm}
\end{small}
The details for channel projections are given later in Sec~\ref{sec:projection}.

\hide{
For our running example, the seller and two buyers
protocol 
can be projected into the three channels, as 
follows:

{\footnotesize
  \centering
  \arraycolsep=1pt
  \[ \vspace*{-2mm}
    \begin{array}{ll}
      \code{\prot{\prj}B_1{\defeq}} & 
      \code{\sendc{s}{}{Order} \seq
      \blueord{\assume{\event{B_1}{1}}} \seq
      \recvc{b_1}{}{Price} \seq
      \blueord{\assume{\event{B_1}{2}}} \seq
      \sendc{b_2}{}{Amt} \seq                          
      \blueord{\assume{\event{B_1}{4}}} \seq
      \blueord{\guard{\lthb{3}{4}}_{B_1}} } \\
      & \code{\Downarrow (\text{per channel projection})}\\
      & \begin{array}{lll}
          \code{\perchanprj{\prot}{B_1}{s}} & \code{\defeq}&\code{\send{}{Order} \seq
                     \blueord{\assume{\event{B_1}{1}}}}\\
          \code{\perchanprj{\prot}{B_1}{b_1}} & \code{\defeq}& \code{\blueord{\guard{\event{B_1}{1}}} \seq
                       \recv{}{Price} \seq
                       \blueord{\assume{\event{B_1}{2}}}}\\
          \code{\perchanprj{\prot}{B_1}{b_2}} & \code{\defeq} & \code{\blueord{\guard{\event{B_1}{2}}} \seq
                       \send{}{Amt} \seq                          
                       \blueord{\assume{\event{B_1}{4}}} \seq
                       \blueord{\guard{\lthb{3}{4}}_{B_1}} 
                       } 
      \end{array} \\            
      \code{\prot{\prj}B_2{\defeq}} & 
      \code{\recvc{b_2}{}{Price} \seq
      \blueord{\assume{\event{B_2}{3}}} \seq
      \recvc{b_2}{}{Amt} \seq
      \blueord{\assume{\event{B_2}{4}}} \seq
      \blueord{\guard{\lthb{3}{4}}_{B_2}}  \seq}\\
      &\code{
       ((
        \sendc{s}{}{No} \seq                                             
        \blueord{\assume{\event{B_2}{5}}} \seq
        \blueord{\guard{\lthb{1}{5}}_{B_2}}
        ) \vee
        (\sendc{s}{}{Yes} \seq                                             
         \blueord{\assume{\event{B_2}{6}}} \seq
         \blueord{\guard{\lthb{1}{6}}_{B_2}} \seq
         \sendc{s}{}{Addr} \seq                                             
         \blueord{\assume{\event{B_2}{7}}} \seq
         \blueord{\guard{\lthb{6}{7}}_{B_2}}
        )
       )
     } 
      \\
      & \code{\Downarrow (\text{per channel projection})}\\
      & \begin{array}{lll}
          \perchanprj{\prot}{B_2}{b_2} & \code{\defeq} & \code{
                       \recv{}{Price} \seq
                       \blueord{\assume{\event{B_2}{3}}} \seq
                       \recv{}{Amt} \seq
                       \blueord{\assume{\event{B_2}{4}}} \seq
                       \blueord{\guard{\lthb{3}{4}}_{B_2}}
                                  }\\
          \perchanprj{\prot}{B_2}{s} & \code{\defeq} & \code{
                                  \blueord{\guard{\event{B_2}{4}}} \seq
                                  ((
                                  \send{}{No} \seq                                             
                                  \blueord{\assume{\event{B_2}{5}}} \seq
                                  \blueord{\guard{\lthb{1}{5}}_{B_2}}
                                  ) \vee
                       (\send{}{Yes} \seq                                             
                       \blueord{\assume{\event{B_2}{6}}} \seq
                       \blueord{\guard{\lthb{1}{6}}_{B_2}} \seq
                       \send{}{Addr} \seq                                             
                       \blueord{\assume{\event{B_2}{7}}} \seq
                       \blueord{\guard{\lthb{6}{7}}_{B_2}}
                       )
                       )    
                                  }\\
          
      \end{array} \\          
      \code{\prot{\prj}S{\defeq}} & 
      \code{\recvc{s}{}{Order} \seq
      \blueord{\assume{\event{S}{1}}} \seq
      (\sendc{b_1}{}{Price} \seq
      \blueord{\assume{\event{S}{2}}})
      \useq
      (\sendc{b_2}{}{Price} \seq
      \blueord{\assume{\event{S}{3}}}) \seq
      \blueord{\guard{\lthb{3}{4}}_S}  \seq }\\
      &\code{
       ((
        \recvc{s}{}{No} \seq                                             
        \blueord{\assume{\event{S}{5}}} \seq
        \blueord{\guard{\lthb{1}{5}}_S}
        ) \vee (\recvc{s}{}{Yes} \seq                                             
         \blueord{\assume{\event{S}{6}}} \seq
         \blueord{\guard{\lthb{1}{6}}_S} \seq
         \recvc{s}{}{Addr} \seq                                             
         \blueord{\assume{\event{S}{7}}} \seq
         \blueord{\guard{\lthb{6}{7}}_S}
        )
       )
      }\\
      & \code{\Downarrow (\text{per channel projection})}\\
      & \begin{array}{lll}
          \code{\perchanprj{\prot}{S}{b_1}} & \code{\defeq} & \code{
                                 \blueord{\guard{\event{S}{1}}} \seq
                                  \send{}{Price} \seq
                                  \blueord{\assume{\event{S}{2}}}
                                  }\\
          \code{\perchanprj{\prot}{S}{b_2}} & \code{\defeq} & \code{
                                  \blueord{\guard{\event{S}{1}}} \seq
                                  \send{}{Price} \seq
                                  \blueord{\assume{\event{S}{3}}}
                                  \seq
                                  \blueord{\guard{\event{S}{2}}} \seq
                                  \blueord{\guard{\lthb{3}{4}}_S} 
                                  }\\
          \code{\perchanprj{\prot}{S}{s}} & \code{\defeq} & \code{
                                \recv{}{Order} \seq
                                \blueord{\assume{\event{S}{1}}} \seq
                                (\blueord{\guard{\event{S}{2}}} \useq
                                \blueord{\guard{\event{S}{3}}}) \seq
                                }\\
          &&\code{
             ((
             \recv{}{No} \seq                                             
             \blueord{\assume{\event{S}{5}}} \seq
             \blueord{\guard{\lthb{1}{5}}_S}
             ) \vee (\recv{}{Yes} \seq                                             
            \blueord{\assume{\event{S}{6}}} \seq
            \blueord{\guard{\lthb{1}{6}}_S} \seq
            \recv{}{Addr} \seq                                             
            \blueord{\assume{\event{S}{7}}} \seq
            \blueord{\guard{\lthb{6}{7}}_S}
            )
            )
        }\\
          
        \end{array} \\ 
      
    \end{array}
  \]
}

Each of these per-channel specification
may actually 
contain event guards, such as
\code{\blueord{\guard{\event{S}{1}}}}. These guards on events 
are important for ensuring that
send/receive operations over different
channels but within the same party
are correctly synchronized. 
For example, we require that \code{\recv{s}{Order}}
be executed before \code{\send{b_1}{Price}} 
within party \code{S}, and this is
ensured by placing
\code{\blueord{\guard{\event{S}{1}}}}
before \code{\send{b_1}{Price}}.
}
\hide{
\begin{verbatim}
B1:
 B1!s;+(B1^1);B1?(b1);+(B1^2);
 B1!b2;+(B1^4);-(3 <_HB 4)_{B1}
=channel=>
 s: B1!s;+(B1^1));
 b1: -(B1^1);B1?(b1);+(B1^2);
 b2: -(B1^2)B1!b2;+(B1^4);-(3 <_HB 4)_{B1};

B2:
 B2?b2;+(B2^3);
 B2?b2;+(B2^4);-(3 <_HB 4)_B2;
(B2!s;+(B2^5);-(1 <_HB 5)_B2;
 \/B2!s;+(B2^6);-(1 <_HB 6)_B2;B2!s;+(B2^7);-(6 <_HB 7)_B2)
=channel=>
 b2: B2?b2;+(B2^3);B2?b2;+(B2^4);-(3 <_HB 4)_B2;
 s:  -(B2^4);
      (B2!s;+(B2^5)-(1 <_HB 5)_B2; 
      \/B2!s;+(B2^6);-(1 <_HB 6)_B2;B2!s
         ;+(B2^7);-(6 <_HB 7)_B2)

S:
 S?s;+(S^1);
 (S?b1;+(S^2)*S?b2;+(S^3));+(3 <_HB 4);
(S?s;+(S^5);-(1 <_HB 5)_{S} 
 \/S?s;+(S^6);-(1 <_HB 6)_{S};S?s;+(S^7);-(6 <_HB 7)_{S})
=channel=>
 b1: -(S^1);(S?b1;+(S^2));
 b2: -(S^1);(S?b2;+(S^3));-(S^2);+(3 <_HB 4);
 s: S?s;+(S^1);-(S^2)*-(S^3);
    (S?s;+(S^5);-(1 <_HB 5)_{S} 
     \/S?s;+(S^6);-(1 <_HB 6)_{S};S?s;+(S^7);-(6 <_HB 7)_{S})
\end{verbatim}
}

\subsection{Automated Local Verification}
Once we have the refined specification that had been 
suitably projected on a per party and per channel
basis, we can proceed to
use these specifications for automated local verification.
The shared global orderings can be immediately added to the 
initial program state.  The per-channel specification 
are released during the verification on each of the respective
program codes, namely \code{(Code_S ~||~ Code_{B_1} ~||~ Code_{B_2})},
for the different parties. For our running example,
the initial and final program states are:
\hide{%
\begin{verbatim}
{ Common(G#All) * Party<S,G#S> * Party<B1,G#B1> 
  * Party<B2,G#B2> }
   (Code_S ~||~ Code_{B1} ~||~ Code_{B2}) 
{ Party<S,emp> * Party<B1,emp> * Party<B2,emp> }
\end{verbatim}
}%
{\footnotesize
\[ \vspace*{-2mm}
\begin{array}{c}
\{\commonspec{{\prot}{\prj}{\All}} ~{\sep}~\partyspec{S}{{\prot}{\prj}{S}} ~{\sep}~\partyspec{B_1}{{\prot}{\prj}{B_1}} 
~{\sep}~\partyspec{B_2}{{\prot}{\prj}{B_2}}\}
\\
\code{(Code_S ~||~ Code_{B_1} ~||~ Code_{B_2})}
\\
\{\partyspec{S}{\emp} ~{\sep}~\partyspec{B_1}{\emp} 
~{\sep}~\partyspec{B_2}{\emp}\}
\end{array}
\]
}

The \code{\emp} in the final state of each
party can help to confirm the completion of
all transmissions in
the global protocol. The abstract predicate \code{\partyspec{S}{{\prot}{\prj}{S}}}
associates each party with its corresponding specification. 
Our approach supports both events that are either
{\em immutable} or {\em mutable}. 
Each immutable event (seen earlier)
is 
labelled uniquely, while mutable
events can be updated flow-sensitively.
To signify each executing party \code{P}, we use
a mutable event \code{\peer{\rolevar}} that is
updated when \code{\rolevar} is executing. Such mutable events are
added to 
as ghost specifications, as shown below.
\hide{
We can also guide 
and restrict the use of each party's specification
by the use on
guards and assumptions on each party, 
such as \code{\guard{B_1}} and \code{\assume{B_1}}
for party \code{B_1},
as shown below.
}
\hide{
\asay{can we only keep one initial state in the below? (with either
  \code{\commonspec{{\prot}{\prj}{\All}}}
  or
   \code{{\prot}{\prj}{\All}} ?
   )}
\asay{
   Why is +(S), +(B1), etc part of the code?}
\asay{
   It looks like the orders disappeared from the poststate.}
\asay{
 I wonder whether we should support an operation which filters the
   non-orders from a spec, not to be used in reasoning, but
   just for illustration purpose, to emphasizing the poststate of S
   for example: \code{({\prot}{\prj}{S}) \% O \wedge {\prot}{\prj}{\All}} }
\asay{
  Should I add +(P).. as part of the language?
}
}
\hide{
\begin{verbatim}
{ Common(G#All) * Party<S,-S;G#S> * Party<B1,-B1;G#B1> 
  * Party<B2,-B2;G#B2> }
{ Party<S,-G;G#S> * Party<B1,-B1;G#B1> * Party<B2,-B2;G#B2> & G$All }
 ( {Party<S,-G;G#S> & G#All} +S;Code_S {Party<S,emp> & ?}
  ~||~ 
   {Party<B_1,-B1;G#B_1> & G#All} 
       +B1;Code_{B1}  {Party<B_1,emp> & ?}
  ~||~ 
   {Party<B_1,-B2;G#B_2> & G#All} 
       +B2;Code_{B2}  {Party<B_2,emp> & ?}
 ) 
{ Party<S,emp> * Party<B1,emp> * Party<B2,emp> & ? }
{ Party<S,emp> * Party<B1,emp> * Party<B2,emp> }
\end{verbatim}
}

{\footnotesize
\[ \vspace*{-2mm}
\begin{array}{c}
\{\commonspec{{\prot}{\prj}{\All}} ~{\sep}~\partyspec{S}{{\prot}{\prj}{S}} ~{\sep}~\partyspec{B_1}{{\prot}{\prj}{B_1}} 
~{\sep}~\partyspec{B_2}{{\prot}{\prj}{B_2}}\}
\\
\{\partyspec{S}{{\prot}{\prj}{S}} ~{\sep}~\partyspec{B_1}{{\prot}{\prj}{B_1}} 
~{\sep}~\partyspec{B_2}{{\prot}{\prj}{B_2}} ~{\wedge}~ \code{{\prot}{\prj}{\All}} 
\}
\\ \\
\left( \begin{array}{l}
\code{ \{\partyspec{S}{{\prot}{\prj}{S}}~{\wedge}~ \code{{\prot}{\prj}{\All}} \} 
\code{~\;\blueord{\assume{\peer{S}}};Code_S~}
\{\partyspec{S}{\emp} 
\} }
\\
\code{~||~}
\\
\code{ \{\partyspec{B_1}{{\prot}{\prj}{B_1}}~{\wedge}~ \code{{\prot}{\prj}{\All}} \} 
\code{~\blueord{\assume{\peer{B_1}}};Code_{B_1}}
\{\partyspec{B_1}{\emp} 
\} }
\\
\code{~||~}
\\
\code{ \{\partyspec{B_2}{{\prot}{\prj}{B_2}}~{\wedge}~ \code{{\prot}{\prj}{\All}} \} 
\code{~\blueord{\assume{\peer{B_2}}};Code_{B_2}}
\{\partyspec{B_2}{\emp} 
\} }
\end{array}
\right)
\\ \\
\{\partyspec{S}{\emp} ~{\sep}~\partyspec{B_1}{\emp} 
~{\sep}~\partyspec{B_2}{\emp}\}
\end{array} 
\]
}\\

Note that \code{\commonspec{{\prot}{\prj}{\All}}} denotes pure global information that can
be duplicated and propagated.

\scnay{Can we explain notations  \code{\commonspec{{\prot}{\prj}{\All}}}, \code{\hov_{S}}, \code{\hov_{B_1}}, \code{\hov_{B_2}}?}

\wnnay{Lemma to move out of overview and
into tech sec of paper}
\hide{
Lastly, to support local verification, we allow the per-channel
specification to be released by
lemmas of the form:
\wnnay{should we use \code{G\#B1\#b1} instead of \code{G\#B1<b1>}}
\scnay{the macro perchanprj is changed to reflect this}
\hide{%
\begin{verbatim}
 Party<B_1,G#B_1> <=> Channel<s,B_1,G#B_1<s>> 
    * Channel<b1,B_1,G#B_1<b1>> 
    * Channel<b2,B_1,G#B_1<b2>>
\end{verbatim}
}%
{\footnotesize
\[
\code{\partyspec{B_1}{{\prot}{\prj}{B_1}} ~{\Leftrightarrow}~ \chanspec{s}{B_1}{\perchanprj{\prot}{B_1}{s}} ~{\sep}~
\chanspec{b_1}{B_1}{\perchanprj{\prot}{B_1}{b_1}} ~{\sep}~
\chanspec{b_2}{B_1}{\perchanprj{\prot}{B_1}{b_2}}}
\]\vspace*{-2mm}
}

On completion, we also have a set of 
lemmas over the \code{emp} state to signify completion
of protocol. An example is shown below:
\hide{%
\begin{verbatim}
 Party<B_1,emp> <=> Channel<s,B_1,emp> 
    * Channel<b1,B_1,emp> * Channel<b2,B_1,emp>
\end{verbatim}
}%
\begin{footnotesize}
\[\vspace*{-2mm}
\code{
\partyspec{B_1}{\emp} ~{\Leftrightarrow}~ \chanspec{s}{B_1}{\emp} ~{\sep}~
\chanspec{b_1}{B_1}{\emp} ~{\sep}~
\chanspec{b_2}{B_1}{\emp}}
\]
\end{footnotesize}

With the help of these lemmas, we can now 
allow local verification to proceed, as follows:
\hide{%
\begin{verbatim}
   {Party<B_1,G#B_1> & G#All} 
   {Channel<s,B_1,G#B_1<s>> * Channel<b1,B_1,G#B_1<b1>> 
    * Channel<s,B_1,G#B_1<b2>> & G#All} 
       Code_{B_1}  
   {Channel<s,B_1,emp> * Channel<b1,B_1,emp 
    * Channel<s,B_1,emp> & ?} 
   {Party<B_1,emp> & ?} 
   {Party<B_1,emp>}
\end{verbatim}
}%
\begin{footnotesize}
\[ \vspace*{-2mm}
\begin{array}{c}
\{\partyspec{B_1}{{\prot}{\prj}{B_1}} ~{\wedge}~ \code{{\prot}{\prj}{\All}} \}
\\
\code{\{\chanspec{s}{B_1}{\perchanprj{\prot}{B_1}{s}} ~{\sep}~
\chanspec{b_1}{B_1}{\perchanprj{\prot}{B_1}{b_1}} ~{\sep}~
\chanspec{b_2}{B_1}{\perchanprj{\prot}{B_1}{b_2}}
~{\wedge}~ \code{{\prot}{\prj}{\All}} \}}
\\
\code{\blueord{\assume{\peer{B_1}}};\code{Code_{B_1}}}
\\
\code{\{ \chanspec{s}{B_1}{\emp} ~{\sep}~
\chanspec{b_1}{B_1}{\emp} ~{\sep}~
\chanspec{b_2}{B_1}{\emp}
~{\wedge}~ \hov_{B_1} \}}
\\
\code{\{\partyspec{B_1}{\emp} ~{\wedge}~ \hov_{B_1}\}}
\\
\code{\{\partyspec{B_1}{\emp}\}}
\end{array} 
\]
\end{footnotesize}
}
\scnay{We still have ? in the above}
\wnnay{Do we need to talk about how assumptions are released}
\wnnay{Do we need to indicate a flow-sensitive event on parties}

\hide{
\begin{verbatim}

{Initial State} S || B1 || B2 {Final State}




Channels are resources, but events
(that have occurred) and orderings (past and future) are pure info
that can be freely duplicated.

Final state would ensure that every channel
has been completed, and thus fully
complied with the global protocol.

Though we can handle also deadlock-freedom, this
aspects is not discussed in the current paper.

Release of resources and guards..

Local verification on send/receive ..

\end{verbatim}
}






\wnnay{well-formed disjunction need to go at a later point}

\section{Global Protocols}
\label{sec:global}
We now formalize our proposal into a logical
system 
called \sessionlogic, whose 
syntax is depicted
in \figref{fig.global_syntax_prot}. We first list down the
elements of the protocol and their properties before
studying the properties of the whole protocol. 
\hide{
Though it 
only uses  binary operators to
compose multiple transmissions into a single global protocol,
there is no ambiguity in how the protocol
gets created or analyzed, since associativity holds for all
operators as highlighted below. As expected, \code{\useq} and
\code{\vee} are commutative, while \code{\emp} acts as
the identity element of \code{\useq}, or is 
ignored in a sequence if it is not the last element: 
}

\noindent{\bf {Communication model.}}
To support a wide range of communication interfaces, the current
session logic is designed 
for a permissive communication model,
where:
\begin{itemize}
\item
  The transfer of a message dissolves \emph{asynchronously},
  that is to say that sending is non-blocking while receiving is blocking.
\item
  The communication interface of choice manipulates \emph{FIFO channels}
  that are \emph{shareable}, in the sense that each channel
  can serve two or more participants.
\item
  For simplicity, the communication assumes unbounded buffers. However, 
extension to bounded buffer is possible due to our projection to a
per-channel specification. 
\end{itemize}

\noindent{\bf {Transmission.}} As depicted in
\figref{fig.global_syntax_prot}, 
a \emph{transmission}
\code{\atransmit{S}{R}{v\,{\cdot}\,\fmsg}{\chanvar}{\atid}}
involves a sender \code{S} and a receiver \code{R}
transmitting a message \code{v} expressed in logical form \code{\fmsg}
over a buffered channel \code{\chanvar}. This transmission
is uniquely identified by a label \code{\atid}. In the subsequent
we often use only the unique label \code{\atid} to refer to
a particular transmission. To access the components of a transmission 
we define the following auxiliary functions:~
\code{\tsend{\atid}  \eqdef \event{S}{\atid}},
\code{\trecv{\atid}  \eqdef \event{R}{\atid}},
\code{\tchan{\atid}  \eqdef \chanvar} and 
\code{\tmsg{\atid}   \eqdef {v\,{\cdot}\,\fmsg}}.
A transmission is irreflexive, since it would make
no sense for the sending and the receiving
to be performed by the same peer. 
We define a function \code{\TR{\prot}} which decomposes a given
protocol to collect a set of all
its constituent transmissions, and a function \code{\first{\prot}} to
return the set of all possible first transmissions:
\begin{figure}[H]
\centering
\captionsetup[subfigure]{labelformat=empty,farskip=-4pt,captionskip=-5pt}  
\subfloat[][]
{$
\begin{array}{ll}
\code{\TR{\atransmit{S}{R}{v\,{\cdot}\,\fmsg}{\chanvar}{\atid}}} &
  \code{\eqdef \{\atransmit{S}{R}{v\,{\cdot}\,\fmsg}{\chanvar}{\atid}\}}\\
\code{\TR{\prot_1 \seq \prot_2}} & \code{ \eqdef
                                   \TR{\prot_1}\cup\TR{\prot_2}}\\
\end{array}
$}\hfill
\subfloat[][]
{$
\begin{array}{ll}
\code{\TR{\prot_1 \useq \prot_2}} & \code{ \eqdef
  \TR{\prot_1} \cup \TR{\prot_2}}\\
\code{\TR{\prot_1 \vee \prot_2}} & \code{\eqdef \TR{\prot_1}\cup\TR{\prot_2}}\\
\end{array}
$}\hfill
\subfloat[][]
{$
\begin{array}{ll}
 \code{\TR{\assume{\racefreeas}}} & \code{ \eqdef \emptyset}\\
\code{\TR{\guard{\racefreeas}}}   & \code{\eqdef \emptyset}
\end{array}
$}\\
\subfloat[][]
{$
\begin{array}{ll}
\code{\first{\atransmit{S}{R}{v\,{\cdot}\,\fmsg}{\chanvar}{\atid}}} &
  \code{\eqdef \{\atransmit{S}{R}{v\,{\cdot}\,\fmsg}{\chanvar}{\atid}\}}\\
\code{\first{\prot_1 \seq \prot_2}} & \code{ \eqdef
                                   \first{\prot_1}}\\
\end{array}
$}\hfill
\subfloat[][]
{$
\begin{array}{ll}
\code{\first{\prot_1 \useq \prot_2}} & \code{ \eqdef
  \first{\prot_1} \cup \first{\prot_2}}\\
\code{\first{\prot_1 \vee \prot_2}} & \code{\eqdef \first{\prot_1}\cup\first{\prot_2}}\\
\end{array}
$}
\end{figure}

\noindent{\bf {Event.}} An \emph{event} \code{\ev} is a pair 
\code{\event{\rolevar}{\atid}} where \code{\rolevar \in \roletyp} is
the sending or receiving party of a transmission identified by 
\code{\atid \in \labeltyp}. Given the uniqueness of the
identifier \code{\atid}, an event uniquely
identifies a send or receive operation.
We denote by \code{\evpeer{\ev}} and \code{\evlbl{\ev}} the elements
of an event, e.g. \code{\evpeer{\event{\rolevar}{\atid}} \eqdef \rolevar} and
\code{\evlbl{\event{\rolevar}{\atid}} \eqdef  \atid}. 
The following
function
collects the set of all the events within a protocol:
\begin{figure}[H]
\centering
\captionsetup[subfigure]{labelformat=empty}
\subfloat[][]
{$
\begin{array}{ll}
\code{\EV{\prot_1 \seq \prot_2}} & \code{ \eqdef \EV{\prot_1}\cup\EV{\prot_2}}\\
\code{\EV{\prot_1 \useq \prot_2}} & \code{ \eqdef
  \EV{\prot_1}\cup\EV{\prot_2}}\\
\code{\EV{\prot_1 \vee \prot_2}} & \code{\eqdef \EV{\prot_1}\cup\EV{\prot_2}}
\end{array}
$}\hfill
\subfloat[][]
{$
  \begin{array}{ll}
\code{\EV{\event{\rolevar}{\atid}}} &  \code{\eqdef \{\event{\rolevar}{\atid}\}}\\
\code{\EV{\assume{\racefreeas}}} & \code{ \eqdef \EV{\racefreeas}}\\
\code{\EV{\guard{\racefreeas}}}   & \code{\eqdef \EV{\racefreeas}}\\
\end{array}
$}\hfill
\subfloat[][]
{$
\begin{array}{ll}
\code{\EV{\atransmit{S}{R}{v\,{\cdot}\,\fmsg}{\chanvar}{\atid}}} &
\code{\eqdef \{\event{S}{\atid},\event{R}{\atid}\}}\\
\code{\EV{\lthb{\event{\rolevar_1}{\atid_1}}{\event{\rolevar_2}{\atid_2}}}} &
  \code{\eqdef \{{\event{\rolevar_1}{\atid_1}},{\event{\rolevar_2}{\atid_2}}\}}\\
\code{\EV{\ltcb{\event{\rolevar_1}{\atid_1}}{\event{\rolevar_2}{\atid_2}}}}
  & \code{ \eqdef
  \{{\event{\rolevar_1}{\atid_1}},{\event{\rolevar_2}{\atid_2}}\}}.
\end{array}
$}
\end{figure}

The messages of 
two arbitrary but distinct transmissions, say \code{\atid_1} and \code{\atid_2},
are said to be \emph{disjoint},
denoted \code{\mdisj{\tmsg{\atid_1}}{\tmsg{\atid_2}}}
if \code{UNSAT(\fmsg_{i_1} \wedge [v_1/v_2]\fmsg_{i_2})},
where 
\code{\tmsg{\atid_1}{=}{v_1\,{\cdot}\,\fmsg_1}} and
\code{\tmsg{\atid_2}{=}{v_2\,{\cdot}\,\fmsg_2}}.

We abuse the set membership symbol, \code{\in}, to denote the
followings:
\begin{figure}[H]
\centering
\captionsetup[subfigure]{labelformat=empty,farskip=-5pt,captionskip=-10pt}  
\subfloat[][]
{$
\begin{array}{llll}
(\code{\in_\text{transm.}}) & \code{\atid \in \prot} & \code{\Leftrightarrow} &\code{\atid \in
                                                  \TR{\prot}}\\
(\code{\in_\text{channel}}) & \code{\chanvar \in \atid} & \code{\Leftrightarrow}
                                                    &\code{\tchan{\atid}=\chanvar}\\
(\code{\in_\text{channel}}) &\code{\chanvar \in \prot} & \code{\Leftrightarrow} &\code{\exists
                                                         \atid \in
                                                         \prot \cdot
                                                         \chanvar
                                                         \in
                                                         \atid}\\
(\code{\in_\text{party}}) & \code{\rolevar \in \atid} & \code{\Leftrightarrow}
                                                    &\code{\evpeer{\tsend{\atid}}=\rolevar
                                                      \text{~or~}}\\
&&& \code{\evpeer{\trecv{\atid}}=\rolevar}\\
(\code{\in_\text{party}}) & \code{\rolevar \in \prot} & \code{\Leftrightarrow}
                                                    &\code{\exists
                                                      \atid \in \prot
                                                      \cdot \rolevar
                                                      \in \atid}\\
\end{array}
$}
\end{figure}

Correspondingly, \code{\notemploy} is used 
to denote the negation of the above.


Transmissions are organized into a global protocol using
a combination of parallel composition \code{\prot_1 \useq \prot_2},
disjunction \code{\prot_1 \vee \prot_2} and
sequential composition \code{\prot_1 \seq \prot_2}.
The parallel composition of global protocols forms a commutative monoid
\code{(\prot,\useq,\emp)}
with \code{\emp} as identity element, while
disjunction and sequence form semigroups, \code{(\prot,\vee)}
and \code{\code{(\prot,\seq)}}, 
with the former also
satisfying commutativity. \code{\emp} acts the
left identity element for sequential composition:

\begin{figure}[H]
\centering
\captionsetup[subfigure]{labelformat=empty,farskip=-5pt}
\subfloat[][]
{$
\begin{array}{rcl}
\code{(\prot_1~\seq~\prot_2)~\seq~\prot_3} &{\equiv}&
\code{\prot_1~\seq~(\prot_2~\seq~\prot_3) } \\
\code{(\prot_1\useq\prot_2)\useq\prot_3} &{\equiv}&
\code{\prot_1\useq(\prot_2\useq\prot_3) } \\
\code{(\prot_1\vee\prot_2)\vee\prot_3} &{\equiv}&
\code{\prot_1\vee(\prot_2\vee\prot_3) } 
\end{array}
$}\hfill
\subfloat[][]
{$
\begin{array}{rcl}
\code{\prot_1\useq\prot_2} &{\equiv}& \code{\prot_2\useq\prot_1} \\
\code{\prot_1~\vee~\prot_2} &{\equiv}& \code{\prot_2~\vee~\prot_1}
\end{array}
$}\hfill
\subfloat[][]
{$
\begin{array}{rcl}
\code{\prot \useq \emp} &{\equiv}& \code{\prot} \\
\code{\emp~\seq~\prot } &{\equiv}& \code{\prot}
\end{array}
$}
\end{figure}

\noindent
Sequential composition is not commutative, unless it
satisfies certain disjointness properties:
\[
  {\arraycolsep=0pt
\begin{array}{c}
    \code{
      \prot_1~\seq~\prot_2~{\equiv}~\prot_2~\seq~\prot_1
      \text{~when~}}\\
    \code{
    \forall c_1{\in}\prot_1, c_2{\in}\prot_2 \implies
    c_1{\neq}c_2
    \text{~and~} \forall \rolevar_1{\in}\prot_1,\rolevar_2 {\in}
    \prot_2 \implies  \rolevar_1{\neq}\rolevar_2}
\end{array}
}\]
  
\noindent{\bf {Sub-protocol.}} A protocol \code{\prot'}
is said to be a sub-protocol of \code{\prot}
when  \code{\prot'} is a  decomposition of \code{\prot},
where the set of all possible decompositions of \code{\prot}
is recursively defined on the structure of \code{\prot}
 as follows:

\begin{figure}[H]
\centering
\arraycolsep=0pt
\captionsetup[subfigure]{labelformat=empty,farskip=-5pt }
\subfloat[][]
{$
  \begin{array}{ll}
    \code{\sub{\atransmit{S}{R}{v\,{\cdot}\,\fmsg}{\chanvar}{\atid}}} & \code{ \eqdef \sub{\atransmit{S}{R}{v\,{\cdot}\,\fmsg}{\chanvar}{\atid}}}\\
    \code{\sub{\assume{\racefreeas}}} & \code{ \eqdef \emptyset}\\
    \code{\sub{\guard{\racefreeas}}}   & \code{\eqdef \emptyset}\\
  \end{array}
$}\hfill
\subfloat[][]
{$
  \begin{array}{ll}
    \code{\sub{\prot_1 \seq \prot_2}} & \code{\eqdef \{\prot_1 \seq \prot_2\} \cup\sub{\prot_1}\cup\sub{\prot_2}}\\
    \code{\sub{\prot_1 \vee \prot_2}} & \code{\eqdef \{\prot_1 \vee \prot_2\} \cup\sub{\prot_1}\cup\sub{\prot_2}}\\
    \code{\sub{\prot_1 \useq \prot_2}} & \code{\eqdef \{\prot_1 \useq \prot_2\} \cup\sub{\prot_1}\cup\sub{\prot_2}}\\
\end{array}
$}\hfill

\end{figure}

\noindent{\bf {Graph ordering.}}
We capture global protocol in terms
of a directed acyclic graph
\code{\graph{\prot}{\eqdef}(V,O)},
where the set of vertices is the set of all the transmissions in
\code{\prot}, \code{V{=}\TR{\prot}}, and the edges
represent the sequence relation between these nodes, 
\code{P{=}\ord{\prot}}, where \code{\ord{}} is defined as follows:
\begin{figure}[H]
\centering
\arraycolsep=0pt
\captionsetup[subfigure]{labelformat=empty,farskip=-5pt}
\subfloat[][]
{$
  \begin{array}{ll}
  \code{\ord{\prot_1\seq\prot_2}} & \code{\eqdef~} \code{\{(\atid_1,\atid_2)  |
    \atid_1{\in}\TR{\prot_1} {\wedge}\atid_2{\in} \TR{\prot_2}\}
         {\cup}\ord{\prot_1}{\cup}\ord{\prot_2}
  }.\\
 \code{\ord{\assume{\racefreeas}}} & \code{\eqdef \emptyset~}.\quad
  \code{~\ord{\guard{\racefreeas}} \eqdef \emptyset}. \quad
  \code{\code{\ord{}(\atransmit{S}{R}{v\,{\cdot}\,\fmsg}{c}{\atid}})} \code{~\eqdef~}  \code{\emptyset~}.
   \end{array}
$}\hfill
\subfloat[][]
{$
  \begin{array}{ll}
    \code{\ord{\prot_1{\useq} \prot_2}} &\code{~\eqdef~} \code{\ord{\prot_1}{\cup}\ord{\prot_2}}.\\
    \code{\ord{\prot_1{\vee} \prot_2}} & \code{~\eqdef~} \code{\ord{\prot_1}{\cup} \ord{\prot_2}}.\\
\end{array}
$}\hfill

\end{figure}

Two transmissions, \code{\atid_1} and \code{\atid_2},
are \emph{sequenced} in a protocol \code{\prot}, denoted \code{\ltseq{\atid_1}{\atid_2}},
if there is a path in
\code{\graph{\prot}} from \code{\atid_1} to \code{\atid_2}.
Two  transmissions, \code{\atid_1} and \code{\atid_2}, are \emph{adjacent} 
in \code{\prot} if they share the same channel \code{\chanvar}, they are
sequenced in \code{\graph{\prot}},
{and there are no other transmission on \code{\chanvar}
in between \code{\atid_1} and \code{\atid_2}}.
And lastly, two transmissions
are \emph{linked} if they are sequenced and they share the same
channel.
These relations are formally described in \figref{fig.fig_def}.
Since these relations are defined as edges of a directed acyclic group
it is straightforward to show that they are
irreflexive and antisymetric. The transitivity of sequenced
also follows directly from the reachability relation of DAGs
which is the transitive closure of the edges in \code{\graph{\prot}},
A simple case analysis on the definition of linked transmissions
shows that the linked relation is also transitive.

\begin{figure}
\centering
\footnotesize
\captionsetup[subfigure]{labelformat=empty,farskip=-5pt,captionskip=-15pt}  
\subfloat[][]
{$
  \begin{array}{l}
    \code{\textbf{Sequenced}} \\ 
    \qquad\code{
    (\atid_1,\atid_2) \in \ord{\prot}} \text{~is denoted by~} \code{\ltseq{\atid_1}{\atid_2}} \\
    \code{\textbf{Linked}}\\
    \qquad\code{
    \link{\atid_1}{\atid_2} \eqdef
    \adj{\atid_1}{\atid_2} \vee
    (\exists \atid' \cdot \adj{\atid_1}{\atid'} \wedge \link{\atid'}{\atid_2})
    }.\\
    \code{\textbf{Adjacent}}\\
    {\qquad\code{
        \adj{\atid_1}{\atid_2} \eqdef
        \tchan{\atid_1}{=}\tchan{\atid_2}
    \wedge \ltseq{\atid_1}{\atid_2}\wedge}}\\
    {\qquad \qquad \qquad \qquad \code{
    \neg\exists \atid'\cdot 
       ( \tchan{\atid_1}{=}\tchan{\atid'} \wedge
         \ltseq{\atid_1}{\atid'}\wedge
        \ltseq{\atid'}{\atid_2}
        ).
    }}\\
  \end{array}
$}

\vspace*{3mm}
\hrule
\vspace*{-3mm}
\caption{Transmission sequencing with respect to a given protocol G}\label{fig.fig_def}
\vspace*{-3mm}
\end{figure}


\subsection{Well-Formedness}
\noindent \emph{\bf Concurrency}. The \code{\useq} operator offers
support for arbitrary-ordered (concurrent) transmissions, where the completion order is not
important for the final outcome.

\hide{
\begin{defn}[Employed]
  Given a transmission
  \code{\atransmit{S}{R}{v\,{\cdot}\,\fmsg}{\chanvar}{\atid}}:
  \begin{enumerate}[(a)]
  \item a communication instrument \code{\chanvar'} is said to be
  employed by transmission \code{\atid} if and only if
  \code{\chanvar'{=}\chanvar};
  \item a role \code{\rolevar} is said to be
    employed by transmission \code{\atid} if and only if
      \code{\rolevar{=}S} or \code{\rolevar{=}R};
  \end{enumerate}
  More generally,
  \code{\chanvar} (or \code{\rolevar}) is employed by a 
  protocol \code{\prot}, if at least one of the
  transmissions specified by \code{\prot} employs \code{\chanvar} (or
  \code{\rolevar}, respectively).
\end{defn}
We abuse the set membership symbol, \code{\employ}, to indicate that
a communication instrument \code{\chanvar} is employed by
a transmission \code{\atid}, \code{\chanvar{\employ}\atid},
or by a protocol \code{\prot},
\code{\chanvar{\employ}\prot},
etc. Correspondingly, \code{\notemploy} is used 
to denote that a certain channel is not employed by the specified 
communication component. The same discussion also applies to roles,
where a role \code{\rolevar} might:
\code{\rolevar \employ \atid}, \code{\rolevar \employ \prot},
\code{\rolevar \notemploy \atid}, or \code{\rolevar \notemploy \prot}.
}

\begin{defn}[Well-Formed Concurrency]
  A protocol specification, \code{\prot_1 \useq \prot_2}, is said to
  be well-formed with respect to \code{\useq} if and only if
  \code{\forall \chanvar{\employ}\prot_1 \implies
    \chanvar{\notemploy}\prot_2},
  and vice versa.
\end{defn}

This restriction is to avoid non-determinism 
from concurrent communications
over the same channel.

\noindent \emph{\bf Choice}. The \code{\vee} operator is essential 
for the expressiveness of \sessionlogic, but its usage must
be carefully controlled:
  
\wnnay{need to define first possible transmission, more
  formally}\anay{Done, using function \code{\first{}}}
\begin{defn}[Well-Formed Choice]
  A disjunctive protocol specification, \code{\prot_1 \vee \prot_2},
  is said to
  be well-formed with respect to \code{\vee} if and only if all of the
  following conditions hold, where
  \code{T_1} and \code{T_2} account for 
  all first transmissions of 
  \code{\prot_1} and \code{\prot_2}, respectively,
  \code{T_1{=}\first{\prot_1}} and \code{T_2{=}\first{\prot_2}}:
  \begin{enumerate}[(a)]
    \item
      (same first channel)
      \code{\forall \atid_1,\atid_2 \in T_1 \cup T_2
        \implies \tchan{\atid_1}=\tchan{\atid_2};
      }
    \item
      (same first sender) \;
      \code{\forall \atid_1,\atid_2 \in T_1 \cup T_2
        \implies \evpeer{\tsend{\atid_1}}=\evpeer{\tsend{\atid_2}};
      }
    \item
      (same first receiver)
      \code{\forall \atid_1,\atid_2 \in T_1 \cup T_2
        \implies \evpeer{\trecv{\atid_1}}=\evpeer{\trecv{\atid_2}};
      }
    \item
      (mutually exclusive ``first'' messages)
      \code{\forall \atid_1,\atid_2 \in T_1 \cup T_2
        \implies \mdisj{\tmsg{\atid_1}}{\tmsg{\atid_2}}
        \vee
        \atid_1{=}\atid_2;
      }
    \item
      (same peers)
      \code{\forall \atid_1,\atid_2 \in \prot_1 \cup \prot_2 \implies}\\
      \code{
        \{\evpeer{\tsend{\atid_1}},\evpeer{\trecv{\atid_1}}\}
        {=}
        \{\evpeer{\tsend{\atid_2}},\evpeer{\trecv{\atid_2}}\};
      } 
    \item
      (recursive well-formedness) \code{\prot_1} and \code{\prot_2}
      are well-formed with respect to \code{\vee}.
    \end{enumerate}
\end{defn}

\begin{defn}[Well-Formed Protocol] A protocol \code{\prot} is said to
  be well-formed, if and only if \code{\prot} contains only
  well-formed concurrent sub-protocols, and well-formed choices.
\end{defn}

To ensure the correctness of our approach, \sessionlogic~
disregards as unsound any usage of \code{\useq} or \code{\vee} which is not
well-formed. 

\subsection{Protocol Safety with Refinement}\label{sec.global.safety}
\hide{
Since multiparty protocols involve transmissions over channels
which may be shared among multiple asynchronous senders and
multiple receivers,
designing such communication becomes a
struggle in employing 
only race-free shared channels.
Even elementary protocols which involve only two
transmissions could easily form
flimsy communication such as the one below:\\
\centerline{\code{\transmit{A}{B}{1}{\chanvar} ~\seq~ \transmit{A}{C}{\text{``foo''}}{\chanvar}}}\\
Though the two transmissions initiated by \code{A} are ordered,
because sending is asynchronously there is
no guarantee which peer reads first, \code{B} or \code{C},
leading thus to a race on channel \code{\chanvar}.
}
As highlighted in the overview of this paper,
even simple protocols which only involve two transmissions
can easily lead to flimsy communication, where
either the receivers, (\code{\transmit{A}{C}{}{} ~\seq~
  \transmit{B}{C}{}{}}), 
or the senders, (\code{\transmit{A}{B}{}{} ~\seq~
  \transmit{A}{C}{}{}}), 
race for the same channel.

To avoid such race conditions it is essential
to study the linearity of the communication. A communication is said
to be linear when all the shared channels are used in a linear
fashion,
or in other words when a send and its corresponding receive
are temporally ordered. 
Since the communication implementation 
is guided by the communication protocol, it is therefore essential
for the protocol to satisfy this safety principle as well.
In the subsequent we proceed in:
\begin{itemize}
  \item defining a minimal set of causality relations relevant in the
    study of linearity and build an ordering system to reason
    about these relations.
  \item defining race-freedom \wrt these causality relations
    in the context of asynchronous communication.
  \item transform any given protocol into a race-free protocol
    via a refinement phase which adds assumptions about the 
    causal relation between ordered events,
    and guards to enforce race-free communication.
\end{itemize}

\hide{
  Solve,
  collect
  update protocol
}


\begin{figure}
  \begin{center}
  \setcounter{subfigure}{0}
  \begin{subfloat}[][Syntax of the ordering-constraints language]    
    { \label{ALsyntax}
      \footnotesize
      $ \arraycolsep=1pt
      \begin{array}{llll}
        Send/Recv~Event~& \code{\ev} & {~ ::=} &\code{\rl{P}{\atid}}\\
        Ordering~Constraints~ &\code{\orderas} & {~ ::=} &
                                                     \code{\ltcb{\ev}{\ev}}
                                                     ~|~ \code{\lthb{\ev}{\ev}}\\
        Race-Free~Assertions~ &\code{\racefreeas} & {~ ::=}  &
                                                      \code{\ev}
                                                     ~|~ \code{\notop{\ev}}
                                                     ~|~ \code{\orderas}
                                                     ~|~ \code{\racefreeas \wedge \racefreeas} ~|~ 
              \code{\ev \implies\racefreeas}
      \end{array}
      $
    }
  \end{subfloat}\hfill
  \begin{subfloat}[][Constraint propagation rule]
    {\label{ALrules}
       \footnotesize
    $
    \begin{array}{lll}
      \code{\lthb{\ev_1}{\ev_2} \wedge
      \lthb{\ev_2}{\ev_3}} & \code{\implies
                            \lthb{\ev_1}{\ev_3}
                            } & \code{\tiny\text{[HB-HB]}} \\
      \code{\ltcb{\ev_1}{\ev_2} \wedge
      \lthb{\ev_2}{\ev_3}} & \code{\implies
                                                     \lthb{\ev_1}{\ev_3}}
                                                       &
                                                         \code{\tiny\text{[CB-HB]}} \\
\hide{      \code{\lthb{\rl{P_1}{\atid_1}}{\rl{P_2}{\atid_2}} \wedge
      \ltcb{\rl{P_2}{\atid_2}}{\rl{P_3}{\atid_2}}} & \code{\implies
                                                     \lthb{\rl{P_1}{\atid_1}}{\rl{P_3}{\atid_2}}}
                                                       &
                                                         \code{\tiny\text{(HB-CB)}} \\}
    \end{array}$
  }\end{subfloat}\\
  
  \begin{subfloat}[][Semantics of race-free assertions]
    {\label{ALsemantics}
      \footnotesize
      $
      \begin{array}{ll}
        \code{\racefree{\morders}{\ev}
        }  & \text{iff~} \code{ \ev \in \morders}
        \\
        \code{\racefree{\morders}{\notop{\ev}}
        }  & \text{iff~} \code{ \ev \notin \morders}
      \end{array} \hspace{+1.5em}
      \begin{array}{ll}
        \code{\racefree{\morders}{{\ev {\implies}\racefreeas}}}
        & \text{iff~}
          \code{\notop{\racefree{\morders}{\ev }}}
          \text{~or~}
          \code{\racefree{\morders}{\racefreeas }}\\
        \code{\racefree{\morders}{{\racefreeas_1} \wedge
        {\racefreeas_2}}}     &
                                \text{iff~}
                                \code{\racefree{\morders}{\racefreeas_1}}
                                \text{~and~} 
                                \code{\racefree{\morders}{\racefreeas_2}}\\
        \code{\racefree{\morders}{\lthb{\ev_1}{\ev_2}
        }}
        & \text{iff~} \code{
          (\bigwedge\limits_{\racefreeas \in \morders}{\racefreeas})
          \implies^*  {\lthb{\ev_1}{\ev_2} }}\\
      \end{array} 
      $
      
    }\end{subfloat} \\
  \hrule
  \vspace*{-3mm}
  \caption{\sessionlogic~III: The Ordering-Constraints Language}
  \label{fig.ordering_assertions} \vspace{-3mm}
\end{center}
\end{figure}

\noindent \emph{\bf Ordering Constraint System.}
Given a set of events \code{\eventtyp} and
a relation \code{\mathcal{R} \subseteq \eventtyp \times \eventtyp},
we denote by \code{\ev_1 \prec_{\mathcal{R}} \ev_2}
 the fact that  \code{(\ev_1,\ev_2) \in \mathcal{R} }.
We next distinguish between two kinds of relations
the so called
\emph{``happens-before''} relations to reflect
the temporal ordering between events,
and \emph{``\cb''} relations
to relate communicating  peers.

\anay{can we define HB below more formally?}
\wnnay{Have we defined E.id?}\scnay{Yes, mentioned in previous section under Event.}


\begin{defn}[Communicates-before]
  A {\cb} relation \code{\CB~} is defined as:\\
        \centerline  { 
      \code{\{(\ev_1,\ev_2)~|~ \exists~ \atid~ \cdot~ \evlbl{\ev_1}{=}\evlbl{\ev_2}{=}\atid
      \text{~and~} \ev_1{=}\tsend{\atid} ~\text{and}~ \ev_2{=}\trecv{\atid}\}}.
}
\end{defn}

The {\CB} relation is not transitive,
since its purpose is to simply assist the transitivity of {\HB}
by relating events which involve different peers.
The {\CB} is however irreflexive since we do not allow
for a message to be sent and received by the same party, and
antisymmetric since each event is unique (described by the
\code{\roletyp \times \labeltyp}
pair), that is, it only occurs within a single transmission. 

\begin{defn}[Happens-before]
  Given a global protocol \code{G}, 
  two events \code{\ev_1 \in \EV{G}} and \code{\ev_2 \in \EV{G}}
  are said to be in a happens-before relation in \code{G} 
   if  \code{(\evlbl{\ev_1},\evlbl{\ev_2}) \in \ord{G}}
or it can be derived via a closure using propagation rules
\code{\text{[HB-HB]}} and \code{\text{[CB-HB]}}.
  This is denoted by
  \code{\lthb{\ev_1}{\ev_2}}.
\end{defn}

Similar to the happened-before a la Lamport \cite{lamport1978time},
the \HB~ relation is transitive, irreflexive  and asymmetric,
therefore \HB~ is a strict partial order over events. 
However, one {\em novel} aspect of our approach is the use
of \code{\text{[CB-HB]}} propagation lemma. We use this
exclusively to prove race-freedom for each pair of adjacent transmissions
over some common channel. Consider this pair to be
\code{\ltcb{S_1}{R_1}} and \code{\ltcb{S_2}{R_2}}.
Assume we have already established \code{\lthb{R_1}{S_2}},
and that no other transmissions with the same
channel are under consideration.
  From this scenario, we are unable to conclude
  \code{\lthb{R_1}{R_2}} since
  the following sequence
  \code{R_2;S_1;R_1;S_2} is possible
  which
  is consistent \code{\lthb{R_1}{S_2}} but yet violate
  \code{\lthb{R_1}{R_2}}.
  However, we can conclude
  \code{\lthb{S_1}{S_2}} since
  for \code{\lthb{R_1}{S_2}} to hold, it must be the case that
  \code{\lthb{S_1}{R_1}}. By transitivity of
  \code{\lthb{}{}}, we can therefore conclude \code{\lthb{S_1}{S_2}}.

To reason about these orderings, we propose
a constraint ordering
system whose syntax is depicted in \figref{ALsyntax}.
The send and receive events are related using either a
{\HB} or a {\HB} relation, while the 
ordering constraints are composed using either \code{\wedge}  or
\code{\vee}. The language supports \code{\notop{\ev}} to indicate
that an (immutable) event has not occurred yet.
\wnnay{Do we need to distinguish mutable 
vs immutable events}.

The semantics of the race-free assertions and ordering constraints
is given in \figref{ALsemantics}, where the proof context 
is a
set of orderings, \code{\morders},
with elements from 
\figref{ALsyntax}.
The implication, \code{\implies^*} is solved by
repeatedly applying the 
propagation rules (\code{\footnotesize\text{[CB-HB],~[HB-HB]}}) from \figref{ALrules}.

\noindent \emph{\bf Race-free protocol.}
For brevity of the presentation,
the transmitted messages are ignored in the
rest of this subsection, that is, a transmission
is described as a tuple
\code{ \roletyp \times \roletyp \times \idtyp \times \chantyp}.


\hide{
Since transmission can be uniquely identified by its \code{ID},
send or receive operations can also be uniquely identified by the role-ID
pair \code{\rl{P}{\atid}} (of type \code{\roletyp \times \idtyp}).
A set of operations associated with the same
role is represented by \code{\rl{P}{\atloc} \in \roletyp \times
  \idtyp^*}.
}

\anay{
  we should\\
  1. define RF transmissions, RF channel \\
  2. define adjacent transmissions wrt a channe; \\
  3. state and prove the rf theorem: a channel is RF if any 2 adjacent
  transmissions are RF}

\begin{defn}[Race-free adjacent transmissions]
  \label{def:race_free_trans}
  Given a protocol \code{\prot}, two adjacent transmissions
  \code{\atid_1\in\prot} and
  \code{\atid_2\in\prot}
  are said to be race-free \wrt to each other,
  denoted by \code{\racefreerel{\atid_1}{\atid_2}}, 
  only when they are in a \HB~ relation: \\
  \centerline{\code{
      \forall {\atid_1},{\atid_2} \in \prot \cdot (\adj{\atid_1}{\atid_2}
      \implies \lthb{\atid_1}{\atid_2}) 
}}\\
where  {\code{
    \lthb{\atid_1}{\atid_2}}} stands for 
{\code{
    \lthb{\tsend{\atid_1}~}{\tsend{\atid_2}}
    \wedge
    \lthb{\trecv{\atid_1}~}{\trecv{\atid_2}}
  }}.
\end{defn}

\begin{defn}[Race-free protocol]
  \label{def:race_free_prot}
  A protocol \code{\prot} is race-free, denoted by
  \code{\racefreerel{\prot}{}},
  when all the linked transmissions are in a \HB~ relation:\\
    \centerline{ 
        \code{
          \forall \atid_1,\atid_2 \in \prot \cdot
          (\link{\atid_1}{\atid_2}
          \implies
          \lthb{\atid_1}{\atid_2})
        .
     }
  }
\end{defn}

\begin{theorem}[Race-free protocol]
  \label{thm:race-free}
  A protocol is race-free  if every pair of adjacent transmissions are race free:\\
  \centerline{ 
    \code{
      (\forall \atid_1,\atid_2 \in \prot \implies 
      \racefreerel{\atid_1}{\atid_2}) \implies \racefreerel{\prot}{}.
   }}
\end{theorem}

\begin{proof}
  Using inductive proving on the definition of linked transmissions
  and the transitivity of \HB. See Appendix, \ref{proof-race-free}. 
\end{proof}
\hide{
\begin{proof} Since we carefully designed our relations as partial
  order
  relations, the proof of this theorem follows directly from that
  of partial order relation properties. We design the proof as a case analysis on
  the possible channel sharing scenarios.
  \emph{Case 1.}
  None of the transmissions in \code{\prot} share a common channel:
  \code{\forall \atid_1,\atid_2 \in \prot \implies
    (\tchan{\atid_1}{\neq}\tchan{\atid_2})}, therefore there are no
  linked transmission. It then follows directly 
  from the definition of  \code{\racefreerel{\atid_1}{\atid_2}} and
  that of 
  \code{\racefreerel{\prot}{}} that there is no race in \code{\prot}.
  \emph{Case 2.}
  If all the channels in \code{\prot} are shared by at most two
  transmissions, then these two transmissions are adjacent:
  \code{\forall \atid_1,\atid_2 {\in} \prot \implies
    (\tchan{\atid_1}{\neq}\tchan{\atid_2}) \vee
    (\adj{\atid_1}{\atid_2} \wedge \neg(\exists \atid'\cdot
    \adj{\atid'}{\atid_1} \vee \adj{\atid_2}{\atid'})) }.
  Assuming that all the adjacent transmissions are also race-free:
  \code{\forall \atid_1,\atid_2 {\in} \prot \cdot \adj{\atid_1}{\atid_2}
    \implies \lthb{\atid_1}{\atid_2}},
  then by the base case definition of 
  linked transmissions (\figref{fig.fig_def})
  we can conclude that
  \code{\forall \atid_1,\atid_2 {\in} \prot \cdot \link{\atid_1}{\atid_2}
    \implies \lthb{\atid_1}{\atid_2}}, that is 
  \code{\racefreerel{\prot}{}} holds.
  \emph{Case 3.} All the channels in \code{\prot} are shared by
  an arbitrary number
  of transmissions. It is enough to show that
  if any tree adjacent transmissions sharing the same channel are
  race-free then \code{\prot} is race-free. We know that
  \code{\forall \atid_1,\atid_2,\atid_3 {\in} \prot
    \cdot \adj{\atid_1}{\atid_2} \wedge \adj{\atid_2}{\atid_3}
    \implies \lthb{\atid_1}{\atid_2} \wedge \lthb{\atid_2}{\atid_3} },
    which by the definition of linked transmissions means that    
    \code{\forall \atid_1,\atid_2,\atid_3 {\in} \prot
    \cdot \link{\atid_1}{\atid_3} 
    \implies \lthb{\atid_1}{\atid_2} \wedge \lthb{\atid_2}{\atid_3}}.
  Using the transitivity of \HB~ results in
  \code{\forall \atid_1,\atid_2,\atid_3 {\in} \prot
    \cdot \link{\atid_1}{\atid_3} 
    \implies \lthb{\atid_1}{\atid_3}}, therefore
  according to the definition of \code{\racefreerel{\prot}{}},
  \code{\prot} is race-free.
\end{proof}
}

\hide{
\begin{defn}[Race-free channel]
  \label{def:race_free}
  A channel \code{\chanvar} is said to be race-free with respect to a
  given protocol \code{\prot}
  if for any two adjacent 
  transmissions on this channel, say
  \code{\atransmit{S_1}{R_1}{}{\chanvar}{\atid_1}} and
  \code{\atransmit{S_2}{R_2}{}{\chanvar}{\atid_2}} in this order, 
  the following assertion holds:
  {\footnotesize
    \begin{align}
      \code{ {\lthb{\rl{S_1}{\atid_1}}{\rl{S_2}{\atid_2}} \wedge
      \lthb{\rl{R_1}{\atid_1}}{\rl{R_2}{\atid_2}}}}  
      \label{rf2a} 
    \end{align}
  } 
\end{defn} 

\noindent
Race-freedom is guaranteed by the FIFO property of channels.

\begin{defn}[Race-free protocol]
  \label{def:race_free_protocol}
  A protocol \code{\prot} is said to be race-free if
  all the channels employed for 
  communication
  are race-free. 
\end{defn}
}

As mentioned before, we are assuming a hybrid communication
system, where both message passing and other explicit
synchronization mechanisms are used. Therefore,
a simple analysis to check whether the protocol is 
linear is insufficient. Instead of analyzing the protocol
for racy channels, we designed
an algorithm - \code{\bf Algorithm~ 1} - which inserts the race-free
constraints as explicit proof obligations 
within the protocol.
We rely on the program verification to detect
when these constraints are not satisfied.

\def\CCol{collect}

\newcommand{\raceFree}{generateRaceFreeProt}
\newcommand{\addfidelity}{addFidelity}
\newcommand{\addsafety}{addGuards}
\newcommand{\addassumptions}{addAssumptions}
\newcommand{\insertes}{insertES}

 \IncMargin{1em}
\begin{algorithm}[]
\SetKwProg{Fn}{Function}{}{}
\SetKwFunction{addFidelityAssrt}{\addfidelity}
\SetKwFunction{addSafetyAssrt}{\addsafety}
\SetKwFunction{addAssumptions}{\addassumptions}
\SetKwData{Left}{left}\SetKwData{This}{this}\SetKwData{Up}{up}
\SetKwFunction{Union}{Union}\SetKwFunction{FindCompress}{FindCompress}
\SetKwInOut{Input}{input}\SetKwInOut{Output}{output}
\let\oldnl\nl
\newcommand{\nonl}{\renewcommand{\nl}{\let\nl\oldnl}}
\SetAlCapHSkip{0pt}
{
  {\Indm  
    \Input{\code{\prot} - a global multi-party protocol }
    \Output{\code{\prot'} - refined \code{\prot}
      }}
  \BlankLine
  \bind{\code{\os}}{\code{\CCol(G)}}
  \\
  \bind{\code{\prot'}}{\addSafetyAssrt{\code{\prot},\code{\osgrd{\os}} }}
  \\
  \bind{\code{\prot'}}{\addAssumptions{\code{\prot'},\code{\osasm{\os}}
    }}
    \\
  \Return{\code{\prot'}}
}
\caption{Decorates a protocol with ordering assumptions, and race-free guards}
\end{algorithm} 


The algorithm is pretty straightforward:
it first collects the necessary ordering assumptions and 
race-free guards, and then inserts them within
the global protocol using a simple scheme guided by
the unique transmission identifier. The methods used
for insertion are not described here for limit of space,
but also because they do not represent any 
theoretical or technical interest. As an intuition though,
the insertion follows these principles:
\wnnay{last line is ambiguous}
   \code{\assume{\event{\rolevar}{\atid}}}
  is inserted immediately after
  transmission \code{\atid}; 
  \code{\assume{\ltcb{\event{S}{\atid}}{\event{R}{\atid}}}}
  is inserted immediately after
  transmission \code{\atid};
  \code{\assume{\lthb{\event{P_1}{\atid_1}}{\event{P_2}{\atid_2}}}}
  and \code{
    \guard{\lthb{\event{P_1}{\atid_1}}{\event{P_2}{\atid_2}}}}
  are inserted immediately after
  transmission \code{\atid_2}.
  According to \code{\bf Algorithm~ 1}
  the assumptions are added to the protocol after inserting the guards,
  therefore, in the refined protocol,
  the assumptions precede the guards relative to the same
  transmission.


\begin{figure}
  \centering
  {\captionsetup[subfigure]{labelformat=empty}  
  \begin{subfloat}[][]
    {\footnotesize
      $ \arraycolsep=1pt
      \begin{array}{llll}
        Base~Element~ &\code{\cbformula~\,a} & {~ ::=} & \code{a} ~|~
                                                       \code{(\cbformula~a)\useq(\cbformula~a)
                                                       } \\
        Border~Element~&\code{\cformula~\,a} & {~ ::=} &
                                                       \code{\bot} ~|~
                                                       \code{\cbformula~a}~|~ 
                                                       \code{(\cformula~a)\vee(\cformula~a)} \\
         Border~Event~
        &\code{\oborder} & {~ ::=} & \code{\cformula~\,\rl{P}{\atid}}\\
         Border~Transmission~
        &\code{\tborder} & {~ ::=} & \code{\cformula~\,
                                                         \atransmit{P}{P}{}{\chanvar}{\atid}}
      \end{array}
      $
    }
  \end{subfloat} }\\  
  {   \captionsetup[subfigure]{labelformat=empty,farskip=0pt,captionskip=1pt}  
  \subfloat[][]
    {\footnotesize
      $ \arraycolsep=0pt
      \begin{array}{ll}
        \code{\text{\emph{(Operation~Map)}}} & \code{ {RMap} \eqdef \roletyp
        \rightarrow \oborder}  \\
        \code{\text{\emph{(Transmission~Map)}}\quad}  & \code{ {CMap} \eqdef \chantyp
        \rightarrow \tborder} \\
        \code{\text{\emph{(Boundary)}}}  & \code{ Border \eqdef RMap \times CMap}
      \end{array}
      $
    }}
  \\ \vspace*{-2mm}
\hide{
\subfloat[][Properties of guards]
{\footnotesize
  \arraycolsep=2pt
$
      \begin{array}{ll}
        \code{\guard{\racefreeas_1 {\wedge} \racefreeas_2} {\wedge} \guard{\racefreeas_3}} &
        \code{{=} ~ \guard{\racefreeas_1} {\wedge} \guard{ \racefreeas_2 {\wedge} \racefreeas_3}}\\
        \code{\guard{\racefreeas_1 {\wedge} \racefreeas_2} } &
        \code{{=} ~ \guard{\racefreeas_2 {\wedge}  \racefreeas_1}}\\
        \code{\guard{\racefreeas_1 {\wedge} \racefreeas_2} } &
        \code{{=} ~ \guard{\racefreeas_1} {\wedge}
                                                               \guard{\racefreeas_2}}\\
        \code{\guard{\racefreeas_1 {\vee} \racefreeas_2} {\vee} \guard{\racefreeas_3}} &
        \code{{=} ~ \guard{\racefreeas_1} {\vee} \guard{ \racefreeas_2 {\vee} \racefreeas_3}}\\
        \code{\guard{\racefreeas_1 {\vee} \racefreeas_2} } &
        \code{{=} ~ \guard{\racefreeas_2 {\vee}  \racefreeas_1}}\\
        \code{\guard{\racefreeas_1 {\vee} \racefreeas_2} } &
        \code{{=} ~ \guard{\racefreeas_1} {\vee} \guard{\racefreeas_2}}
      \end{array} 
$
}
\subfloat[][Properties of assumptions]
{\footnotesize
  \arraycolsep=2pt
$
      \begin{array}{ll}
        \code{\assume{\racefreeas_1 {\wedge} \racefreeas_2} \wedge \assume{\racefreeas_3}} &
        \code{{=} ~ \assume{\racefreeas_1} {\wedge} \assume{ \racefreeas_2 {\wedge} \racefreeas_3}}\\
        \code{\assume{\racefreeas_1 {\wedge} \racefreeas_2} } &
        \code{{=} ~ \assume{\racefreeas_2 {\wedge}  \racefreeas_1}}\\
        \code{\assume{\racefreeas_1 {\wedge} \racefreeas_2} } &
        \code{{=} ~ \assume{\racefreeas_1} {\wedge}  \assume{\racefreeas_2}}\\        
        \code{\assume{\racefreeas_1 {\vee} \racefreeas_2} \vee \assume{\racefreeas_3}} &
        \code{{=} ~ \assume{\racefreeas_1} {\vee} \assume{ \racefreeas_2 {\vee} \racefreeas_3}}\\
        \code{\assume{\racefreeas_1 {\vee} \racefreeas_2} } &
        \code{{=} ~ \assume{\racefreeas_2 {\vee}  \racefreeas_1}}\\
        \code{\assume{\racefreeas_1 {\vee} \racefreeas_2} } &
        \code{{=} ~ \assume{\racefreeas_1} {\vee}  \assume{\racefreeas_2}}\\        
      \end{array} 
$
}
}
\hrule
\vspace*{-3mm}
\caption{Elements of the boundary summary.}\label{fig.boundarysum}
\vspace*{-3mm}
\end{figure}



The \code{\CCol} method derives the
ordering relations necessary for the refinement
of a user-defined protocol into a race-free protocol.
In a top-down approach, \code{\CCol} compiles 
protocol summaries for each compositional
sub-protocol, until each sub-protocol is reduced to a single
transmission. It then gradually merges each such summary
to obtain the global view of the all orderings within the initial protocol.

\noindent{\bf Protocol Summary.} The summary of a protocol \code{\prot} 
is a tuple
\code{\os{=}{\langle}\mback,\mfront,\massum,\mguards{\rangle}},
where \code{\mback} is the left boundary, also called a \emph{backtier},
mapping roles to their first occurrence and channels to their first transmissions
within \code{G} \scnay{why not `protocol'
  but `sub-protocol' here? A bit confusing to me; similar later}
\anay{indeed, i changed},
\code{\mfront} is the right boundary, also called \emph{frontier},
mapping roles to their last occurrence and channels to their last transmissions
within \code{G},
\code{\massum} is a set of ordering assumptions
which reflect the implicit synchronization ordering relations, and
\code{\mguards} is a set of ordering proof obligations 
needed to ensure the race-freedom of \code{\prot}.
We denote by
\code{\osbck{\os}}, \code{\osfrt{\os}}, \code{\osasm{\os}}
and \code{\osgrd{\os}} the elements of the summary.
The \code{\osasm{\os}} and \code{\osgrd{\os}} sets
are the result of fusing together 
the communication summaries of each compositional protocol.

\noindent{\bf Protocol Boundary.} Each boundary, whether left or right, 
is a pair of maps \code{{\langle}\rolemap,\chanmap{\rangle}},
where \code{\rolemap} relates protocol roles to events
and \code{\chanmap} relates channels to transmission.
The fact that a map is
not defined for a particular input
is indicated by \code{\atzero}.
The constituent elements of these maps are formally described in
\figref{fig.boundarysum}, where \code{\oborder} is used to populate
the events map -
\code{\rolemap}, and the \code{\tborder} elements populate the
transmissions map -
\code{\chanmap}. We denote by \code{\bordrmap{\mfront}} and 
\code{\bordcmap{\mfront}} the elements of a boundary \code{\mfront}.

Finally, the decomposition of the protocol into sub-protocols and
the building of summaries is recursively 
defined as below, where the
helper functions, \code{\composeseqop}, \code{\composestarop} and
\code{\composeveeop},
merge the summaries specific to each
protocol composition operator:

\hide{
\noindent \code{\rolemap : {RMap} \eqdef \roletyp \rightarrow \roletyp
  \times \idtyp}
maps roles to transmission identifiers. The fact that \code{\rolemap} is
not defined for a particular role \code{P} is indicated by \code{\rolemap(R){=}\atzero}.\\
\code{\chanmap : {CMap} \eqdef \chantyp \rightarrow \roletyp \times \roletyp \times
  \idtyp} maps channels to transmissions.
The fact that \code{\chanmap} is
not defined for a particular channel \code{\chanvar}  is indicated by \code{\chanmap(R){=}\atzero}.\\
\noindent \code{\mback, \mfront : Border \eqdef {RMap} \times {CMap}} the back and
front boundaries of a protocol.\\
\noindent \code{\locof: \roletyp \times \roletyp \times \idtyp
  \rightarrow \idtyp}, such that \code{\locof(\rrl{P_1}{P_2}{i}){=}\atid}.\\
\noindent \code{\CCol : L^a(G) \rightarrow Border \times Border \times
  {\mathcal{O}^*} \times {\mathcal{E}^*}} collects a set of proof
obligations of
type \code{\mathcal{E}} in order to identify the
locations where the implicit
synchronization is not sufficient for ensuring communication safety. 
}

\begin{figure}[H]
 \centering
 \captionsetup[subfigure]{labelformat=empty,farskip=-5pt}

 \subfloat[][]
 {\footnotesize
  \arraycolsep=2pt
 $
 \begin{array}{lll}
  \code{{\CCol}(\prot_1 \seq \prot_2)} & \code{\eqdef} &
  \code{\composeseq{\CCol(\prot_1)}{\CC(\prot_2)}} 
  \\
  \code{{\CCol}(\prot_1 \useq \prot_2)} & \code{\eqdef} &
  \code{\composestar{\CCol(\prot_1)}{\CCol(\prot_2)}} \\
  \code{{\CCol}(\prot_1 \vee \prot_2)} & \code{\eqdef} &
  \code{\composevee{\CCol(\prot_1)}{\CCol(\prot_2)} }
  \\
 \end{array}
 $
} \hfill
{ \subfloat[][]
{ \footnotesize
  \arraycolsep=2pt
 $
 \begin{array}{lll} 
  \code{{\CCol}(\atransmit{S}{R}{}{\chanvar}{\atid})} & \code{\eqdef} &
     \code{\langle {\langle \rolemap,\chanmap \rangle},~
            {\langle \rolemap,\chanmap \rangle},~\{\assume{\atransmit{S}{R}{}{}{\atid}}\},
  ~{\emptyset}\rangle} 
  \\
   \multicolumn{3}{l}{\qquad\text{where~} \code{\rolemap{=}(S:\event{S}{\atid},R:\event{R}{\atid})} \text{~and~}
   \code{\chanmap{=}(\chanvar:\atransmit{S}{R}{}{}{\atid})}}\\
 \end{array}
 $
}\vfill}
\end{figure}

\begin{figure}[H]
{\footnotesize
  \arraycolsep=2pt
  $
\begin{array}{lll}
  \code{\composeseq{\os_1}{\os_2}}  & \code{\eqdef} &
                  \code{\langle \osbck{\os}_1  \mergeseq \osbck{\os}_2 ,
                         \osfrt{\os}_2  \mergeseq \osfrt{\os}_1,
                         \osasm{\os}_1\cup\osasm{\os}_2\cup \massum_3,
                         \osgrd{\os}_1\cup\osgrd{\os}_2\cup \mguards_3
                                                                  \rangle}\\
  && \text{where~}
     \code{\langle \massum_3,\mguards_3 \rangle {~=~}\mergeseqtier{\osfrt{\os}_1}{\osbck{\os}_2}}
  \\
  \code{\composestar{\os_1}{\os_2}}  & \code{\eqdef} &
  \code{ \langle \osbck{\os}_1 \mergestar \osbck{\os}_2 ,
         \osfrt{\os}_1 \mergestar \osfrt{\os}_2 ,
         \osasm{\os}_1\cup\osasm{\os}_2,
         \osgrd{\os}_1\cup\osgrd{\os}_2
        \rangle}
  \\
  \code{\composevee{\os_1}{\os_2}}  & \code{\eqdef} &
  \code{\langle \osbck{\os}_1 \mergevee \osbck{\os}_2 ,
         \osfrt{\os}_1 \mergevee \osfrt{\os}_2 ,
         \osasm{\os}_1\cup\osasm{\os}_2,
         \osgrd{\os}_1\cup\osgrd{\os}_2
       \rangle}
\end{array}
$
}
\end{figure}

\hide{
(M1,C1) \oplus (M2,C2) =
   let M3 = [i:A:r  | (i:A:r in M1) or (0:A in M1 and i:A:r in M2)] in
   let C3 = [i:A:r  | (i:A->B(k) in C1) or (0(k) in C1 and i:A->B(k) in C2)] in
   (M3, C3)

(M1,C1) \+ (M2,C2) =
   let M3 = [i:A:r  | (i:A:r in M1) or (0:A in M1 and i:A:r in M2)] in
   let C3 = [i:A:r  | (i:A->B(k) in C1) or (0(k) in C1 and i:A->B(k) in C2)] in
   (M3, C3)
 }
 
 \noindent where the \code{\bordermerge{op}} operator (over-loaded over boundaries
 and maps' elements)
 represents the  fusion
specific to each protocol connector:

\begin{figure}[H]
\centering
\captionsetup[subfigure]{labelformat=empty,farskip=-5pt}
\subfloat[][]
{\footnotesize
  $ 
  \arraycolsep=2pt
  \begin{array}{l}
  \code{{\langle \rolemap_1,\chanmap_1  \rangle} \bordermerge{op} {
    \langle \rolemap_2,\chanmap_2  \rangle}}   \code{~\eqdef~} 
                                                                                                 \code{
    \langle \rolemap,\chanmap  \rangle}\\
  \text{~where~} 
    \code{\rolemap{=} \rolemap_1 [P:\rolemap_1(P) \bordermerge{op}
    \rolemap_2(P)]_{\forall P \in dom(\rolemap_2)}}
  \\                                                      
   \quad\qquad\code{\chanmap{=} \chanmap_1 [ \chanvar:\chanmap_1(\chanvar) \bordermerge{op} \chanmap_2(\chanvar)]_{\forall \chanvar \in dom(\chanmap_2)}} 
\end{array}
$}\hfill
{\footnotesize
  $ 
  \arraycolsep=2pt
  \begin{array}{ll}
  \code{{\eborder_1} \mergeseq {\eborder_2}} &
  \code{ \eqdef
    \begin{cases}
      \code{{\eborder_2}} & \text{if~} \code{\eborder_1{=}\atzero} \\
      \code{{\eborder_1}} & \text{otherwise}\\
    \end{cases}}\\
    \code{{\eborder_1} \mergevee {\eborder_2}} &
    \code{\eqdef
    {\eborder_1} \vee {\eborder_2} 
  }
  \end{array}
$}\hfill
\subfloat[][]{\footnotesize
  $ 
  \code{{\eborder_1} \mergestar {\eborder_2}
    \eqdef
    \begin{cases}
      \code{{\eborder_1}} & \text{if~} \code{\eborder_2{=}\atzero} \\
      \code{{\eborder_2}} & \text{if~} \code{\eborder_1{=}\atzero} \\
      \code{{\eborder_1} \useq {\eborder_2}}
      & \text{~otherwise}
    \end{cases}
  } 
  $
}
\end{figure}

\noindent \code{\rolemap[P:\oborder]} denotes an update to \code{\rolemap},
such that the value corresponding to \code{P} is
updated to \code{\oborder}, even if \code{\rolemap} was previously not defined
on \code{P}. Similarly,
\code{\chanmap[\chanvar:\tborder]}
indicates an update to \code{\chanmap} such that \code{\chanvar}
is mapped to \code{\tborder}.

In the case of backtier fusion for sequence,
\code{\osbck{\os}_1  \mergeseq \osbck{\os}_2}
ensures that the resulted backtier  
reflects all the first transmissions
for each channel employed by either \code{\prot_1} or \code{\prot_2},
whichever first,
and all possible first events for each role in the sequence.  
Dually, \code{\osfrt{\os}_2  \mergeseq \osfrt{\os}_1},
ensures
all last transmission and events with respect to the considered
sequence
will be captured by the newly derived summary.

One of the key points of this phase is captured by
the merge of adjacent boundaries, \code{\osfrt{\os}_1}  and
\code{\osbck{\os}_2}, respectively, merge which generates
a set of assumptions, \code{\massum_3},  and a set of proof obligations, \code{\mguards_3}.
The assumptions 
capture the happens-before relation between
adjacent events, namely 
the last event
of \code{\prot_1} and the first event of \code{\prot_2} with respect to
a particular role. Similarly, the guards capture 
the race-free conditions for all the adjacent transmissions
sharing a common channel. Formally this merged is defined by
the \code{\mergeset} function:

\begin{figure}[H]
\centering
\noindent
{\footnotesize
  \arraycolsep=2pt
  $
  \begin{array}{ll}
    \multicolumn{2}{l}{\code{\mergeseqtier{\langle\rolemap_1,\chanmap_1  \rangle}
    {\langle \rolemap_2,\chanmap_2  \rangle}}   \code{~\eqdef~} 
    \code{\langle \rolemap_1 \mergemap \rolemap_2,
    \chanmap_1 \mergemap \chanmap_2 \rangle}} \\
    \text{\quad where} &
                         \code{Map_1 \mergemap Map_2 {=}
                         \bigcup\limits_{Key\in Keys}
                         \ltset{{Map_1(Key)}}{{Map_2(Key)}}}\\
    \text{\quad and} &
                 \code{Keys {=}dom(Map_1) \cup dom(Map_2)}.
                 
  \end{array}
  $
}
\end{figure}

In the following, the recursive function \code{merge} is overloaded
such that it can cater to both event merging (first base-case definition)
and transmission merging (second base-case definition):

\begin{figure}[H]
\centering
\captionsetup[subfigure]{labelformat=empty,farskip=-5pt}
\subfloat[][]
{\footnotesize
  $
\begin{array}{lll}
  \code{\ltset{{\eborder_1\useq\eborder_2}}{{\eborder}}}  & \code{\eqdef} &
  \code{\ltset{{\eborder_1}}{{\eborder}} \cup \ltset{{\eborder_2}}{ {\eborder}}}\\
  \code{\ltset{{\eborder}}{{\eborder_1\useq\eborder_2}}}  & \code{\eqdef} &
  \code{\ltset{{\eborder}}{{\eborder_1}} \cup \ltset{{\eborder}}{{\eborder_2}}}\\
  \code{\ltset{{\eborder_1\vee\eborder_2}}{{\eborder}}}  & \code{\eqdef} &
  \code{\ltset{{\eborder_1}}{{\eborder}} \cup \ltset{{\eborder_2}}{ {\eborder}}}\\
  \code{\ltset{{\eborder}}{{\eborder_1\vee\eborder_2}}}  & \code{\eqdef} &
  \code{\ltset{{\eborder}}{{\eborder_1}} \cup \ltset{{\eborder}}{{\eborder_2}}}\\
\end{array}
$
}\hfill
\subfloat[][]
{\footnotesize
  $
\begin{array}{lll}
  \code{\ltset{\rl{P}{\atid_1}}{ \rl{P}{\atid_2}}}  & \code{\eqdef} &
                      \code{\assume{\lthb{\rl{P}{\atid_1}}{\rl{P}}{\atid_2} }}\\
  \code{\ltset{\atid_1}{\atid_2}} & \code{\eqdef} & \code{\guard{\lthb{\atid_1}{\atid_2}}}\\
\end{array}
$
}
\end{figure}



\section{Local Projection} 
\label{sec:projection}
Based on the communication interface, but also
on the verifier's requirements,
the projection
of the global protocol to local specifications
could go through a couple of automatic projection
phases before being used by the verification process.
This way, the projection 
could
describe how each party is contributing to
the communication, 
or it could be more granular
describing how each communication instrument
is used with respect to a communicating party.

\noindent\textbf{Projection Language}.
\figref{fig.proj_syntax} describes the two
kinds of specification mentioned above. 
The per party specification language 
is depicted in \figref{fig.proj_syntax_per_party}. Here, each send and receive
specification name the communication
instrument \code{\chanvar} along with a message \code{v}
described by a formula \code{\St}. In the per channel specifications,
\figref{fig.proj_syntax_per_channel}, the
communication instrument is implicit.
The congruence of all the compound terms described in
\secref{sec:global} holds for 
the 
projected languages as well, with the exception
of sequential commutativity since the disjointness conditions for
the latter do not hold (eg. either the peer or the channel
are implicitly the same for the entire projected specification).


\begin{figure}
\begin{center}
{
  \captionsetup[subfigure]{labelformat=empty,farskip=1pt,captionskip=1pt}
\subfloat[][]
{\footnotesize
  $\arraycolsep=1pt
    \begin{array}{l}
      Local~protocol \\
      Send/Receive/Transmission \\
      HO~ variable\\
      Concurrency \\
      Choice \\
      Sequence \\
      Guard/ Assumption 
    \end{array}
  $
}}
{\setcounter{subfigure}{0}
\subfloat[][Per party ]
{\footnotesize
  \label{fig.proj_syntax_per_party}
  $\arraycolsep=1pt
    \begin{array}{ll}
      \code{\quad \projp} {~ ::=} &\\
      & ~\code{\,\; \sendc{\chanvar}{v}{\St}}  
       ~|~ \code{\recvc{\chanvar}{v}{\St}}  \\
      & ~|~ \code{\hovars} \\
      & ~|~ \code{\projp {\useq} \projp }  \\
      & ~|~ \code{\projp {\vee} \projp }  \\
      & ~|~ \code{\projp{\seq} \projp}\\
      & ~|~ \code{\guard{\racefreeas}} 
       ~|~ \code{\assume{\racefreeas}}
    \end{array}
  $
}}
\subfloat[][Per endpoint]
{\footnotesize
  \label{fig.proj_syntax_per_endpoint}
  $\arraycolsep=1pt
    \begin{array}{ll}
      \code{\quad \projc} {~ ::=} &\\
      & ~\code{\,\; \send{v}{\St}}  
       ~|~ \code{\recv{v}{\St}}  \\
      & ~|~ \code{\hovars} \\
      & \\
      & ~|~ \code{\projc {\vee} \projc }  \\
      & ~|~ \code{\projc{\seq}\projc}\\
      & ~|~ \code{\guard{\racefreeas}}
       ~|~ \code{\assume{\racefreeas}} 
    \end{array}
  $
}
\subfloat[][Per channel]
{\footnotesize
  \label{fig.proj_syntax_per_channel}
  $\arraycolsep=1pt
    \begin{array}{ll}
      \code{\quad \cproj} {~ ::=} &\\
      & ~\code{{\atransmit{\rolevar_1}{\rolevar_2}{v\,{\cdot}\,\fmsg}{}{\atid}}}   \\
      & \\
      & ~|~ \code{\cproj{\useq}\cproj} \\
      & ~|~ \code{\cproj {\vee} \cproj }  \\
      & ~|~ \code{\cproj{\seq}\cproj}\\
      & ~|~ \code{\guard{\racefreeas}}
       ~|~ \code{\assume{\racefreeas}} 
    \end{array}
  $
}
\hrule
\vspace*{-3mm}
\caption{\sessionlogic~ IV: The Projection Language}
\label{fig.proj_syntax} \vspace{-3mm}
\end{center}
\end{figure}

\noindent\textbf{Automatic projection}.
Using different projection granularities should not
permit event re-orderings (modulo \code{\useq} composed events).
\newtheorem{prop}{Proposition}
\begin{prop}[Projection Fidelity]
The projection to a decomposed specification, such as
global protocol to per party, or per party to per channel,
does not alter the communication pattern specified before projection.
\end{prop}

To support the above proposition, we have designed a set of 
structural projection rules, described in \figref{fig.projection}.
\figref{fig.projection.party}, describing
per party projection rules, is
quite self-explanatory, with the exception of the guard projection
rule. The latter distinguishes between the projection on the party which
needs to prove the guarded happens-before ordering 
before assuming it, from the party which can soundly
assume the ordering without prior proof. 

As expected, the per channel projection rules, \figref{fig.projection.channel},
strips the channel information from the per party specifications,
since
it will be implicitly available. 
Furthermore, inserting a guard \code{\guard{\event{\rolevar}{\atid}}} 
between adjacent transmissions on different channels with a common sender
ensures that the order of events
at the sender's site 
is accurately inherited 
from the corresponding per party specification
across
different channels. To emphasize this behavior
we consider the following sequence of receiving
events captured by a per-party specification,
\code{\project{\rolevar}{\prot}}:

\begin{figure}
\begin{center}
{\captionsetup[subfigure]{farskip=1pt,captionskip=-5pt}
  \subfloat[][global spec \code{\rightarrow} per party spec]{
  \label{fig.projection.party}
  \footnotesize
  \arraycolsep=3pt
  \captionsetup[subfigure]{labelformat=empty,farskip=1pt,captionskip=0pt}
  \shortstack{
    \subfloat[][]{
    $
    \begin{array}{ll}
    \code{\project{\rolevar}{\atransmit{\resid}{\R}{\St}{\chanvar}{\atid}}} &~:=~
    \left\{
         \begin{array}{ll}
           \code{\sendc{\chanvar}{v}{\St}} & \code{\text{if\,}
                                             \rolevar{=}\resid}\\
           \code{\recvc{\chanvar}{v}{\St}} & \code{\text{if\,}
                                             \rolevar{=}\R}\\
           \code{\emp} & \code{\text{otherwise}}
         \end{array}
     \right.\\
      \code{\project{\rolevar}{\prot_1{\useq}\prot_2}} & ~:=~
      \code{\project{\rolevar}{\prot_1} \useq \project{\rolevar}{\prot_2}} \\
      \code{\project{\rolevar}{\prot_1{\vee}\prot_2}} &
      ~:=~\code{\project{\rolevar}{\prot_1} \vee \project{\rolevar}{\prot_2}}\\
      \code{\project{\rolevar}{\prot_1{\seq\,}\prot_2}} & ~:=~
      \code{\project{\rolevar}{\prot_1}\, \seq \,\project{\rolevar}{\prot_2}} \\
      \code{\project{\rolevar}{\assume{\event{\rolevar_1}{\atid}}}} & ~:=~
      \left\{  
         \begin{array}{ll}
           \code{\assume{\event{\rolevar}{\atid}}} & \code{\text{if\,}
                                             \rolevar{=}\rolevar_1}\\
           \code{\emp} & \code{\text{otherwise}}
    \end{array}
    \right.
    \\
    \code{\project{\rolevar}{\guard{\lthb{\event{\rolevar_1}{\atid_1}}{\event{\rolevar_2}{\atid_2}}} }} & ~:=~
    \left\{
         \begin{array}{ll}
           \code{\guard{\lthb{\event{\rolevar_1}{\atid_1}}{\event{\rolevar_2}{\atid_2}}}
           } & \code{\text{if\,}
                                             \rolevar{=}\rolevar_2}\\
           \code{\assume{\lthb{\event{\rolevar_1}{\atid_1}}{\event{\rolevar_2}{\atid_2}} }} & \code{\text{otherwise}}
         \end{array}
   \right. 
  \end{array}
  $
}} \setcounter{subfigure}{1}}}\\
{\captionsetup[subfigure]{farskip=1pt,captionskip=-5pt}
\subfloat[][per party spec \code{\rightarrow} per endpoint spec]{
  \label{fig.projection.channel}
  \footnotesize
  \arraycolsep=1pt
  \captionsetup[subfigure]{labelformat=empty,farskip=1pt,captionskip=0pt}
  \shortstack{
  \subfloat[][]{
  $
  \begin{array}{ll}
  \code{\project{\chanvar}{\sendc{\chanvar_1}{v}{\St}}} &~:=~
  \left\{
         \begin{array}{ll}
           \code{\send{v}{\St}} & \code{\text{if\,}
                                             \chanvar{=}\chanvar_1}\\
           \code{\emp} & \code{\text{otherwise}}
         \end{array}
  \right.
  \\
  \code{\project{\chanvar}{\recvc{\chanvar_1}{v}{\St}}} &~:=~
  \left\{
         \begin{array}{ll}
           \code{\recv{v}{\St}} & \code{\text{if\,}
                                             \chanvar{=}\chanvar_1}\\
           \code{\emp} & \code{\text{otherwise}}
         \end{array}
  \right. \\
    \code{\project{\chanvar}{\projp_1{\useq}\projp_2}} & ~:=~
    \left\{
         \begin{array}{ll}
           \code{\code{\project{\chanvar}{\projp_j}}}
           & \code{\text{if~}
                                             \chanvar{\employ}\projp_j,j{=}1~or~2
             }\\
           \code{\emp} & \code{\text{otherwise}}
         \end{array}
   \right.
    \\
  \code{\project{\chanvar}{\projp_1{\vee}\projp_2}} &
  ~:=~\code{\project{\chanvar}{\projp_1} \vee \project{\chanvar}{\projp_2}}\\
  \code{\project{\chanvar}{\projp_1{\seq\,}\projp_2}} & ~:=~
  \code{\project{\chanvar}{\projp_1}\, \seq\, \project{\chanvar}{\projp_2}}  \\
  \code{\project{\chanvar}{\assume{\event{\rolevar}{\atid}}}} & ~:=~
   \left\{
         \begin{array}{ll}
           \code{\assume{\event{\rolevar}{\atid}}} & \code{\text{if\,}
                                                     \chanvar{\employ}\atid}\\
           \code{\guard{\event{\rolevar}{\atid}}} & \code{\text{otherwise}}
         \end{array}
   \right.
  \\
  \code{\project{\chanvar}{\guard{\lthb{\event{\rolevar_1}{\atid_1}}{\event{\rolevar_2}{\atid_2}}} }} & ~:=~
   \left\{
         \begin{array}{ll}
           \code{\guard{\lthb{\event{\rolevar_1}{\atid_1}}{\event{\rolevar_2}{\atid_2}}}
           } & \code{\text{if\,} \chanvar{\employ}\atid_2}\\
           \code{\emp} & \code{\text{otherwise}}
         \end{array}
                         \right. \\
  \code{\project{\chanvar}{\assume{\lthb{\event{\rolevar_1}{\atid_1}}{\event{\rolevar_2}{\atid_2}}} }} & ~:=~
   \left\{
         \begin{array}{ll}
           \code{\assume{\lthb{\event{\rolevar_1}{\atid_1}}{\event{\rolevar_2}{\atid_2}}}
           } & \code{\text{if\,} \chanvar {\employ} \atid_2}\\
           \code{\emp} & \code{\text{otherwise}}
         \end{array}
                         \right.
  \end{array}
  $
} } \setcounter{subfigure}{2} }}\\

{\captionsetup[subfigure]{farskip=1pt,captionskip=-5pt}
  \subfloat[][global spec \code{\rightarrow} per channel spec]{
  \label{fig.projection.party.chan}
  \footnotesize
  \arraycolsep=3pt
  \captionsetup[subfigure]{labelformat=empty,farskip=1pt,captionskip=0pt}
  \shortstack{
    \subfloat[][]{
    $
    \begin{array}{ll}
    \code{\project{\chanvar}{\atransmit{\resid}{\R}{v\,{\cdot}\,\fmsg}{\chanvar}{\atid}}} &~:=~
      \code{\atransmit{\resid}{\R}{v\,{\cdot}\,\fmsg}{}{\atid}}
            \\
      \code{\project{\chanvar}{\prot_1{\useq}\prot_2}} & ~:=~
     \left\{
         \begin{array}{ll}
           \code{\project{\chanvar}{\prot_1}} & \code{\text{if\,}
                                                     \chanvar \in \prot_1}\\
          \code{\project{\chanvar}{\prot_2}} & \code{\text{if\,}
                                                     \chanvar \in \prot_2}\\
          \code{\emp} & \code{\text{otherwise}}
         \end{array}
     \right.
      \\
      \code{\project{\chanvar}{\prot_1{\vee}\prot_2}} &
      ~:=~\code{\project{\chanvar}{\prot_1} \vee \project{\chanvar}{\prot_2}}\\
      \code{\project{\chanvar}{\prot_1{\seq\,}\prot_2}} & ~:=~
      \code{\project{\chanvar}{\prot_1}\, \seq \,\project{\chanvar}{\prot_2}} \\
      \code{\project{\chanvar}{\assume{\event{\rolevar_1}{\atid}}}} & ~:=~
      \left\{  
         \begin{array}{ll}
           \code{\assume{\event{\rolevar}{\atid}}} & \code{\text{if\,}
                                                     \chanvar \in \atid}\\
           \code{\guard{\event{\rolevar}{\atid}}} & \code{\text{if\,}
                                                     \chanvar \notin \atid}\\
    \end{array}
    \right.
    \\
    \code{\project{\chanvar}{\guard{\lthb{\event{\rolevar_1}{\atid_1}}{\event{\rolevar_2}{\atid_2}}} }} & ~:=~
    \left\{
         \begin{array}{ll}
           \code{\guard{\lthb{\event{\rolevar_1}{\atid_1}}{\event{\rolevar_2}{\atid_2}}}
           } & \code{\text{if\,}
                                             \chanvar \in \atid_2}\\
           \code{\emp} & \code{\text{otherwise}}
         \end{array}
   \right. 
  \end{array}
  $
}} \setcounter{subfigure}{3}}}\\

{
\subfloat[][global spec \code{\rightarrow} shared spec]{
\label{fig.projection.global}
\footnotesize
$
\begin{array}{ll}
  \code{\project{\All}{\assume{\ltcb{\event{\rolevar_1}{\atid}}{\event{\rolevar_2}{\atid}}} }} & ~:=~
  \code{\assume{\ltcb{\event{\rolevar_1}{\atid}}{\event{\rolevar_2}{\atid}} }}\\
  \code{\project{\All}{\assume{\lthb{\event{\rolevar_1}{\atid_1}}{\event{\rolevar_2}{\atid_2}}} }} & ~:=~
  \code{\assume{\lthb{\event{\rolevar_1}{\atid_1}}{\event{\rolevar_2}{\atid_2}} }} \\
  \end{array}
  $
}}
\vspace*{2mm}

\end{center}
\hrule
\vspace*{-3mm}
\caption{Projection rules} \label{fig.projection}
\vspace*{-5mm}
\end{figure}

\begin{center}
\noindent{\footnotesize
\arraycolsep=0pt
$
\begin{array}{lclccclcl}
  \code{\project{\rolevar}{\prot}}:
  &  {\code{\recvc{c_1}{v}{\fmsg_1}}\;\,}
  &\code{\seq \blueord{\assume{\event{\rolevar}{1}}} \seq ~}
  &  \code{\recvc{c_2}{v}{\fmsg_2}  \seq}
  & \code{\blueord{\assume{\event{\rolevar}{2}}}
    \seq}
  & \code{ \recvc{c_2}{v}{\fmsg_3}}
  &\code{\seq \blueord{\assume{\event{\rolevar}{3}}} \seq ~}
  & {\code{\recvc{c_1}{v}{\fmsg_4}}}
&\\
\hline
  \code{\project{\rolevar,\chanvar_1}{\prot}:}
  & {\code{\recvc{c_1}{v}{\fmsg_1}}}
  & \code{\seq \blueord{\assume{\event{\rolevar}{1}}} \seq }
  &
  & \code{\seq \grayord{\guard{\event{\rolevar}{2}}} \seq}
  &
  &\code{\seq \textcolor{red}{\guard{\event{\rolevar}{3}}} \seq }
  & \fcolorbox{red}{white}{%
  \minipage[t]{\dimexpr0.13\linewidth-2\fboxsep-2\fboxrule\relax}
  \centerline {\code{\recvc{c_1}{v}{\fmsg_4}}}
  \endminipage}
&
  \\
  \code{\project{\rolevar,\chanvar_2}{\prot}:}
  & 
  &\code{\seq \textcolor{red}{\guard{\event{\rolevar}{1}}} \seq }
  & \fcolorbox{red}{white}{%
  \minipage[t]{\dimexpr0.13\linewidth-2\fboxsep-2\fboxrule\relax}
  \centerline {\code{\recvc{c_2}{v}{\fmsg_2}}}
  \endminipage}
  &
    \code{\seq \blueord{\assume{\event{\rolevar}{2}}} \seq}
  & \code{ \recvc{c_2}{v}{\fmsg_3}}
  &\code{\seq \blueord{\assume{\event{\rolevar}{3}}} \seq }
  & 
& 
\end{array}
$
}
\end{center}

The above local specification snapshot
highlights how local fidelity is secured:
the events marked with red boxes are guarded
by their immediately preceding events, since
they are handled by different channels. A subsequent refinement
removes redundant guards, grayed in the example above,
since 
adjacent same channel events need to guard only
the last event on the considered channel.

Given the congruence of global protocols and
local specifications, the projection is an isomorphism
courtesy to the unique labelling and ordering relations 
carefully
inserted after each transmission.
Given two protocols \code{\prot_1} and \code{\prot_2},
with
\code{\rolevar_1..\rolevar_n \in \prot_1} and
\code{\neg(\exists \rolevar \in \prot_1 \cdot \rolevar \notin
  \{\rolevar_1..\rolevar_n\})},
and  \code{\chanvar_1..\chanvar_m \in \prot_1} and
\code{\neg(\exists \chanvar \in \prot_1 \cdot \chanvar \notin
  \{\chanvar_1..\chanvar_m\})}:
 \code{\prot_1 {\equiv}\prot_2 {\Leftrightarrow}
  \{\project{\rolevar_j}{\prot_1}\}_{j=1..n} {\equiv}
  \{\project{\rolevar_j}{\prot_1} \}_{j=1..n}}.
This condition holds even for the more granular specifications:
\code{\prot_1 {\equiv}\prot_2 {\Leftrightarrow}
  \{\project{\rolevar_j,\chanvar_k}{\prot_1}\}_{j=1..n,{k=1..m}} {\equiv}
  \{\project{\rolevar_j,\chanvar_k}{\prot_1} \}_{j=1..n,{k=1..m}}}.
Detailed proofs of the isomorphism under all operators 
to be provided in the detailed technical report.

And lastly, as discussed in \secref{sec.overview}, the communicated-before and
happened-before assumptions are projected into a shared store, so that each
party can benefit from them (\figref{fig.projection.global}).



 \section{Verification of C-like Programs}
The user provides the global protocol which is then automatically
refined according to the methodology described in \secref{sec:global}.
The refined protocol is automatically projected onto a per party
specification, followed by a per channel endpoint basis
as described in \secref{sec:projection}.
Using such a modular
approach where we provide a specification for each channel
endpoint adds natural support for delegation, where a channel (as well
as its specification) could be delegated to a third party. 
These communication specifications are made available
in the program abstract state using a combinations of
ghost assertions and release lemmas (detailed in the subsequent).
The verification could then automatically check whether a certain implementation
follows the global protocol, after it  had first bound 
the \emph{program elements} (threads and channel endpoints)
to the \emph{logical ones} (peers and channels).

\noindent\emph{\bf Language.}
\figref{fig.lang.syntax} gives the syntax of a core language with
support for communication primitives.
We omit the details of the Boolean and arithmetic expression
and focus on the language support for asynchronous
message passing via channels.
{\footnotesize
\vspace*{-5mm}
\begin{figure}[htb!]
\begin{center}\[
\begin{array}{lcl}
                e{\in}\btt{\Prog} &::= &
 \btt{open~()~with~(\lchanvar,\{\rolevar_1,..,\rolevar_n\})}~|~ 
                                         \btt{close}(\echanvar)\\
  &&
                                         ~|~ \btt{send}(\echanvar,\code{v})
                                         ~|~ \btt{recv}(\echanvar)
                                         ~|~ \code{notifyAll(\ev)}
                                         ~|~ \code{wait(\ev)}
  \\
  && 
     ~|~ f~(e^*)~|~ e||e~|~e;e~|~\myif{b}{e}{e} ~|~\wait{\atid_1}{\atid_2}

\end{array}
\]
\hrule
\end{center}
\vspace*{-3mm}
\caption{A Core Language}\label{fig.lang.syntax}
\vspace*{-5mm}
\end{figure}
}
\anay{what about malloc and dispose?}

\noindent\emph{\bf Concurrent Separation Logics.}
Due to its expressive power and elegant proofs, we choose
to integrate our session logic on top of concurrent
  separation logic. Separation logic is an attractive
extension of Hoare logic in which assertions are interpreted
\wrt to some relevant portion of the heap. Spatial conjunction,
the core operator of separation logic, 
\code{P \sep Q} divides the heap between two disjoint heaps described
by assertions \code{P} and \code{Q}, respectively.
The main benefit of this approach is the local reasoning:
the specifications of a program code
need only mention the portion of the resources which it uses,
the rest are assumed unchanged.

\figref{fig.fig_semantics_assertions} defines the state model and the semantics of state assertions. 
Our approach of layered  abstractions permits us to build on top of
the traditional storage model for heap manipulating
programs with a minimal, yet important extension to account for
the race-free assertions. Therefore, we define the program state as
the triple comprising
a stack \code{s {\in} Stack} which is a total mapping
from local and logical variables \code{Var} to primitive values
\code{Val} or memory locations \code{Loc};
a heap \code{h {\in}  Heap} which is a finite partial mapping
from memory locations to data structures stored
in the heap, \code{DVal}; and an ordering relations store
\code{\mord {\in} \morders} which is a set of assumed events
and events relations. 

The semantics of the state assertions
\figref{fig.fig_semantics_assertions}
are similar to those of
separation logic, with the exception of the conjunction
with the order relations which is evaluated in the relations
store \code{\mord}.
We also benefit from the main axioms of separation logic,
that of the frame rule and disjoint concurrency, where \code{fv}
returns the set of free variables within an expression
or logical formula:

{\footnotesize
  \renewcommand{\arraystretch}{3.5}
  $
  \begin{array}{l@{\hskip 10pt}l}
    \ebpverif{}{{
    \Hypo{{\{\constr_1 \} ~e~ \{ \constr_2  \}}}
    \Infer1[]{{\{\constr_1 \sep \constr\}~ e~ \{ \constr_2 \sep \constr \}}}      
    }} &
         \ebpverif{}{{
         \Hypo{\code{{fv(\constr) \cap modif(e)= \emptyset}}}
         }}
    \\
    \ebpverif{}{{
    \Hypo{{\{\constr_1 \} ~e~ \{ \constr_2  \}}}
    \Hypo{{\{\constr_1' \} ~e'~ \{ \constr_2'  \}}}
    \Infer2[]{{\{\constr_1 \sep \constr_1'\}~ e ~||~ e'~ \{ \constr_2 \sep \constr_2' \}}}      
    }} &
         \ebpverif{}{{
         \Hypo{\code{{(fv(\constr_1) \cup fv(\constr_2))\cap modif(e')=\emptyset}}}
         \Infer[no rule]1{\code{{(fv(\constr_1') \cup fv(\constr_2'))\cap modif(e)=\emptyset}}}
    }} 
  \end{array}
  $
}

\begin{figure*}
\vspace*{-5mm}
\centering
\footnotesize
\captionsetup[subfigure]{labelformat=empty,farskip=-5pt,captionskip=-15pt}  
\vspace*{5mm}
\subfloat[][]
{$
  \begin{array}{lll}
    \code{\satisf{\smstack}{\smheap,\mord}{\pure}} &\code{~\text{iff}~} & \code{\llbracket \pure\rrbracket{=}\code{true}} \qquad\qquad\qquad
    \code{\satisf{\smstack}{\smheap,\mord}{\emp}} \quad\code{~\text{iff}~}\quad  \code{true} \\
    \code{\satisf{\smstack}{\smheap,\mord}{\code{\node{v}{d}{v_1,..,v_n}}}} &\code{~\text{iff}~} &
                                                       \code{struct~d\{t_1~f_1;
                                                       .. ;t_n~f_n\}
                                                       \in
                                                       \Prog}\text{~and} ~
  \code{\smheap{=}[\smstack(v) \mapsto d[f_1 {\mapsto} \smstack(v_1), .. ,
                                f_n {\mapsto} \smstack(v_n)]]} \\
  \code{\satisf{\smstack}{\smheap,\mord}{\heap_1 \sep \heap_2}} &\code{~\text{iff}~} &
                          \code{\exists ~\smheap_1,\smheap_1 \cdot (
                                                                   \smheap_1    \disjoint
                                                                   \smheap_2
                                                                    ~and~
                                                                   \smheap_1  \uplus
                                                                   \smheap_2 =
                                                                   \smheap
                                                                   )
                                                                   }
                                                                   \text{~and}~
  \code{\satisf{\smstack}{\smheap_1,\mord}{\heap_1} \text{~and~}
     \satisf{\smstack}{\smheap_2,\mord}{\heap_2}} \\
  \code{\satisf{\smstack}{\smheap,\mord}{\heap \wedge \pure \wedge \racefreeas}} &\code{~\text{iff}~} &
        \code{\satisf{\smstack}{\smheap,\mord}{\heap}} \text{~and~}
                                                                                          \smstack \implies \llbracket{\pure}\rrbracket
      \text{~and~} \racefree{\morders}{\racefreeas}
        \\
  \code{\satisf{\smstack}{\smheap,\mord}{\constr_1 \vee \constr_2}} &\code{~\text{iff}~} &
       \code{\satisf{\smstack}{\smheap,\mord}{\constr_1}} \text{~or~}
       \code{\satisf{\smstack}{\smheap,\mord}{\constr_2}}\\
  
  \end{array}
$}\hfill

\vspace*{5mm}
\subfloat[][]
{$
  \begin{array}{lll}
    \code{State \defeq Stack \times Heap \times \morders} &&
    \quad\qquad\code{\smheap_1 \disjoint \smheap_2 }  \code{\Leftrightarrow}
    \code{dom(\smheap_1) \cap dom(\smheap_2) = \emptyset}\\

    \code{Stack \defeq Var \rightarrow  Val \cup Loc} &
    \code{Heap \defeq Loc \rightharpoonup_{fin}   DVal}&

   \quad\qquad \code{\smheap = \smheap_1 \uplus \smheap_2 }  \code{\Leftrightarrow}
                                        \code{
                                          dom(\smheap_1) \cup
                                          dom(\smheap_2) = dom(\smheap) }
\end{array}
$}\hfill

\vspace*{3mm}
\hrule
\vspace*{-3mm}
\caption{The semantics of the state assertions}\label{fig.fig_semantics_assertions}
\vspace*{-3mm}
\end{figure*}

\noindent\emph{\bf Verification.}
To check whether a user program follows the stipulated communication
scenario, a traditional analysis would
need to reason about the
program's behaviour using the operational semantics of
the primitives' implementation.
Since our goal is to emphasize on the benefits
of implementing a protocol guided communication,
rather than deciding the correctness of the primitives machinery,
we adopt a specification strategy using
abstract predicates
\cite{parkinson2005separation,Dinsdale-Young:2010:CAP:1883978.1884012}
to describe the behavior of
the program's primitives. Provided that the primitives respect
their abstract specification, developers could then choose
alternative communication libraries, without
the need to re-construct the correctness proof of their underlying
program.

The verification process follows the traditional
forward verification rules, where
the pre-conditions are checked for each method call, and if the check succeeds it adds
their corresponding  postcondition to the poststate. The verification of the
method definition starts
by assuming its precondition as the initial abstract state,
and then inspects whether the postcondition holds after
progressively checking each
of the method's body instructions.

{\footnotesize
\begin{figure*}
	\begin{center}
        {\renewcommand{\arraystretch}{3.5}
\subfloat[][Annotated communication primitives.]{
  \label{fig.commprimitives}
          
$
	\begin{array}{cll}
          \ebpverif{OPEN}{
          \Hypo{
          \code{\triple{~\init(\lchanvar)~}{open()~ with~ (\lchanvar,P^*)}{~\opened{\lchanvar}{\rolevar^*}{\res}}~}
          }}
          \qquad\qquad\qquad 
          \hspace{-5mm}
          \ebpverif{CLOSE}{{
          \Hypo{
          \code{\triple{~\emptyc{\lchanvar}{\echanvar}
          }
          {close(\echanvar)}
          {~\emp~}}}
      }}\\ \hspace{-5mm}
          \ebpverif{SEND}{{
          \Hypo{\code{inv \defeq \peer{\rolevar} \sep
          \opened{\lchanvar}{\rolevar^*}{\echanvar} \wedge \rolevar
          {\in} \rolevar^*}}
          \Infer1{\code{\triple{\chani{\lchanvar}{\rolevar}{{\color{red}
          \send{v}{V(v)}} {\seq} {\proj}} {\sep}
          {\color{red}V(x)} {\sep} inv} 
          {send(\echanvar,x)}
          {\chani{\lchanvar}{\rolevar}{\proj} 
          {\sep} {inv}}
          }}
    }}~~
           \ebpverif{RECV}{{
           \Hypo{\code{inv \defeq \peer{\rolevar} \sep
           \opened{\lchanvar}{\rolevar^*}{\echanvar} \wedge \rolevar
           {\in} \rolevar^*}}
           \Infer1{\code{\triple{
           \chani{\lchanvar}{\rolevar}{{\color{red}
           \recv{v}{V(v)}} {\seq} {\proj}} {\sep} inv}
           {recv(\echanvar)}
           {\chani{\lchanvar}{\rolevar}{\proj} {\sep} {\color{red}V(\res)} {\sep} inv}
           }}
	 }}\\ 
	\end{array}
        $
      }}\\
    \subfloat[][Splitting lemmas]{
      \label{fig.splitlemmas}
      {\footnotesize
        \tabcolsep=0pt
        $\hspace{-3mm}
        \begin{array}{l@{\hskip 2pt} l@{\hskip 2pt} l}
          \code{\prot(\{\rolevar_1 .. \rolevar_n\} ,\lchanvar^*)}
          &\code{\implies}
          & \code{
            \partyspec{\rolevar_1,\lchanvar^*}{\project{\rolevar_1}{\prot}}
            \sep ... \sep
            \partyspec{\rolevar_n,\lchanvar^*}{\project{\rolevar_n}{\prot}}
            ~\sep
            ~initall(\lchanvar^*).
            }\\
          \code{\partyspec{\rolevar,\{\lchanvar_1 .. \lchanvar_m\}}{\project{\rolevar}{\prot}}
          } & \code{\implies }
            & \code{
              \chani{\lchanvar_1}{\rolevar}
              {\project{\rolevar,\lchanvar_1}{\prot}}
              \sep ... \sep
              \chani{\lchanvar_m}{\rolevar}{\project{\rolevar,\lchanvar_m}{\prot}} 
            \sep~ \bindv{\rolevar}{\{\lchanvar_1 .. \lchanvar_m\}}}.\\
          \code{initall(\{\lchanvar_1 .. \lchanvar_m\})}
          & \code{\implies}
          & \code{init(\lchanvar_1) \sep ...  \sep init(\lchanvar_m).} \\
	\end{array}
        $
      }}\\
    \subfloat[][Joining simpagation rules]{
      \label{fig.joinlemmas}
      {\footnotesize
        \tabcolsep=0pt
        $\hspace{-3mm}
        \begin{array}{l}
          \code{              
            \chani{\lchanvar}{\rolevar_1}{\emp} \sep ... \sep
            \chani{\lchanvar}{\rolevar_n}{\emp}
            ~\simpag~
            \opened{\lchanvar}{\{\rolevar_1 .. \rolevar_n\}}{\echanvar}
            }
          \code{~\implies~}
          \code{\emptyc{\lchanvar}{\echanvar}}
          \\
          \code{              
          \chani{\lchanvar_1}{\rolevar}{\emp} \sep ... \sep
          \chani{\lchanvar_m}{\rolevar}{\emp}
          ~\simpag~
          \bindv{\rolevar}{\{\lchanvar_1 .. \lchanvar_m\}}
          }
          \code{~\implies~}
          \code{\partyspec{\rolevar,\lchanvar^*}{\emp}}
        \end{array}
        $  
      }}\hfill 
    \subfloat[][Lemmas to handle orders]{
      \label{fig.orderslemmas}
      {\footnotesize
        \tabcolsep=0pt
        $\hspace{-3mm}
        \begin{array}{l@{\hskip 2pt} l@{\hskip 2pt} l}
          \code{\chani{\lchanvar}{\rolevar}{\assume{\racefreeas}\seq{\proj}}
          } & \code{\implies}
          & \code{
            \chani{\lchanvar}{\rolevar}{{\proj}} \wedge {\racefreeas}
            }\\
          \code{\chani{\lchanvar}{\rolevar}{\guard{\racefreeas}\seq{\proj}}
          \wedge {\racefreeas}
          } & \code{\implies}
          & \code{              
            \chani{\lchanvar}{\rolevar}{{\proj}} 
            }\\
        \end{array}
        $  
      }}
        \hrule
	\end{center}
	\vspace*{-3mm}
	\caption{Communication primitives} \label{fig.primitives}
	\vspace*{-3mm}
\end{figure*}
}



\noindent\emph{\bf Abstract Specification.}
\hide{
To emphasize the benefits of using \sessionlogic~ in reasoning about
communication-centered programs we have plugged it into an
existing automated verification system \cite{Chin:2012:AVS:2221987.2222283}
expressive enough to specify and verify
communication primitives
for a C-like language. 
}
We define a set of abstract predicates  to support session
specification of different granularity. Some of these predicates
have been progressively introduced across the paper, but for brevity
we have omitted certain details. 
We resume their presentation here with more details:
\code{\bf {\partyspec{\rolevar,\lchanvar^*}{\projp}}}
    - associates a local protocol projection \code{\projp} to its
    corresponding party
    \code{\rolevar} and the set of channels \code{\lchanvar^*} used by \code{\rolevar};
   g\code{\bf \chani{\lchanvar}{\rolevar}{\projc}}
   - associates an endpoint specification \code{\projc}
   to its corresponding  party \code{\rolevar} and
   channel \code{\lchanvar};
   \code{\bf \commonspec{{\prot}{\prj}{\All}}}
    - comprises the ordering assumptions shared among all the parties;
   \code{\bf \peer{\rolevar}} - flow-sensitively
   tracks the executing
    party, since the execution of
    parties can either be in
    parallel or
    sequentialized;
    \code{\bf \bindv{\rolevar}{\lchanvar^*}} - binds a
    party \code{\rolevar} to all the channels \code{\lchanvar^*}
    it uses.
  \anay{explain logical channels. when writing the protocol, the communication designer uses logical channel to describe the communication. The developer then uses program channels, and in order to verify the code against the protocol, it links the code to the spec by:
1. identifying program codes to communication peers
2. binding a program channel to a logical channel from the protocol. 
  }
To cater for each verification phase, the session specifications
with the required granularity
are made available in the program's abstract state
via the lemmas in \figref{fig.primitives}.
\hide{
 \noindent
{\footnotesize
$
\begin{array}{ll}
\code{\gloabspec{\{\rolevar_1,..,\rolevar_n\}}{C
}{\prot}
  } &\code{\implies
  \partyspec{\rolevar_1,C}{\perpartyprj{\prot}{\rolevar_1}}
  \sep ... \sep \partyspec{\rolevar_n,C}{\perpartyprj{\prot}{\rolevar_n}}
  }\\
  \code{\partyspec{\rolevar,\{\echanvar_1,..,\echanvar_m\}}{\prot}
  } & \code{\implies
  \chani{\echanvar_1}{\rolevar}{\perchanprj{\prot}{\rolevar}{\echanvar_1}}
      \sep ... \sep
      \chani{\echanvar_m}{\rolevar}{\perchanprj{\prot}{\rolevar}{\echanvar_m}}
      }\\
  \code{\chani{\echanvar}{\rolevar}{\assume{\racefreeas}\seq{\proj}}
  } & \code{\implies
      \chani{\echanvar}{\rolevar}{{\proj}} \wedge {\racefreeas}
      }\\
  \code{\chani{\echanvar}{\rolevar}{\guard{\racefreeas}\seq{\proj}}
  \wedge {\racefreeas}
  } & \code{\implies
      \chani{\echanvar}{\rolevar}{{\proj}} 
  }\\ 
\end{array}
$  
}}

Channel endpoint creation and closing
described by the \code{[\underline{{\bf \scriptstyle }\rulen{OPEN}}]}
and \code{[\underline{{\bf \scriptstyle }\rulen{CLOSE}}]} triples
in \figref{fig.primitives},
have mirrored specification: \code{open}
associates the specification of a
channel \code{\lchanvar} to its corresponding program endpoint
\code{\echanvar}. 
\code{close} regards the closing of a channel endpoint
as safe
only when all the parties
have finished their communication \wrt 
the closing endpoint. 

To support send and receive operations, we decorate
the corresponding methods with dual generic specifications.
The precondition of
\code{[\underline{{\bf \scriptstyle }\rulen{SEND}}]}
ensures that indeed a send operation is expected,
\code{\send{v}{V(v)}}, where the message \code{v}
to be transmitted is described by a higher-order relation over
\code{v}. Should this be confirmed, to ensure
memory safety, the verifier also
checks whether the program state indeed
owns the message to be transmitted and that it
adheres to the properties described by
the freshly discovered
relation, \code{V(x)}.
Dually, \code{[\underline{{\bf \scriptstyle
    }\rulen{RECV}}]} ensures that
the receiving state gains the ownership of the
transmitted message. Both specifications guarantee
that the transmission is consumed by the expected
party, \code{\peer{\rolevar}}.
\hide{
\begin{verbatim}
a detailed main program
+
a detailed role code
+
a detailed function call check
\end{verbatim}
}

The 
proof obligations generated by this verifier
are discharged to a Separation Logic solver
in the form of enatailment checks, detailed in the subseqent.

\noindent\emph{\bf Entailment.}
{ \footnotesize
\begin{figure*}
  \begin{center}
    \renewcommand{\arraystretch}{4.0}
    \tabcolsep=1pt
    \begin{tabular}{c}
      \ebpentail{CHAN-MATCH}{{
          \Hypo{\code{\St_a {\implies} \lchanvar_1 {=} \lchanvar_2}}
	  \Hypo{\code{\entailS{\chani{\lchanvar_1}{\rolevar_1}{\proj_a}}{\chani{\lchanvar_2}{\rolevar_2}{\proj_c}}{\resid_1}}}
	  \Infer[no rule]2{\code{\resid_2 = \{\pure_i^e ~|~ \pure_i^e
      {\in} \resid_1 \text{~and~} \code{SAT}(\St_a {\wedge}
      \pure_i^e) \text{~and~} \code{SAT}( \St_c {\wedge}
      \pure_i^e)\} \quad   
	  \entailS{\bigvee \limits_{\pure^e {\in} \resid_2}(\St_a \wedge \pure^e)}{\St_c}{\resid}}}
	  \Infer1{\code{\entailS{\chani{v_1}{\rolevar}{\proj_a} * \St_a}{\chani{v_2}{\rolevar}{\proj_c} * \St_c}{\resid}}}
      }}      
      \\
     \ebpentail{CHAN}{{
          \Hypo{\code{\rolevar_1 {=} \rolevar_2}}
	  \Hypo{\code{\entailS{{\proj}_a}{{\proj}_c}{\resid'}}}
	  \Hypo{\code{\resid{=}\{\pure_i^e | \pure_i^e {\in} \resid' \}}}
	  \Infer3{\code{\entailS{\chani{\lchanvar}{\rolevar_1}{\proj_a}} {\chani{\lchanvar}{\rolevar_2}{\proj_c}}{\resid}}}
      }}
      \ebpentail{RHS-PVAR}{{
	  \Hypo{\code{\resid{=}\{\emp{\wedge}\hov{=}{\proj}_a\}}}
	  \Infer1{\code{\entailS{{\proj}_a} {\hov}{\resid}}}
      }}
      \ebpentail{RECV}{{
	  \Hypo{\code{\entailS{\St_a}{[v_{1}/v_{2}]\St_ c}{\resid'}}	}
	  \Hypo{\code{{\resid{=}\{\pure_i^e | \pure_i^e{\in}\resid' \}}}}
	  \Infer2{\code{\entailS{\recv{v_1}{\St_a}} {\recv{v_2}{\St_c}}{\resid}}}
      }}\\
      \ebpentail{SEND}{{
	  \Hypo{\code{\entailS{[v_{1}/v_{2}]\St_c}{\St_a}{\resid'}}}
	  \Hypo{\code{{\resid{=}\{\pure_i^e | \pure_i^e{\in}\resid' \}}}}	
	  \Infer2{\code{\entailS{\send{v_{1}}{\St_a}} {\send{v_{2}}{\St_c}}{\resid}}}
      }}
      \ebpentail{SEQ}{{
	  \Hypo{\code{\entailS{\unkbox_a}{\unkbox_c}{\resid_1}}}
	  \Hypo{\code{\entailS{{\proj}_a}{{\proj}_c}{\resid_2}}}
	  \Hypo{\code{\text{where} ~\unkbox := \recv{v}{\St} ~|~ \send{v}{\St} 
      }}
	  \Infer3{\code{\entailS{\unkbox_a {\seq} {\proj}_a} {\unkbox_c {\seq} {\proj}_c}{\{\emp{\wedge}\pure_1{\wedge}\pure_2~|~ \pure_1{\in}\resid_1 {\text{~and~}} \pure_2{\in}\resid_2\}}}}
      }}  \\
      \ebpentail{LHS-HO-VAR}{{
          \Hypo{\code{\hov \notin \text{fv}(\St_c)}}
	  \Hypo{\code{{SAT}(\St_c)}}
          \Hypo{\code{\text{fresh}~w}}
	  \Hypo{\code{\resid{=}\{\emp {\wedge}\hov(w){=}[w/v]\St_c \}}}
	  \Infer4{\code{\entailS{\hov(v)}{\St_c}{\resid}}}
      }}
      \ebpentail{LHS-OR}{{
	  \Hypo{\code{\entailS{\proj_i \seq \proj_a}{\proj_c}{\resid_i}}}
	  \Hypo{\code{\resid=\{ \bigvee_i \St_i  ~|~ \St_i {\in} \resid_i \}}	}
	  \Infer2{\code{\entailS{(\bigvee_i {\proj}_i ) \seq \proj_a} {{\proj}_c}{\resid}}}
      }} 
      \\
      \ebpentail{RHS-HO-VAR}{{
          \Hypo{\code{\hov \notin \text{fv}(\St_a)}}
	  \Hypo{\code{\entailS{\St_a}{\St_c}{\resid'}}}
          \Hypo{\code{\text{fresh}~w}}
	  \Hypo{\code{\resid{=}\{\emp {\wedge}\hov(w){=}[w/v]\St_i | \St_i {\in} {\resid'}\}}}
	  \Infer4{\code{\entailS{\St_a}{\hov(v) * \St_c}{\resid}}}
      }} 
      \ebpentail{RHS-OR}{{
	  \Hypo{\code{\entailS{\proj_a}{\proj_i \seq {\proj_c} }{\resid_i}}}
	  \Hypo{\code{\resid=\bigcup \resid_i }	}
	  \Infer2{\code{\entailS{{\proj}_a } {(\bigvee_i {\proj}_i) \seq {\proj_c}}{\resid}}}
      }}  
    \end{tabular}
  \end{center}
  \vspace*{1mm}
  \hrule
  \vspace*{-3mm}
  \caption{Selected entailment rules: 
    \code{\pure^e} is a shorthand for \code{\emp {\wedge} \pure},
    \code{fv(\St)} returns all free variables in \code{\St}, and
    \code{fresh} denotes a fresh variable.
  } \label{fig.entailment}
  \vspace*{-3mm}
\end{figure*}
}

\asay{explain more about CHAN-MATCH and the need for sat. also
  should i add the unfold here?}

Traditionally, the logical entailment 
between formalae written in the symbolic
heap fragment of separation logic is expressed as follows: 
$\entail{\St_a}{\St_c * \St_r}$, where $\St_r$  
comprises those residual resources 
described by $\St_a$, but not by $\St_c$.
Intuitively, a valid entailment suggests that
the resource models described by $\St_a$ are sufficient
to 
conclude 
the availability of those described by $\St_c$.

Since the proposed logic is tailored to support 
reasoning about communication primitives with
generic protocol specifications, the entailment should
also be able to interpret and instantiate such generic specifications.
Therefore we equip the entailment checker 
to reason about 
formulae which contain second-order variables. 
Consequently, the proposed entailment is designed 
to support the instantiation of 
such variables. However, the instantiation might
not be unique, so we collect the candidate instantiations
in a set of residual states. The entailment has thus
the following form: 
${\entailS{\St_a}{\St_c}{\resid}}$, where $S$ is the 
set of possible residual states. Note that $S$ is derived and
its size should be of at least 1 in order to consider the
entailment as valid.

The entailment rules needed to accommodate session reasoning
are given in  \figref{fig.entailment}.
Other rules used for the manipulation of general resource
predicates are adapted from Separation Logic
\cite{reynolds2002separation}.



To note also how \code{\entrulen{RECV}} and \code{\entrulen{SEND}}
are soundly designed to be the dual of each other: while the former
checks for covariant subsumption of the communication models,
the latter enforces contravarinat subsumption since
the information should only flow from a stronger constraint towards 
a weaker one.

Considering the example below, a context expecting to read an integer greater than
or equal to 1 could engage a channel 
designed
with a more relaxed specification \textit{(a)}. However, a
context expecting to transmit an integer greater than or equal to 1
should only be allowed to engage a more specialized channel,
such as one which can solely transmit the exact number 1
\textit{(b)}.

{\footnotesize \vspace*{-3mm}
\begin{center}\[
\begin{array}{c}
  {\hspace{-2mm}%
    \begin{prooftree}
    \Hypo{\code(a)}
      \Infer[no rule]1{\code{\entail{v_1{\geq}1}{[v_1/v_2]v_2{\geq}0}}}
      \Infer1[\code{\entrule{RECV}}]{\code{\entail{\recv{v_1}{v_1{\geq}1}} {\recv{v_2}{v_2{\geq}0}}}}
      \Infer1[\code{\entrule{CHAN}}
      ]{\code{\entail{\chani{c}{\rolevar}{\recv{v_1}{v_1{\geq}1}}} {\chani{c}{\rolevar}{\recv{v_2} {v_2{\geq}0}}}}}
    \end{prooftree}
  } \\
  
{
    \begin{prooftree}
     \Hypo{\code(b)}
      \Infer[no rule]1{\code{\entail{[v_1/v_2]v_2{=}1}{v_1{\geq}1}}}
      \Infer1[\code{\entrule{SEND}}]{\code{\entail{\send{v_1}{v_1{\geq}1}} {\send{v_2}{v_2{=}1}}}}
      \Infer1[\code{\entrule{CHAN}}
      ]{\code{\entail{\chani{c}{\rolevar}{\send{v_1}{v_1{\geq}1}}} {\chani{c}{\rolevar}{\send{v_2} {v_2{=}1}}}}}
    \end{prooftree}
}\end{array}\]
\end{center}
}


\noindent\emph{\bf Soundness.}
The soundness of our verification rules is defined with respect to the
operational semantics by proving progress and preservation. For lack
of space we defer the corresponding theorems and proofs to the
appendix.

\hide{
\def\TS{\btt{TS}}
\def\CS{\btt{CS}}
\def\Prot{\btt{Prot}}
\def\SStore{\btt{SStore}}
\def\PS{\btt{PS}}

{\footnotesize
\begin{figure*}
	\begin{center}
        {\renewcommand{\arraystretch}{3.5}
\subfloat[][Machine Reduction]{
  \label{fig.semantics.machine}
          
$
	\begin{array}{c}
          \hspace{-5mm}
          \ebpsem{MACHINE}{{
          \Hypo{\code{
          {\langle}\tstate,\cstate,\sstore{\rangle} \ttrans  {\langle}\tstate',\cstate',\sstore'{\rangle}
          }}
          \Infer1{\code{
          {\langle}\pstate[\atid\,{\mapsto}\tstate],\cstate,\sstore{\rangle} \ttrans  {\langle}\pstate[\atid\,{\mapsto}\tstate'],\cstate',\sstore'{\rangle}
          }}
	}}
	\\  
	 \hspace{-5mm}
          \ebpsem{PAR}{{
          \Hypo{\code{
          \tstate{=}{\langle}\sigma,P,(e_1||e_2);e{\rangle} 
          \quad  \textit{fresh}~ \atid_1, \atid_2 \quad \tstate'{=} {\langle}\emp,P,(\wait{\atid_1}{\atid_2});e{\rangle} }}
          \Infer[no rule]1{\code{\sigma {=}\sigma_1 \uplus \sigma_2 \quad \pstate'{=}\pstate[\atid\,{\mapsto}\tstate'][\atid_1{\mapsto} {\langle}\sigma_1,P,e_1{\rangle}][\atid_2{\mapsto} {\langle}\sigma_2,P,e_2{\rangle}]
          }}
          \Infer1{\code{
          {\langle}\pstate[\atid\,{\mapsto}\tstate],\cstate,\sstore{\rangle} \ttrans  
          {\langle}\pstate',\cstate,\sstore{\rangle}
          }}
	}}\quad 
          \ebpsem{JOIN}{{
          \Hypo{\code{
          \pstate'[\atid_1{\mapsto} {\langle}\sigma_1,\_,\_{\rangle}][\atid_2{\mapsto} {\langle}\sigma_2,\_,\_{\rangle}] = \pstate}}
          \Infer[no rule]1{\code{
          \tstate{=} {\langle}\sigma,P,(\wait{\atid_1}{\atid_2});e{\rangle} \quad \sigma' {=} \sigma {\uplus} \sigma_1 {\uplus} \sigma_2 
          }}
          \Infer1{\code{
          {\langle}\pstate[\atid\,{\mapsto} \tstate],\cstate,\sstore{\rangle} \ttrans  
          {\langle}\pstate'[\atid\,{\mapsto}{\langle}\sigma',P,e{\rangle}],\cstate,\sstore{\rangle}
          }}
	}}\\ 
	\end{array}
        $
     }}\\
    {\renewcommand{\arraystretch}{3.5}
    \subfloat[][Per-Thread Reduction]{
      \label{fig.semantics.thread}
        \tabcolsep=0pt
    $
	\begin{array}{c}
          \hspace{-5mm}
          \ebpsem{SEND}{{
          \Hypo{\code{
          \sigma {=}\sigma' {\uplus} \{v\}
          \quad 
          (\prot,\css) {=}\sstore \quad
          \chanvar,\_=\cstate(\echanvar)
          \quad
          \cproj {=}
          \css(\chanvar)
          }}
          \Infer[no rule]1{\code{
            (\SAFE,\cproj') = \safesend(\cproj,P,v) \quad
          (\SAFE,\prot') = \fid(\prot,\chanvar,P) \quad
          \sstore = (\prot', \css[\chanvar{\mapsto}\cproj'])
        }}
          \Infer1{\code{
           {\langle} {\langle}\sigma,P, send(\echanvar,v);e{\rangle},\cstate,\sstore{\rangle} 
           \ttrans  {\langle}{\langle}\sigma',P, e{\rangle},\cstate,\sstore{\rangle}
          }}
	}}\\
	  \hspace{-5mm}
          \ebpsem{RECV}{{
          \Hypo{\code{
         \sigma' {=} \sigma {\uplus}\{\res\}
         \quad 
          (\prot,\css) {=}\sstore \quad
          \chanvar,\_=\cstate(\echanvar) \quad
          \cproj {=} \css(\chanvar) 
          }}
          \Infer[no rule]1{\code{
            (\SAFE,\cproj',\res) = \saferecv(\cproj,P) \quad
          (\YES,\prot') = \fid(\prot,\chanvar,P) \quad
          \sstore = (\prot', \css[\echanvar{\mapsto}\cproj'])
        }}
          \Infer1{\code{
           {\langle} {\langle}\sigma,P, recv(\echanvar);e{\rangle},\cstate,\sstore{\rangle} 
           \ttrans  {\langle}{\langle}\sigma',P, e{\rangle},\cstate,\sstore{\rangle}
          }}
	}}\\
	  \hspace{-5mm}
          \ebpsem{RECV-BLOCK}{{
          \Hypo{\code{
          (\prot,\css) {=}\sstore \quad
                    \chanvar,\_=\cstate(\echanvar) \quad
          \cproj {=} \css(\chanvar) 
        \quad
            (\BLOCK,\_,\_) = \saferecv(\cproj,P) \quad
          (\YES,\_) = \fid(\prot,\chanvar,P)         
          }}
          \Infer1{\code{
           {\langle} {\langle}\sigma,P, recv(\echanvar);e{\rangle},\cstate,\sstore{\rangle} 
           \ttrans  {\langle}{\langle}\sigma,recv(\echanvar);P, e{\rangle},\cstate,\sstore{\rangle}
          }}
	}}\\
	\hide{
                 (Prot,_) = SStore            HasHappen(E,Prot)
---------------------------------------------------------------------------------------------- [NOTIFY]
             <<\phi,P,(notify e);s>,CS,SStore> --> <<\phi,P,s>,CS,SStore>

                 (Prot,_) = SStore            not(HasHappen(E,Prot))  
---------------------------------------------------------------------------------------------- [WAIT]
             <<\phi,P,(wait e);s>,CS,SStore> --> <<\phi,P,s>,CS,SStore>
	}
	\hspace{-5mm}
          \ebpsem{NOTIFYALL}{{
          \Hypo{\code{
          (\prot,\_) {=}\sstore \quad
                   \btt{HasHappen}(\ev,\prot)
           }}
           \Infer[no rule]1{\code{ \tstate{=}{\langle}\sigma,P, notifyAll(\ev);e{\rangle}
        }}
          \Infer1{\code{
           {\langle} \tstate,\cstate,\sstore{\rangle} 
           \ttrans  {\langle}{\langle}\sigma,P, e{\rangle},\cstate,\sstore{\rangle}
          }}
	}}\quad
          \ebpsem{WAIT-BLOCK}{{
          \Hypo{\code{
          (\prot,\_) {=}\sstore \quad
                  \neg \btt{HasHappen}(\ev,\prot) 
           }}
           \Infer[no rule]1{\code{ \tstate{=}  {\langle}\sigma,P, wait(\ev);e{\rangle}
        }}
          \Infer1{\code{
           {\langle}\tstate,\cstate,\sstore{\rangle} 
           \ttrans {\langle}\tstate,\cstate,\sstore{\rangle}
          }}
	}}\quad
          \ebpsem{WAIT}{{
          \Hypo{\code{
          (\prot,\_) {=}\sstore \quad
                  \btt{HasHappen}(\ev,\prot) 
           }}
           \Infer[no rule]1{\code{ \tstate{=}  {\langle}\sigma,P, wait(\ev);e{\rangle}
        }}
          \Infer1{\code{
           {\langle}\tstate,\cstate,\sstore{\rangle} 
           \ttrans {\langle}{\langle}\sigma,P, e{\rangle},\cstate,\sstore{\rangle}
          }}
	}}\	
	\end{array}
        $     
        }}\\
         {\renewcommand{\arraystretch}{3.5}
        \subfloat[][Ghost Transition]{
      \label{fig.semantics.ghost}
        \tabcolsep=0pt
    $
	\begin{array}{c}
          \hspace{-5mm}
          \ebpsem{ASSERT-PEER}{{
        \Hypo{}
          \Infer[no rule]1{\code{
           {\langle} {\langle}\sigma,\_, assert~Peer(P);e{\rangle},\cstate,\sstore{\rangle} 
           \ttrans  {\langle}{\langle}\sigma,P, e{\rangle},\cstate,\sstore{\rangle}
          }}
	}}\\
            \hspace{-5mm}
          \ebpsem{OPEN}{{
            \Hypo{}
          \Infer[no rule]1{\code{
           {\langle} {\langle}\sigma,\_, open()~with~ (\lchanvar,\rolevar^*);e{\rangle},\cstate,\sstore{\rangle} 
           \ttrans  {\langle}{\langle}\sigma,\_, e{\rangle},\cstate[\res{\mapsto}(\lchanvar,\rolevar^*)],\sstore{\rangle}
          }}
	}}\\
            \hspace{-5mm}
          \ebpsem{CLOSE}{{
        \Hypo{\code{
        \cstate{=}\cstate'[\echanvar{\mapsto}(\lchanvar,\rolevar^*)] 
        \quad 
        (\_,\css){=}\sstore \quad \cproj{=}\css(\lchanvar) \quad \SAFE{=}\btt{EMPTY}(\cproj,\rolevar^*)
        }}
          \Infer1{\code{
           {\langle} {\langle}\sigma,\_,close(\echanvar);e{\rangle},\cstate,\sstore{\rangle} 
           \ttrans  {\langle}{\langle}\sigma',\_, e{\rangle},\cstate',\sstore{\rangle}
          }}
	}}\\
	\end{array}
        $     
        }}
         \hrule
	\end{center}
	\vspace*{-3mm}
	\caption{Some Semantic Rules} \label{fig.semantics}
	\vspace*{-3mm}
\end{figure*}

}

\noindent{\bf \emph{Locality.}}
Assuming that each message is characterized by 
a precise formula, each time a thread performs
a read it acquires the resource ownership of exactly
that heap portion needed
to satisfy the formula corresponding to the received message.
Similarly, 
on sending a message the thread releases a resource,
i.e. it transfers the
ownership of exactly that heap portion determined by the message formula.
Since the formulae describing the messages are
precise, a transmission can only modify the state
of the local heap in one way, releasing or acquiring
the resource which is being transmitted, or in other words
there is only one possible local transmission
which is safe, 
as highlighted 
in 

The validity of a triple is inductively defined with respect to the
reduction rules and satisfaction relation as follows:
\begin{defn}[Validity]
  A triple \code{\triple{\St_1}{\expr}{\St_2}} is valid,
written \code{\Valid{\triple{\St_1}{\expr}{\St_2}}}
if:\\
{\code{
  \forall {\itst \in \TS,\CC \in \cconf} ~ \cdot}}\\
\centerline{\code{({\ISatisf{\itst}{\CC}{\St_1}}) ~\wedge~
  ({\itst = \tstate{\_}{\_}{\expr}})  ~\wedge~
  ({\tconfig{\itst}{\CC} \ttrans \tconfig{\itst'}{\CC'}}) ~\wedge~
  ({\itst' = \tstate{\_}{\_}{\expr'}})
  }}\\
\centerline{\code{
    {~\implies~ \exists \St \cdot (\ISatisf{\itst'}{\CC'}{\St})
      ~\wedge~
      ({\triple{\St}{\expr'}{\St_2}})
    }.
}}
\end{defn}

\begin{thm}[Preservation]
  For expression \code{\expr} and states \code{\St_1} and \code{\St_2}, if \code{\htriple{\St_1}{\expr}{\St_2}}
  then \code{\Valid{\triple{\St_1}{\expr}{\St_2}}}.
\end{thm}

\begin{thm}[Progress]
  If \code{\htriple{\St_1}{\expr}{\St_2}} and 
  \code{\exists {\itst \in \TS,\CC \in \cconf} ~\cdot}
  \code{\ISatisf{\itst}{\CC}{\St_1}} then either
  \code{\expr} is a value or 
  \code{\exists {\itst' \in \TS,\CC' \in \cconf} ~\cdot}
  \code{{\tconfig{\itst}{\CC} \ttrans \tconfig{\itst'}{\CC'}}}.
\end{thm}

\noindent{\bf \emph{Interference.}}
We stress on the fact that
the current paper
highlights the effects explicit
synchronization has strictly over the communication.
The effects over the local heap are orthogonal issue
tackled by works such as  
\cite{Brookes:2007:SCS:1235896.1236120,
  Reddy:2012:SCI:2103656.2103695,
  Feng:2007:RCS:1762174.1762193}.
This explains the semantic choice for the programming
model of \figref{fig.semantic_model} which 
separates
the resources owned by a thread
from those owned by a communication channel,
and where the
environment interference
only affects the state of communication and not the local state.

}

\section{Explicit Synchronization}
{\footnotesize
\renewcommand{\arraystretch}{3.0}
\begin{figure}
	\begin{center}
        $
	\begin{array}{c}
          \ebpverif{CREATE}{{
          \Hypo{\code{ \hov {=} \bigwedge\limits_{j\in\{2..n\}} \assume{\ev_j \implies
          {\lthb{\ev_1}{\ev_j} } } }}
          \Infer1{\code{
          \triple
          {emp}
          {\bf create() ~with ~\ev_1,\overline{\ev_2..\ev_n}}
          {NOTIFY(\ev_1,\guard{\ev_1}) * WAIT(\overline{\ev_2..\ev_n},\hov) }
          }}
	}}\\
	\ebpverif{NOTIFY-ALL}{{
          \Hypo{\code{
          \triple
          {NOTIFY(\ev,\guard{\ev}) \wedge \ev }
          {\bf notifyAll(\ev)}
          { NOTIFY(\ev,\emp)} } }
	}}\\ 
          \ebpverif{WAIT}{{
          \Hypo{\code{\hov^{rel} {=} \assume{\ev \implies {\lthb{\ev_1}{\ev} } }}}
          \Infer1{\code{
          \triple
          {WAIT(\ev,\hov^{rel}) \wedge \notop{\ev}}
          {\bf wait(\ev)}
          {WAIT(\ev,\emp) \sep \hov^{rel} } }}
          }}
        \end{array}
        $
        \renewcommand{\arraystretch}{2.0}
        $        
        \begin{array}{ll}
        \text{\it(Wait~ lemma)} &
        \code{\assume{\ev_2 \implies
                               {\lthb{\ev_1}{\ev_2}
                               } } \wedge \ev_2 \implies
                            {\lthb{\ev_1}{\ev_2}}} \\
        \text{\it(Distribute-waits)} &
        \code{WAIT(\overline{\ev_2..\ev_n}, \bigwedge\limits_{j\in\{2..n\}} \racefreeas_j) 
                               \implies \bigwedge\limits_{j\in\{2..n\}}WAIT(\overline{\ev_2..\ev_n},\racefreeas_j)
                                    } \\
        \end{array}
        $        
        \hrule
	\end{center}
	\vspace*{-3mm}
	\caption{Synchronization Primitives for
          \code{wait}-\code{notifyAll} } \label{fig.synch.primitives}
          \vspace*{-5mm}
\end{figure}
}



Depending on the communication context, and 
on the instrument used for communication we could opt
amongst a few explicit synchronization mechanisms, such
as \code{CountDownLatch}, \code{wait-notifyAll},
\code{barriers}, etc. The choice of these
mechanisms are orthogonal to our approach.
We have experimented some 
examples with both
\code{CountDownLatch} and \code{wait-notifyAll}
and faced similar results.
We chose to formalize the latter since its
simplicity 
suffice for our running example.

The creation of a conditional-variable for
\code{wait}, \code{\code{[\underline{{\bf \scriptstyle
      }\rulen{CREATE}}]}}
in \figref{fig.synch.primitives},
releases the specification 
to verify the calls of \code{notifyAll}
and \code{wait}, respectively, with respect to
the session logic orderings. 

A call to notify is safe only if
the triggering event has occurred already,  \code{\code{[\underline{{\bf \scriptstyle
      }\rulen{NOTIFY-ALL}}]}}.
In other words,
the caller's state should
contain the triggering event
information.
\code{\code{[\underline{{\bf \scriptstyle
        \rulen{WAIT}}}]}} on the other hand,
 releases an ordering
 assumption conditioned by the send/receive
 event which is protected by the current \code{wait}.
 The condition has a double meaning in this context:
 (i) the protected event should have not occurred
 before a call to \code{wait}, and (ii) the ordering
 is released to the state only after proving that the event indeed 
 occurred, facilitated by the {\it Wait~ lemma } in \figref{fig.synch.primitives}.

The ease of detecting a \code{wait-notifyAll} deadlock 
(within a single synchronization object) is
a bonus offered by our logic, since it is simply reduced
to checking whether there is any context in which
a call to \code{wait} terminated without a corresponding \code{notify} call.
Formally, this is captured by:

\centerline{\code{          \code{NOTIFY(\ev,\guard{\ev})
           \sep WAIT(\ev,emp) \implies \errdead}
}}

For more general deadlocks across multiple synchronization
objects, we
will need to build {\em waits-for} graphs amongst these
objects and detect cycles, where possible,  using lemmas
similar to the above. For simplicity, these issues
are ignored in the current presentation.

\section{Modular Protocols}
Modularity is essential in designing and implementing
new software since it often involves reusing and
composing existing components.
Software components are modular
if their composition in larger software
preserves the overall expected computation and,
equally important, the safety properties.
Checking software compatibility is a known hard problem,
even more so when asynchronous communication
across components is involved, a heavily used mechanism
in building distributed system.
In the subsequent subsections, we list a series of extensions to the current
logic which support the design of modular protocols. Modular protocols
are designed and refined once, and then re-used multiple times
to support the design of more complex protocols.

\subsection{\bf Labelling.} Thus far our reasoning was based on the fact that each
transmission label is unique. To ensure
that there is no label clash across
all transmission, even in the presence
of composed protocols and multiple
instantiations of the same protocol,
a label is defined as a composition
between a parameterized label root
and a unique local id:
\code{\atid{\#}\atid_{\%d}}, where such a composed
label is a potential label root
for nested protocol instantiations.
\\
\noindent{\emph {Example~\setExampleNo{firstexample}}}. Let
\code{\gp} be a global protocol which is
assembled using protocol \code{\gp_0}:
\\
$
\arraycolsep=0pt
\begin{array} {lcl}
  \code{\gprot{\gp_0}{A,B,c}{\atid}{}} 
    & \code{\defeq} & \code{\atransmit{A}{B}{}{c}{\atid{\#}1}.} \\
    \code{\gprot{\gp}{A,B,C,c}{\atid}{}} 
    & \code{\defeq} & \code{\atransmit{A}{B}{}{c}{\atid{\#}1} \seq
    \gprot{\gp_0}{B,C,c}{\atid{\#}2}{}}.
\end{array}
$\\
Unrolling \code{\gp_0} within the definition of \code{\gp} results in
the following protocol with unique labels:
\\
$
\arraycolsep=0pt
\begin{array} {l}
  \code{\gprot{\gp}{A,B,C,c}{\atid}{}}
   \code{\defeq 
    \atransmit{A}{B}{}{c}{\atid{\#}1} \seq
    \atransmit{B}{C}{}{c}{\atid{\#}2{\#}1}}.\\
\end{array}
$\\

\subsection{\bf Parameterized Frontier for Previous State.}
As highlighted in \secref{sec.global.safety}
the ordering assumptions and race-free
proof obligations are collected via a 
manipulation of protocol summaries, where a summary consists
of a backtier, a frontier and accumulated ordering
constraints. The orderings are then added to the
global protocol as part of the refinement phase. 
To extend
this approach to composable protocols we
equip each protocol with an extra parameter
meant to
link the current protocol  with a generic previous state
ensuring therefore that wherever plugged-in, the considered protocol
does not create a communication race. Generally speaking,
a protocol is described by a predicate in our session logic
whose parameters represent the communicating peers, the
logical channels used for communication, a root label and
the previous state frontier:
\code{\gprot{\gp}{\rolevar^*,\chanvar^*}{\atid,\mfront}{}}.
The same \code{\mfront} variable is also used
later in Sec~\ref{modular:poststate} to denote the frontier for the next stage,
via \code{\mfront'},

\noindent{\emph {Example~\ref{firstexample} - revisited}}. Using the same simple example
as in the previous paragraph we highlight the
key ideas of parameterized frontier, \code{\mfront}, as below:
\\
\noindent
$
\arraycolsep=0pt
\begin{array}{lcl}
\code{\gprot{\gp_0}{A,B,c}{\atid,\mfront}{}}  &\code{\defeq}&
\code{\atransmit{A}{B}{}{c}{\atid{\#}1}.} \\
\code{\gprot{\gp}{A,B,C,c}{\atid,\mfront}{}} 
& \code{\defeq } &
\code{\atransmit{A}{B}{}{c}{\atid{\#}1} \seq
  \gprot{\gp_0}{B,C,c}{\atid{\#}2, \mfront_0}{}}.
\end{array}
$\\
\noindent
where
\code{ \mfront_0 {=} \mfront \mergeseq \code{\osfrt{\os}_1}}
and
\code{\os_1 {=} \CC(\atransmit{A}{B}{}{c}{\atid{\#}1})},
therefore\\
\code{ \mfront_0 {=} \mfront \mergeseq \langle [A{:}\event{A}{\atid{\#}1},B{:}\event{B}{\atid{\#}1}],
    [c{:}\atransmit{A}{B}{}{}{\atid{\#}1}] \rangle}.
 
  After the refinement phase
  and the proper instantiation of
  \code{\gp_0} in the body of \code{\gp}, we get:
  \\
\noindent
$
\arraycolsep=0pt
\begin{array}{llll}
\code{\gprot{\gp_0}{A,B,c}{\atid,\mfront}{}}
 &\code{\defeq}&
\code{\atransmit{A}{B}{}{c}{\atid{\#}1}} &
  \code{\blueord{
    \seq \assume{\atransmit{A}{B}{}{}{\atid{\#}1}}
    \seq \assume{{\lthb{\bordrmap{\mfront}(A)}{\event{A}{\atid{\#}1}}}}}}\\
  & & & \code{\blueord{    \seq \assume{{\lthb{\bordrmap{\mfront}(B)}{\event{B}{\atid{\#}1}}}}
    \seq \guard{\lthb{\bordcmap{\mfront}(c)}{\atid{\#}1}}.
  }}\\
\code{\gprot{\gp}{A,B,C,c}{\atid,\mfront}{}}
 & \code{\defeq } & 
\code{\atransmit{A}{B}{}{c}{\atid{\#}1}} & 
  \code{ \blueord{
  \seq \assume{\atransmit{A}{B}{}{}{\atid{\#}1}}
  \seq \assume{{\lthb{\bordrmap{\mfront}(A)}{\event{A}{\atid{\#}1}}}}}}\\
  & & & \code{ \blueord{
  \seq \assume{{\lthb{\bordrmap{\mfront}(B)}{\event{B}{\atid{\#}1}}}}
  \seq \guard{\lthb{\bordcmap{\mfront}(c)}{\atid{\#}1}}} \seq}\\
  & \multicolumn{2}{r} {\code{\rho(
  \atransmit{A}{B}{}{c}{\atid{\#}1}}} &
  \code{\blueord{
    \seq \assume{\atransmit{A}{B}{}{}{\atid{\#}1}}
    \seq \assume{{\lthb{\bordrmap{\mfront}(A)}{\event{A}{\atid{\#}1}}}}}}\\
  &&&  \code{\blueord { \seq \assume{{\lthb{\bordrmap{\mfront}(B)}{\event{B}{\atid{\#}1}}}}
    \seq \guard{\lthb{\bordcmap{\mfront}(c)}{{\atid{\#}1}}}})}.

\end{array}
$

\noindent where
\code{\rho{=}[B/A,C/B,\atid{\#}2/\atid,\mfront_0/\mfront]}.
Knowing \code{\mfront_0} and applying the substitution \code{\rho},
\code{\gp} is then normalized to:\\
\noindent
$
\arraycolsep=0pt
\begin{array}{@{}llll}
\code{\gprot{\gp}{A,B,C,c}{\atid,\mfront}{}}
& \code{\defeq } &
\code{\atransmit{A}{B}{}{c}{\atid{\#}1}}& 
 \code{\blueord{
  \seq \assume{\atransmit{A}{B}{}{}{\atid{\#}1}}
  \seq \assume{{\lthb{\bordrmap{\mfront}(A)}{\event{A}{\atid{\#}1}}}}}}\\
& & & \code{\blueord{
  \seq \assume{{\lthb{\bordrmap{\mfront}(B)}{\event{B}{\atid{\#}1}}}}
  \seq \guard{\lthb{\bordcmap{\mfront}(c)}{\atid{\#}1}}} \seq }\\
&\multicolumn{2}{r}{\code{
  \atransmit{B}{C}{}{c}{\atid{\#}2{\#}1}}} &
   \code{\blueord{
     \seq \assume{\atransmit{B}{C}{}{}{\atid{\#}2{\#}1}}
     \seq \assume{{\lthb{\event{B}{\atid{\#}1}}{\event{B}{\atid{\#}2{\#}1}}}}}} \\
&&&   \code{\blueord{\seq \assume{{\lthb{\bordrmap{\mfront}(C)}{\event{C}{\atid{\#}2{\#}1}}}}
     \seq \guard{\lthb{{\atid{\#}1}}{{\atid{\#}2{\#}1}}}.}
 }
\end{array}
$

Note that, for the clarity of this example (and subsequent ones),
we assume that \code{\bordcmap{\mfront}(\chanvar)} returns exactly
one transmission and \code{\bordrmap{\mfront}(\rolevar)} returns
exactly one event, which is actually the case for this example.
However, according to the definition of a
map element, \figref{fig.boundarysum}, the maps are populated with  a
composition of transmissions or events. To handle element maps,
we should simply form a \code{\lthb{}{}} relation for
each event/transmission returned
by the map and add an assumption or a guard for each such created relation.

\subsection{\bf Sufficient condition for implicit synchronization with the
  pre-context.}
\label{cond-impl-sync}
Adding an instantiable 
frontier to link the current protocol with its usage context
enables the refinement process to add the necessary ordering
assumptions and  guards  within the current protocol.
As explained in previous sections, these guards hold either when sufficient
implicit synchronization is provided, or when explicit synchronization mechanisms are used. 
To give system
designer the option to compose only protocols
that are implicitly sychronized
with the environment where they are plugged in,
we could derive a guard which 
enforces the implicit synchronization between the protocol and its pre-usage context.
To this purpose, we consider a diagramatic view of the ordering relations where each edge is 
either an \code{\HB} or \code{\CB} relation, and aim to find those
missing edges which would 
make the race-free guards (involving the pre-state) hold. 

\begin{defn}[Diagramatic Ordering Relations]
  A diagramatic view for a set \code{\massum} of event orderings is a directed acyclic graph 
  \code{\graph{\massum} {=} (V,Edg)}, where the set \code{V} of vertices
  contains all the events in \code{\massum}, 
  \code{V{=}\bigcup\limits_{\racefreeas \in \massum}
    \EV{\racefreeas}},
  and the set \code{Edg} of edges represents  
  all the \code{\HB} and \code{\CB} ordering relations in \code{\massum}, \\
  \code{Edg{=} \{(\ev_1,\ev_2) ~|~
    \lthb{\ev_1}{\ev_2} \in \massum
    ~\text{\emph{or}}~
    \ltcb{\ev_1}{\ev_2}
    \in \massum \}.
  }
\end{defn}

\newcommand{\aline}[1]{line \code{#1}}
The derivation of the precondition for the implicit synchronization is
depicted in \code{\bf Algorithm~ 2}: the generic frontier
is merged with the backtier of the protocol's body (\aline{2})
to generate the \RF guards which ensure
safe composition with the environment. For each such guard
\code{\lthb{\ev_1}{\ev_2} \in \mguards}
to be satisfied, the algorithm searches backwards, starting from
\code{\ev_2},
a way to connect it 
 to \code{\ev_1} using only 
ancestor nodes associated to the generic frontier. 
The \code{ancestors(\ev,\graph{})} method
returns all the ancestors nodes of node \code{\ev} in 
graph \code{\graph{}}. The Cartesian product \code{\times_{HB}}
is used to create \emph{weak} \HB~ relations between \code{\ev_1}
and the ancestors of \code{\ev_2}. To omit redundancies,
the algorithm only considers \code{\ev_2}'s earliest ancestors.
 The Cartesian product between
sets of events is defined as follows:\\
\centerline{
\code{S_1 \times_{HB} S_2 {~\eqdef~}
  \{\lthbw{\ev_1}{\ev_2}
  ~|~ \ev_1 \in S_1 \text{~and~}  \ev_2 \in S_2\}}}
\centerline{
~where~ \code{\lthbw{\ev_1}{\ev_2} {~\eqdef~}
  \lthb{\ev_1}{\ev_2}
  \vee
  \ev_1{=}{~\ev_2}
}
}

The weaker relation \code{\lthbw{}{}} is needed to express that two events might
be derived as identical.
It is easy to notice that the \code{\footnotesize\text{[HB-HB]}}
propagation rule
defined in \figref{ALrules}
can soundly
be extended to account for the \code{\lthbw{}{}} relations as well:
\begin{center}
$
\begin{array}{lll}
  \code{\lthb{\ev_1}{\ev_2} \wedge \lthbw{\ev_2}{\ev_3}} &
  \code{\implies \lthb{\ev_1}{\ev_3}} &
  \code{\tiny\text{[HB-HB(a)]}}\\
  \code{\lthbw{\ev_1}{\ev_2} \wedge  \lthb{\ev_2}{\ev_3}} &
  \code{\implies \lthb{\ev_1}{\ev_3}} &
  \code{\tiny\text{[HB-HB(b)]}}\\
  \code{\lthbw{\ev_1}{\ev_2} \wedge  \lthbw{\ev_2}{\ev_3}} &
  \code{\implies \lthbw{\ev_1}{\ev_3}} &
  \code{\tiny\text{[HB-HB(c)]}}
\end{array}
$
\end{center}

\noindent however, the \code{\footnotesize\text{[CB-HB]}} cannot
be fired in the presence of \code{\lthbw{}{}} since that would
involve changing a \code{\ltcb{}{}} relation into \code{\lthb{}{}}
which would lead to unsoundness.
For brevity, we use the shorthand:
\code{\EV{\bordrmap{\mfront}}} to denote
\code{\bigcup\limits_{\rolevar \in dom(\bordrmap{\mfront})}
  \EV{\bordrmap{\mfront}(\rolevar)}}
and
\code{\EV{\bordcmap{\mfront}}} to denote
\code{\bigcup\limits_{\chanvar \in dom(\bordrmap{\mfront})}
  \EV{\bordcmap{\mfront}(\chanvar)}}.

\IncMargin{1em}
\begin{algorithm}
\SetKwProg{Fn}{Function}{}{}
\SetKwData{Left}{left}\SetKwData{This}{this}\SetKwData{Up}{up}
\SetKwFunction{Union}{Union}\SetKwFunction{FindCompress}{FindCompress}
\SetKwInOut{Input}{input}\SetKwInOut{Output}{output}
\let\oldnl\nl
\newcommand{\nonl}{\renewcommand{\nl}{\let\nl\oldnl}}
\SetAlCapHSkip{0pt}
{
  { \Indm  
    \Input{\code{\gprot{\gp}{\rolevar^*,\chanvar^*}{\atid,\mfront}{} \eqdef \prot} - a global multi-party protocol }
    \Output{\code{\mguards_{pre}} - a set of orderings guards
      }}
  \BlankLine
  \bind{\code{\os}}{\code{\CC(\prot)}} ;\quad
  \bind{\code{\mguards_{pre}}}{\code{\emptyset}} \;
  \bind{\code{\langle{\massum,\mguards}\rangle}}{\code{\mfront \mergeseq {\osbck{\os}}}}
  \tcc*{relations between environment and \code{\gp}}
  \ForEach{\code{\lthb{\ev_1}{\ev_2} \in \mguards} } 
  {\bind{\code{A^{ev}}}{ancestors(\code{\ev_2},\code{\graph{\massum}})
      \code{\cap ~(\EV{\bordrmap{\mfront}} \cup
        \EV{\bordcmap{\mfront}})} }  \tcc*{only frontier ancestor}
    \bind{\code{\overline{A^{ev}}}}{
     \code{ \{\ev ~|~
      ancestors(\ev,\code{\graph{\massum}}) \cap
      {A^{ev}} {=} \emptyset\}
    }}  \tcc*{only earliest ancestors}
    \bind{\code{A^{hb}}}{\code{\{\ev_1\}
        \times_{HB} \overline{A^{ev}}} } \tcc*{candidate HB relations}
    \bind{\code{\overline{A^{hb}}}} 
    {\code{\{ \orderas_{hb} ~|~ \orderas_{hb} \in A^{hb} ~s.t.~ \racefree{\massum \cup \{\orderas_{hb}\}}
               {\lthb{\ev_1}{\ev_2}}
               \} }} \tcc*{keep only useful HB 
           }
           \bind{\code{\mguards_{pre}}}{\code{\mguards_{pre} \cup
               \{\bigvee\limits_{\orderas_{hb} \in \overline{A^{hb}}} \orderas_{hb}\}
             }} \tcc*{updates the pre guards}
 }  
  \Return{\code{\mguards_{pre}}}
}
\caption{Derives the necessary conditions for this protocol to be
  implicitly synchronized with the pre-context}
\end{algorithm} 
\DecMargin{1em}

\noindent{\emph {Example~\ref{firstexample} - revisited}}. Using the same simple example
as in the previous paragraph, we derive the conditions for which
\code{\gp_0} could be implicitly synchronized with the body of
\code{\gp}:\\
$
\arraycolsep=0pt
\begin{array}{ll}
\code{\langle{\massum,\mguards}\rangle}  \code{:=}
\code{
  \langle \{ \event{A}{\atid{\#}1},\event{B}{\atid{\#}1},
  \ltcb{\event{A}{\atid{\#}1}}{\event{B}{\atid{\#}1}},
  {\lthb{\bordrmap{\mfront}(A)}{\event{A}{\atid{\#}1}}},}\\
\code  {\lthb{\bordrmap{\mfront}(B)}{\event{B}{\atid{\#}1}}
  \},}
\code{\;\{ {\lthb{\tsend{\bordcmap{\mfront}(c)}}{\event{A}{\atid{\#}1}}},
     {\lthb{\trecv{\bordcmap{\mfront}(c)}}{\event{B}{\atid{\#}1}}} \}
  \rangle.
  }
\end{array}
$
\noindent

Building the graph corresponding to \code{\massum}, finding the
ancestors of \code{\event{A}{\atid{\#}1}} and
\code{{\event{B}{\atid{\#}1}}}, 
and building the HB candidates yields
\code{\mguards_{pre} =
  \{\lthbw{\tsend{\bordcmap{\mfront}(c)}}{\bordrmap{\mfront}(A)},
  \lthbw{\trecv{\bordcmap{\mfront}(c)}}{\bordrmap{\mfront}(B)}
  \}}  in two iterations of the loop.
Given this precondition for \code{\gp_0} it is easy to observe
with the appropriate instantiation that \code{\gp_0} cannot
be simply plugged into \code{\gp} without proper explicit
synchronization:\\
\code{
  \racefree{
    \{
    \event{A}{\atid{\#}1},\event{B}{\atid{\#}1},
    \ltcb{\event{A}{\atid{\#}1}}{\event{B}{\atid{\#}1}},
    {\lthb{\bordrmap{\mfront}(A)}{\event{A}{\atid{\#}1}}},
    {\lthb{\bordrmap{\mfront}(B)}{\event{B}{\atid{\#}1}}}
    \}}}\\
\hspace*{0pt}\hfill  \code{ {\rho}
  (\lthbw{\tsend{\bordcmap{\mfront}(c)}}{\bordrmap{\mfront}(A)}
  \wedge
}
\code{
    \lthbw{\trecv{\bordcmap{\mfront}(c)}}{\bordrmap{\mfront}(B)}
    ).
  }
\\ \noindent
with \code{\rho{=}[B/A,C/B,\atid{\#}2/\atid,\mfront_0/\mfront]}, the
above involves proving that
\code{\lthbw{\event{A}{\atid{\#}1}}{\event{B}{\atid{\#}1}}
  \wedge \lthbw{\event{B}{\atid{\#}1}}{\bordrmap{\mfront}(C)}}.
This precondition cannot be proved since
there is no sufficient implicit synchronization.

Using \code{\gp_0} in a different context:\\
\centerline{
$
\arraycolsep=0pt
\begin{array}{@{}ll}
\code{\gprot{\gp_1}{A,B,c}{\atid,\mfront}{}} &
\code{\defeq }
\code{\atransmit{A}{B}{}{c}{\atid{\#}1} \seq
  \gprot{\gp_0}{A,B,c}{\atid{\#}2, \mfront_0}{}}.\\
\end{array}
$}\\
yields the following proof obligation:\\
\centerline{
\code{\lthbw{\event{A}{\atid{\#}1}}{\event{A}{\atid{\#}1}}
  \wedge \lthbw{\event{B}{\atid{\#}1}}{\event{B}{\atid{\#}1}}}}\\
which is trivially true. Therefore, \code{\gp_0} can be plugged in
the body of  \code{\gp_1} without any additional synchronization.

\subsection{\bf Predicate summary for post-context.} 
\label{modular:poststate}
The condition for implicit
synchronization  
ensures race-freedom \wrt the pre-usage context.
To also ensure communication safety \wrt its
post-usage context
the predicate's frontier is made available to be plugged-in
for merging with the usage site backtier. 


\noindent{\emph {Example~\setExampleNo{secondexample}}}.
Using a variation of the previous
examples, we highlight the role of a predicate's frontier
at its usage context:\\
$
\arraycolsep=0pt
\begin{array}{lcl}
\code{\gprot{\gp_2}{A,B,C,c}{\atid,\mfront}{}} 
 & \code{\defeq } &
\code{\atransmit{A}{B}{}{c}{\atid{\#}1}~ \seq~}
\code{\atransmit{B}{C}{}{c}{\atid{\#}2}}.\\
\code{\gprot{\gp_3}{A,B,C,c}{\atid,\mfront}{}} 
 & \code{\defeq } &
\code{\gprot{\gp_2}{A,B,C,c}{\atid{\#}1,\mfront}{} ~\seq~}
\code{\atransmit{B}{C}{}{c}{\atid{\#}2}.}
\end{array}
$

The frontier of \code{\gp_2} is derived to be:\\
\centerline{\code{\langle [A{:}\event{A}{\atid{\#}1},B{:}\event{B}{\atid{\#}2},C{:}\event{C}{\atid{\#}2}],
    [c{:}{\atid{\#}2}] \rangle}}.
Using this information within the body
of \code{\gp_3} leads to the following refined protocol:\\
$
\arraycolsep=0pt
\begin{array}{@{}ll}
\code{\gprot{\gp_3}{A,B,C,c}{\atid,\mfront}{}} 
\code{\defeq } &
\code{\gprot{\gp_2}{A,B,C,c}{\atid{\#}1,\mfront}{} ~\seq~}
                 \code{\atransmit{B}{C}{}{c}{\atid{\#}2}} 
  \code{\blueord{\seq \assume{\atransmit{B}{C}{}{}{\atid{\#}2}}}} \\
& \code{\blueord{
  \seq \assume{{\lthb{\event{B}{\atid{\#}1{\#}2}}{\event{B}{\atid{\#}2}}}}
  \seq
  \assume{{\lthb{\event{C}{\atid{\#}1{\#}2}}{\event{C}{\atid{\#}2}}}}}}\\
& \code{\blueord{ 
  \seq \guard{\lthb{\atid{\#}1{\#}2}{\atid{\#}1}}}}.
\end{array}
$\\
Notice that without the frontier of \code{\gp_2}, its usage in the
body of \code{\gp_3} would potentially lead to a race on \code{\chanvar}
since \code{\gp_3}
would otherwise not have enough information to add the ordering relations
and race-free guards within its body (events labelled \code{\atid{\#}1{\#}2}).

\subsection{\bf Recursion.}
For recursive predicates we only consider those predicates
which are self-contained, that is to say
 explicit synchronization
is only supported within the body of the predicate but not across
recursive calls. If this condition is not met,
the recursion would have to pass synchronized objects across recursion,
creating an unnecessary complicated communication model.
\hide{
\IncMargin{1em}
\begin{wrapfigure}{L}{0.5\textwidth}
\begin{minipage}{1.1\linewidth}
\addtolength{\hsize}{-1.5\algomargin}
\begin{algorithm}[H]
\SetKwProg{Fn}{Function}{}{}
\SetKwInOut{Input}{input}\SetKwInOut{Output}{output}
\let\oldnl\nl
\newcommand{\nonl}{\renewcommand{\nl}{\let\nl\oldnl}}
\SetAlCapHSkip{0pt}
{
  {\Indm  
    \Input{\code{\gprot{\gp}{\rolevar^*,\chanvar^*}{\atid,\mfront}{} \eqdef \prot}}
    \Output{ Summary of {\code{\gp}
      }}}
  \BlankLine
  \bind{\code{\os}}{\code{\CC(\prot)}} \;
  \bind{\code{\langle{\massum,\mguards}\rangle}}{\code{\mfront \mergeseq {\osbck{\os}}}}
  \\
  \bind{\code{\massum_{post}}}{\code{\massum \cup {\osasm{\os}}}}\;
  \Repeat{\code{n=m}}
  {
    \bind{\code{n}}{\code{|\massum_{post}|}}\;
    \bind{\code{\massum_{post}}}{\code{\massum_{post} \uplus \{
        \orderas ~|~  {\orderas_1 \wedge \orderas_2 \implies^* \orderas}
  ~\text{and}~}} \code{\text{\qquad\qquad\qquad\qquad} \orderas_1,\orderas_2 \in \massum_{post}\}}\;
    \bind{\code{m}}{\code{|\massum_{post}|}}\;
  }
  \bind{\code{\overline{\massum_{post}}}}{\code{\{\orderas | \orderas
      \in {\massum_{post}} \text{~and~}
      \EV{\orderas} \subseteq \EV{\mfront} \cup \EV{\osfrt{\os}} \}}}\;
  \Return{\code{\langle \overline{\mguards_{post}}, {\mfront \mergeseq\osfrt{\os}}\rangle }}
}
\caption{Derives a protocol summary}
\end{algorithm} 
\end{minipage}%
\end{wrapfigure}
}
To support recursion we:
(1) derive the sufficient condition for implicit synchronization with the
pre-state according to the algorithm described in
\secref{cond-impl-sync};
(2) we derive the summary of the protocol up to the recursive invocation of
the considered predicate, as described in \secref{modular:poststate};
(3) given the summary in (2) as ordering context,
we next check whether the implicit synchronization
condition holds for the recursive invocation. If it does, then the
protocol is implicitly synchronized and could be safely plugged-in
within any environment which satisfies its implicit synchronization
condition.\\
$
\arraycolsep=0pt
\begin{array}{@{}ll}
\code{\gprot{\gp_5}{A,B,C,c}{\atid,\mfront}{}} 
\code{\defeq } &
              \code{\atransmit{A}{B}{}{c}{\atid{\#}1}}
              \seq \code{\atransmit{C}{A}{}{c}{\atid{\#}2}}
              \seq \code{\atransmit{A}{B}{}{c}{\atid{\#}3}}\\
&              \seq \code{\gprot{\gp_5}{A,B,C,c}{\atid,\mfront}{}}. 
\end{array}
$\\
To check whether this protocol is self-contained, the algorithm first
follows the steps described
in \secref{cond-impl-sync} to derive the sufficient condition for implicit synchronization.
This implicit synchronization condition is derived
to be: \\
\code{\lthbw{\tsend{\bordcmap{\mfront}(c)}}{\bordrmap{\mfront}(A)} \wedge \lthbw{\trecv{\bordcmap{\mfront}(c)}}{\bordrmap{\mfront}(B)}}\\
which, when instatiated to the body of \code{\gp_5} it boils down to check that:
\code{\lthbw{\event{A}{\atid{\#}3}}{\event{A}{\atid{\#}3}}
  \wedge \lthbw{\event{B}{\atid{\#}3}}{\event{B}{\atid{\#}3}}},
which is trivially
true.
Despite containing races within its body, \code{\gp_5} is actually
safely synchronized across the recursive invocations.


\hide{
Collecting all the order relations of \code{\gp_2}'s body 
(incl. those \wrt its parameterized frontier) and applying the
propagation rules until the set of relations saturates, results in the
following set:\\
\code{\{
  \ltcb{\event{A}{\atid{\#}1}}{\event{B}{\atid{\#}1}},
  \lthb{\bordrmap{\mfront}(A)}{\event{A}{\atid{\#}1}},
  \lthb{\bordrmap{\mfront}(B)}{\event{B}{\atid{\#}1}},\\
  \ltcb{\event{B}{\atid{\#}2}}{\event{C}{\atid{\#}2}},
  \lthb{\event{B}{\atid{\#}1}}{\event{B}{\atid{\#}2}},     
  \lthb{\bordrmap{\mfront}(C)}{\event{B}{\atid{\#}2}},\\
  \quad
  \lthb{\event{A}{\atid{\#}1}}{\event{B}{\atid{\#}2}},
  \lthb{\bordrmap{\mfront}(A)}{\event{B}{\atid{\#}2}},
  \lthb{\bordrmap{\mfront}(B)}{\event{B}{\atid{\#}2}}
  \}}.
Finally the algorithm filters only those relations
which relate the protocol's frontier with the parameterized
one and results in:
\code{\{
  \lthb{\bordrmap{\mfront}(A)}{\event{A}{\atid{\#}1}},}
\code{  
  \ltcb{\event{B}{\atid{\#}2}}{\event{C}{\atid{\#}2}},
  \lthb{\bordrmap{\mfront}(C)}{\event{B}{\atid{\#}2}},\\
  \lthb{\event{A}{\atid{\#}1}}{\event{B}{\atid{\#}2}},
  \lthb{\bordrmap{\mfront}(A)}{\event{B}{\atid{\#}2}},
  \lthb{\bordrmap{\mfront}(B)}{\event{B}{\atid{\#}2}}
  \}}.
\\
}


\section{Related Work}

{\bf \emph {Behavioral Types approaches to communication protocols.}}
The behavioral types specify the expected interaction
pattern of communicating entities. Most of seminal works 
develop type systems on the $\code{\pi}$-calculus 
\cite{kobayashi2002type, kobayashi2003type} for  
deadlock \cite{kobayashi2006new} and livelock \cite{kobayashi2000type}
detection. 
Later developments \cite{kobayashi2017deadlock} handle deadlock detection with recursion 
and arbitrary networks of nodes for asynchronous CCS processes. By using a special recursive model
deadlock detection reduce to a check of circularity over dependencies.
\hide{
Instead of reasoning on finite approximations of such system,
the compounded processes are associated  
with terms of a basic recursive model, called lam. 
The lam terms are meant for binding the inter-channel dependencies 
which are then used to check for deadlocks by
checking  
}
Linearity is also studied in these systems \cite{padovani2014deadlock}, 
offering a better reasoning about unbounded dependency chains and
recursive types. A generic type system to
express types and type environments as abstract processes is proposed in \cite{IGARASHI:tcs2014}.
Following ideas of process algebras, and 
separation logic \cite{reynolds2002separation}, 
the work of \cite{caires2013type, caires2008spatial} introduces 
a behavioral separation for disciplining the interference of higher-order programs 
in the presence of concurrency and aliasing. 
The most intensely
studied refinement of the linear behavioral type systems are
the session types. 
Initially proposed for systems with interaction between exactly 
two peers \cite{Honda:esop98}, they
have been extended for  an arbitrary number of 
participants \cite{Honda:2008:MAS:1328897.1328472}.
A binary session type describes a protocol from the perspective of only one peer, 
with support for branching 
and selection and even with the possibility to delegate the communication 
to a third party \cite{Honda:esop98, Gay:acta2005}. 
Other extensions of session types include that of adding support:
to handle exceptions across participants \cite{carbone2008structured, Capecchi:msc2014};
for multithreaded functional languages 
\cite{neubauer2004implementation, gay2010linear, lindley2016embedding, Orchard:popl16};
for web service description languages \cite{Carbone:esop07};
for actor-based \hide{programming} languages \cite{Fowler:eptcs16};
for operating systems \cite{Fahndrich:os2006};
for\ MPI \hide{Message Passing Interface} \cite{Lopez:2015};
for \code{C} like languages \cite{Ng:2012};
for OO \hide{object oriented programming} languages 
\cite{Dezani-Ciancaglini:fmco2006, Dezani-Ciancaglini:ecoop06, Gay:popl10};
and for
event-driven systems \cite{kouzapas2016asynchronous}.
\hide{
In the context of message passing paradigm, there are numerous works
which use session types to guarantee statically and/or dynamically
scenario fidelity, communication safety and global progress for two-party
channel,  multi-party channel and multi-channel sessions.
}
Logical foundations of the session types have been studied in the context of
a unified theory \cite{Castagna:ppdp09} and linear logic \cite{Wadler:icfp12, lindley2015semantics, Caires:concur10}. 
Session types are able to express effects, and the effect system is powerful enough in order to
express session types \cite{Orchard:popl16}.
In a multiparty session types calculus \cite{Honda:2008:MAS:1328897.1328472}, the user provides a global 
descriptions of the communication and a projection algorithm 
computes the local view of each communicating party.
Even though it offers general communication channels\hide{ (asynchronous, shared across participants, 
FIFO queue)}, it complicates the formalization due to
possible race conditions.
A less general solution but with a clear formalization to avoid race condition is given in \cite{coppo2016}.
This restricts the number of channels to the number of distinct communicating participants 
pairs in the system. 
Another solution to disambiguate the usage of a single channel across multiple participants 
is to label the communication actions \cite{Caires:esop2009}. \hide{hat a single operation on a
channel shared by multiple participants can be identified by the (channel,label) pair.}
Asynchronous semantics is defined either in terms
of a projection of the global types \cite{Honda:2008:MAS:1328897.1328472}, 
or in terms of an asynchronous communication automata \cite{Denielou:2012}.  
An immediate application of multiparty session types
in that of designing deadlock-free choreographies \cite{Carbone:2013:DMA:2480359.2429101}, 
where even if syntactically written in a sequence, certain transmissions which are not in a causal dependence
can be swapped. 
An attempt to describe the content of the exchanged message by adding support for assertions to session types
is done in \cite{Bocchi:2010}.
\hide{
Multiparty asynchronous session types have been used for checking the systems 
with multiple endpoints \cite{Honda:2008:MAS:1328897.1328472, coppo2016, Denielou:2011:DMS:1926385.1926435}.
Different settings are taken into account, e.g. a fixed number of participants \cite{Honda:2008:MAS:1328897.1328472}, a dynamic number of participants \cite{Denielou:2011:DMS:1926385.1926435}, or 
interleaved  multi-sessions \cite{Bocchi:2012}. 
}
Our paper focuses on race-freedom  \cite{Voung:fse07, Balabonski2014, Yoga:fse16, Kahlon:fse09}
over common channels using both implicit and explicit
synchronizations\cite{Leino:esop10, Maiya:pldi14, Cogumbreiro:2015}. We achieve this abstractly with the help of
both \code{\precx{HB}}-ordering and a 
novel \code{\precx{CB}}-ordering.
Most of the previous works on session types  make the assumption
that the underlying system utilizes
only implicit synchronization using extra channels, where needed.

\noindent{\bf \emph {Concurrent Logics for message passing and synchronization mechanisms.}}
The idea of coupling together the model theory of concurrent
separation logic with that of Communicating Sequential Processes \cite{hoare1978communicating}
is studied in \cite{hoare2008separation}.
The processes are modeled by using trace semantics, drawing an analogy between
channels and heap cells, and distinguishing between separation in space from separation in
time. Our proposal shares the same idea of distinguishing between separation
in space and separation in time, by using the \code{∗} and \code{;} operators, respectively. However, their
model relies on process algebras, while we propose an
expressive logic based on separation logic able to also tackle memory management. Moreover,
our communication model is more general, yet more precise, by
accounting for explicit synchronizations as well.
Heap-Hop \cite{villard2009proving, villard:scp14} is a sound proof system for copyless message passing
managed by contracts. 
The system is integrated within a static analyzer which checks
whether messages are safely transmitted.
Similar with our proposal, this work is also based on separation logic.
\hide{, a natural
choice when dealing with memory management, and more generally when reasoning about
resource .} 
As opposed to our proposal, its communication model is limited to
bidirectional channels, whereas our model is general enough to also capture buffered channels
which may be shared among an arbitrary number of participants.
Our proposed session logic is more expressive 
than previous extensions of separation logic used to express session protocols
\cite{Florin:IECCS15}. It provides more details so that implementation can be safely written.
Moreover our logic is specially designed for multiparty protocols.
Another extension of separation logic \cite{bell2010concurrent}
for reasoning about multithreaded programs 
traces transmitted  messages as sets of ordered histories. Our approach does not need to track
histories of values, since we rely on an event ordering constraint system which ensures not only
the correct transmission order, but also that of explicit synchronization.
The papers \cite{leino2009basis, Leino:esop10} propose  
a verification methodology to prevent deadlocks of concurrent programs communicating via asynchronous shared channels, 
permit copyless message passing and share memory
via locks.\hide{ Deadlock prevention is achieved by enforcing safe usage of channels.} 
This work is similar to ours in the sense that they acknowledge the benefits but also the
complications of combining different synchronization mechanisms
A resource analysis for $\pi$-calculus \cite{turon2011resource}  continues the work of
\cite{hoare2008separation} by adding support for two kinds of channels, namely public and private 
and define two denotational models to
reason about both communication safety and liveness. 
Different concurrent logics based on separation logic have been proposed to reason
in a modular manner about the synchronization mechanisms \cite{Dodds:popl2011, Svendsen:esop2014}.
A permission based logic approach to race-free sharing of heap
between concurrent threads is described in \cite{Bornat:popl2005}.
Recently, the authors of \cite{Jung:popl2015} use monoids to express and invariants to enforce
protocols on shared data in a context of a concurrent separation logic.
Numerous static analyses for multiparty protocols have been 
proposed: deadlock detection \cite{Ng:2016} by synthesis of a global session graph;
minimization of the upper bound of the buffers size \cite{Denielou:2010} by generation 
explicit new messages for synchronization; 
computation of recovery strategies in case of failure \cite{Neykova:2017} by solving
a causal dependency graph; or 
liveness and safety checking \cite{Lange:2017:FOG:3009837.3009847}.
Comparing with these analyses, our approach generates
separation logic  assumptions and
proof obligations that are very general,
and can moreover be proven in a 
cooperative manner.   

\hide{
\noindent{\bf \emph {Causality Analysis.}} 

Seminal work of Lamport introduces sequential consistency \cite{lamport1978time, Lamport79}.
Starting from it a lot of memory models are proposed for 
happens before relation \cite{Zhang:2016, Kang:2017}. 
The comlexity of causal conssitency  is studied in \cite{Bouajjani:popl2017}.

}

\hide{
\begin{verbatim}
Session types add more structure to these channel types: for example, we can describe a (very) simplified version of the SMTP protocol as follows:

SMTPClient = ⊕ {
  EHLO: !Domain.!FromAddress.!ToAddress.!Message.SMTPClient
  QUIT: end
}

SMTPServer = & {
  EHLO: ?Domain.?FromAddress.?ToAddress.?Message.SMTPServer
  QUIT: end
}
# internal choice disjoint on the labels..

\end{verbatim}
}

\hide{
\begin{verbatim}
Intro to Languages with session types:
  http://simonjf.com/2016/05/28/session-type-implementations.html

Basic & unified theory of session type
  http://www.di.unito.it/~dezani/papers/cdgp.pdf

\end{verbatim}

\begin{verbatim}
SILL

(Static, Binary)

A programming language based on the intuitionistic linear logic view of session types.

Resources

From Linear Logic to Session-Typed Concurrent Programming, by Frank Pfenning.
Polarised Substructural Session Types, by Frank Pfenning and Dennis Griffith. In proceedings of FoSSaCS’15.


Erlang

monitored-session-erlang

(Dynamic, Multiparty, Actor-based)

An framework for monitoring Erlang/OTP applications by dynamically verifying communication against multiparty session types. Erlang actors can take part in multiple roles in multiple instances of multiple protocols.

The tool is inspired by the Session Actor framework of Neykova & Yoshida.

Resources

An Erlang Implementation of Multiparty Session Actors, by Simon Fowler. In proceedings of ICE’16.
Monitoring Erlang/OTP Applications using Multiparty Session Types, by Simon Fowler.
Go

DinGo Hunter

(External tool)

A static analyser for Go programs, which can statically detect deadlocks. 

Resources

Static Deadlock Detection for Concurrent Go by Global Session Graph Synthesis, by Nicholas Ng and Nobuko Yoshida. In proceedings of CC’16.
Gong

(External tool)

A more recent static analyser for Go, building on a minimal core calculus for Go called MiGo. MiGo types can be extracted from Go programs using another tool called GoInfer.

Given MiGo types obtained from GoInfer, Gong will check liveness and safety of communications.

Resources

Fencing off Go: Liveness and Safety for Channel-based Programming, by Julien Lange, Nicholas Ng, Bernardo Toninho, and Nobuko Yoshida. In proceedings of POPL’17.
\end{verbatim}
}
\hide{
- multiparty asynchronous session types (mast) \cite{Honda:2008:MAS:1328897.1328472,Denielou:2012,Denielou:2011:DMS:1926385.1926435,Bocchi:2010}

-automata  villard +\cite{Denielou:2012}
-multiparty multi-session logic \cite{Bocchi:2012}
- session interleaving \cite{coppo2016}
framework for verifying processes against assertions \cite{Bocchi:2010}  \cite{Bocchi:2012} villard
- deadlock-freedom by construction  \cite{Carbone:2013:DMA:2480359.2429101}
buffered communication \cite{Denielou:2010}
- static deadlock detection \cite{Ng:2016}
- recovery strategies \cite{Neykova:2017}
- barrier synchronization \cite{Cogumbreiro:2015}
in

- able to detect communication errors and partial deadlocks in a general class of realistic concurrent programs, including those with dynamic channel creation, unbounded thread creation and recursion \cite{Lange:2017:FOG:3009837.3009847}

- mpi verification \cite{Lopez:2015}

- mast for c \cite{Ng:2012}

- effects as sessions popl16 mast with effects 
}

\section{Discussions and Final Remarks}
{\noindent{\bf \emph{Implementation.}}}
We have implemented a prototype for \sessionlogic~ in OCaml
and added support for the communication primitives using the
verifier in \cite{Chin:2012:AVS:2221987.2222283}. As the
current work requires specification to be provided,
we have only been able to experiment with small
examples with our initial experiments.
Even so, message passing flawlessly intertwines with
different explicit synchronization mechanisms, such as
\code{wait-notifyAll} and \code{CountDownLatch}.


{\noindent{\bf \emph{Conclusion.}}}
We have designed an expressive
multi-party session logic that works seemlessly
with both implicit and explicit synchronization to ensure 
correctness of communication-centric programs.
Our approach is built up from first-principle and
is based on the use of
two fundamental ordering constraints, namely {\em ``happens-before''} \code{\precx{HB}}  and {\em ``communicates-before''} \code{\precx{CB}},
to ensure channel race-freedom. Our proof system
supports modular verification with the help
of automatically generated assumptions and proof obligations
from each given global protocol. We have also pioneered
the concept of {\em cooperative proving} amongst a
set of concurrent processes.
As part of future work, we intend to provide
support for the synthesis of explicit synchronization
that can help guarantee race-free communications.
We also intend to go beyond the current limits
of well-formed disjunctions.
\hide{
\begin{verbatim}
Many new venues for research:
Future:
automatically insert the CDL methods into the program code 
infer projections and compose them into a global protocol
create a semantic which accommodates both mailboxes and 
bidirectional channels uniformly 
handle programs with dynamic patterns, such
as spawning new threads after communication started
Immutability: when is it safe to share a data? 
\end{verbatim}
}

%
%
%





\bibliography{ref}
\bibliographystyle{ACM-Reference-Format}
\citestyle{acmauthoryear}


\appendix
\section{Semantics}

\newcommand{\opsemmachine}[1]{
\booltrue{#1}
{
\ifbool{appendix}{\captionsetup[subfigure]{labelformat=empty}}{}
\subfloat[][\ifbool{paper}{Machine Reduction}{}]{
\label{fig.semantics.machine}
$
\begin{array}{c}
  \ebpsem{MACHINE}{{
      \Hypo{\code{
      \tconfig{\ltst}{\CH} \ttrans^*
      \tconfig{\ltst'}{\CH'}
      }}
      \Infer1{\code{
      \mstate{\PS[i\,{\mapsto}\ltst]}{\CH}
      \ttrans
      \mstate{\PS[i\,{\mapsto}\ltst']}{\CH'}
      }}
  }} 
  \\  
  \ebpsem{PAR}{{
      \Hypo{\code{
      \ltst{=}\ltstate{\tst}{(\expr_1||\expr_2)}
      \quad
      \textit{fresh}~ \thdid_1, \thdid_2
      \quad
       \ltst'{:=} \ltstate{\emp_{sh}}{(\joinfunction{\thdid_1}{\thdid_2})}
      }}
      \Infer[no rule]1{\code{
      \tst {=}\tst_1 \uplus \tst_2
      \quad
      \PS'{:=}\PS[\thdid\,{\mapsto}\ltst']
      [\thdid_1{\mapsto} \ltstate{\tst_1}{\expr_1}]
      [\thdid_2{\mapsto} \ltstate{\tst_2}{\expr_2}]
      }}
      \Infer1{\code{
      \mstate{\PS[\thdid\,{\mapsto}\ltst]}{\CH} \ttrans  
      \mstate{\PS'}{\CH}
      }}
  }}\\
  \ebpsem{JOIN}{{
      \Hypo{\code{
      \ltst{=}
      \ltstate{\tst}{\joinfunction{\thdid_1}{\thdid_2}}
      }}
      \Hypo{\code{
      \PS=\PS'[\thdid_1{\mapsto} \ltstate{\tst_1}{\ccskip}]
      [\thdid_2{\mapsto} \ltstate{\tst_2}{\ccskip}]
      }}
      \Hypo{\code{
      \tst' {:=} \tst {\uplus} \tst_1 {\uplus} \tst_2 
      }}
      \Infer3{\code{
      \mstate{\PS[\thdid\,{\mapsto} \ltst]}{\CH} \ttrans  
      \mstate{\PS'[\thdid\,{\mapsto}\ltstate{\tst'}{\ccskip}]}{\CH}
      }}
  }}\\
  \ebpsem{JOIN-SKIP1}{{
      \Hypo{\code{
      \ltst{=} \ltstate{\_}{\joinfunction{\thdid_1}{\thdid_2}}          
      }}
      \Hypo{\code{
      \PS=\PS'[\thdid_1{\mapsto} \ltstate{\_}{\ccskip}]
      [\thdid_2{\mapsto} \ltstate{\_}{\expr}]
      }}
      \Infer2{\code{
      \mstate{\PS[\thdid\,{\mapsto} \ltst]}{\CH} \ttrans
      \mstate{\PS[\thdid\,{\mapsto} \ltst]}{\CH}
      }}
  }}\\
  \ebpsem{JOIN-SKIP2}{{
      \Hypo{\code{
      \ltst{=} \ltstate{\_}{\joinfunction{\thdid_1}{\thdid_2}}          
      }}
      \Hypo{\code{
      \PS=\PS'[\thdid_1{\mapsto} \ltstate{\_}{\expr}]
      [\thdid_2{\mapsto} \ltstate{\_}{\ccskip}]
      }}
      \Infer2{\code{
      \mstate{\PS[\thdid\,{\mapsto} \ltst]}{\CH} \ttrans
      \mstate{\PS[\thdid\,{\mapsto} \ltst]}{\CH}
      }}
  }}\\
  \ifbool{appendix}{
  \ebpsem{JOIN-ERR1}{{
      \Hypo{\code{
      \ltst{=} \ltstate{\_}{\joinfunction{\thdid_1}{\thdid_2}}          
      }}
      \Hypo{\code{
      \PS=\PS'[\thdid_1{\mapsto} \ltstate{\_}{\expr}]
      [\thdid_2{\mapsto} \error ]
      }}
      \Infer2{\code{
      \mstate{\PS[\thdid\,{\mapsto} \ltst]}{\CH} \ttrans
      \error
      }}
  }} \\ }{}
  \ifbool{appendix}{
  \ebpsem{JOIN-ERR2}{{
      \Hypo{\code{
      \ltst{=} \ltstate{\_}{\joinfunction{\thdid_1}{\thdid_2}}          
      }}
      \Hypo{\code{
      \PS=\PS'[\thdid_1{\mapsto} \error]
      [\thdid_2{\mapsto} \ltstate{\_}{\expr}]
      }}
      \Infer2{\code{
      \mstate{\PS[\thdid\,{\mapsto} \ltst]}{\CH} \ttrans
      \error
      }}
  }}  \\ }{}
\end{array}
$
}}}


\newcommand{\opsemthread}[1]{
\booltrue{#1}
\ifbool{appendix}{\captionsetup[subfigure]{labelformat=empty}}{}
\subfloat[][\ifbool{paper}{Per-Thread Reduction}{}]{
\label{fig.semantics.thread}
\tabcolsep=0pt
$
\begin{array}{c}
  \ebpsem{SEND}{{
     \Hypo{\code{
      \channel=\CH(\echanvar)
     }}
     \Hypo{\code{
     \channel' := (\eeval{\smstack}{\expr})::\channel
     }}
     \Hypo{\code{
     \CH' := \CH[\echanvar{\mapsto}\channel']
     }}
     \Infer3{\code{
     {\langle} \ltstate{\tst}{send(\echanvar,\expr)},\CH{\rangle} 
     \ttrans
     {\langle} \ltstate{\tst}{\ccskip},\CH'{\rangle}
     }}
  }}\\
  \ebpsem{RECV}{{
     \Hypo{\code{
      \channel=\CH(\echanvar)
     }}
     \Hypo{\code{
     \channel = \channel'::[\vl]
     }}
     \Hypo{\code{
     \CH' := \CH[\echanvar{\mapsto}\channel']
     }}
     \Infer[no rule]3{\code{
     \qquad  \qquad \qquad
     \tst = \langle \smstack,\smheap \rangle
     \quad
     \tst' := \langle \smstack[\res{\mapsto}\vl],\smheap \rangle
     \qquad  \qquad \qquad
     }}
     \Infer1{\code{
     {\langle} \ltstate{\tst}{\trecv{\echanvar}},\CH{\rangle} 
     \ttrans
     {\langle} \ltstate{\tst'}{\ccskip},\CH'{\rangle}
     }}
  }}\\
    \ebpsem{RECV-BLOCK}{{
     \Hypo{\code{
      \channel=\CH(\echanvar)
     }}
     \Hypo{\code{
     \channel = [~]
     }}
     \Infer2{\code{
     {\langle} \ltstate{\tst}{\trecv{\echanvar}},\CH{\rangle} 
     \ttrans
     {\langle} \ltstate{\tst}{\trecv{\echanvar}},\CH{\rangle} 
     }}
  }}
  \hide{
  \hspace{-5mm}
  \ebpsem{NOTIFYALL}{{
  \Hypo{\code{
  (\prot,\_) {=}\sstore \quad
  \btt{HasHappen}(\ev,\prot) }}
  \Infer[no rule]1{\code{ \ltst{=}{\langle}\tst,\rolevar, notifyAll(\ev){\rangle}
  }}
  \Infer1{\code{
  {\langle} \ltst,\CH,\sstore{\rangle} 
  \ttrans  {\langle}{\langle}\tst,\rolevar, \ccskip{\rangle},\CH,\sstore{\rangle}
  }}
  }}\quad
  \ebpsem{WAIT-BLOCK}{{
  \Hypo{\code{
  (\prot,\_) {=}\sstore \quad
  \neg \btt{HasHappen}(\ev,\prot) 
  }}
  \Infer[no rule]1{\code{ \ltst{=}  {\langle}\tst,P, wait(\ev){\rangle}
  }}
  \Infer1{\code{
  {\langle}\ltst,\CH,\sstore{\rangle} 
  \ttrans
  {\langle}\ltst,\CH,\sstore{\rangle}
  }}
  }}\quad
  \ebpsem{WAIT}{{
  \Hypo{\code{
  (\prot,\_) {=}\sstore \quad
  \btt{HasHappen}(\ev,\prot) 
  }}
  \Infer[no rule]1{\code{ \ltst{=}  {\langle}\tst,P, wait(\ev);e{\rangle}
  }}
  \Infer1{\code{
  {\langle}\ltst,\CH,\sstore{\rangle} 
  \ttrans {\langle}{\langle}\tst,P, e{\rangle},\CH,\sstore{\rangle}
  }}
  }}	}
\end{array}
$     
}}


\newcommand{\opsemghost}[1]{
\booltrue{#1}
\ifbool{appendix}{\captionsetup[subfigure]{labelformat=empty}}{}
\subfloat[][\ifbool{paper}{Ghost Transition}{}]{
\label{fig.semantics.ghost}
\tabcolsep=0pt
$
\begin{array}{c}
  \ebpsem{OPEN}{{
    \Hypo{\code{\CH' = \CH[\res{\mapsto} [~] ] }}
    \Infer1{\code{
    {\langle} {\langle}\tst, open()~with~{\myit{sspec}}{\rangle},\CH{\rangle}
    \ttrans 
    {\langle}{\langle}\tst, \ccskip {\rangle},\CH' {\rangle}
    }}
  }}\\
  \ebpsem{CLOSE}{{
    \Hypo{\code{
    \CH = \CH'[\echanvar{\mapsto}[~] ] 
    }}
    \Infer1{\code{
    {\langle} {\langle}\tst,close(\echanvar){\rangle},\CH{\rangle} 
    \ttrans  {\langle}{\langle}\tst', \ccskip {\rangle},\CH'{\rangle}
    }}
  }}
  \ebpsem{CLOSE-LEAK}{{
    \Hypo{\code{
    \CH(\echanvar) \neq [~] 
    }}
    \Infer1{\code{
    {\langle} {\langle}\tst,close(\echanvar){\rangle},\CH{\rangle} 
    \ttrans
    \error
    }}
  }}
\end{array}
$     
}}

\newcommand{\iopsemmachine}[1]{
\booltrue{#1}
{
\ifbool{appendix}{\captionsetup[subfigure]{labelformat=empty}}{}
\subfloat[][\ifbool{paper}{Machine Reduction}{}]{
\label{fig.semantics.machine}
$
\begin{array}{c}
  \ebpisem{MACHINE}{{
      \Hypo{\code{
      \tconfig{\itst}{\CC} \ttrans^*
      \tconfig{\itst'}{\CC'}
      }}
      \Infer1{\code{
      \mstate{\PS[\thdid\,{\mapsto}\itst]}{\CC}
      \ttrans
      \mstate{\PS[\thdid\,{\mapsto}\itst']}{\CC'}
      }}
  }} 
  \\  
  \ebpisem{PAR}{{
      \Hypo{\code{
      \itst{=}\tstate{\tst}{\rolevar}{(\expr_1||\expr_2)}
      \quad
      \textit{fresh}~ \thdid_1, \thdid_2
      \quad
       \itst'{:=} \tstate{\emp_{sh}}{\rolevar}{(\joinfunction{\thdid_1}{\thdid_2})}
      }}
      \Infer[no rule]1{\code{
      \tst {=}\tst_1 \uplus \tst_2
      \quad
      \PS'{:=}\PS[\thdid\,{\mapsto}\itst']
      [\thdid_1{\mapsto} \tstate{\tst_1}{\rolevar}{\expr_1}]
      [\thdid_2{\mapsto} \tstate{\tst_2}{\rolevar}{\expr_2}]
      }}
      \Infer1{\code{
      \mstate{\PS[\thdid\,{\mapsto}\itst]}{\CC} \ttrans  
      \mstate{\PS'}{\CC}
      }}
  }}\\
  \ebpisem{JOIN}{{
      \Hypo{\code{
      \itst{=}
      \tstate{\tst}{\rolevar}{\joinfunction{\thdid_1}{\thdid_2}}
      }}
      \Hypo{\code{
      \PS=\PS'[\thdid_1{\mapsto} \tstate{\tst_1}{\_}{\ccskip}]
      [\thdid_2{\mapsto} \tstate{\tst_2}{\_}{\ccskip}]
      }}
      \Hypo{\code{
      \tst' {:=} \tst {\uplus} \tst_1 {\uplus} \tst_2 
      }}
      \Infer3{\code{
      \mstate{\PS[\thdid\,{\mapsto} \itst]}{\CC} \ttrans  
      \mstate{\PS'[\thdid\,{\mapsto}\tstate{\tst'}{\rolevar}{\ccskip}]}{\CC}
      }}
  }}\\
  \ebpisem{JOIN-SKIP1}{{
      \Hypo{\code{
      \itst{=} \tstate{\_}{\_}{\joinfunction{\thdid_1}{\thdid_2}}          
      }}
      \Hypo{\code{
      \PS=\PS'[\thdid_1{\mapsto} \tstate{\_}{\_}{\ccskip}]
      [\thdid_2{\mapsto} \tstate{\_}{\_}{\expr}]
      }}
      \Infer2{\code{
      \mstate{\PS[\thdid\,{\mapsto} \itst]}{\CC} \ttrans
      \mstate{\PS[\thdid\,{\mapsto} \itst]}{\CC}
      }}
  }}\\
  \ebpisem{JOIN-SKIP2}{{
      \Hypo{\code{
      \itst{=} \tstate{\_}{\_}{\joinfunction{\thdid_1}{\thdid_2}}          
      }}
      \Hypo{\code{
      \PS=\PS'[\thdid_1{\mapsto} \tstate{\_}{\_}{\expr}]
      [\thdid_2{\mapsto} \tstate{\_}{\_}{\ccskip}]
      }}
      \Infer2{\code{
      \mstate{\PS[\thdid\,{\mapsto} \itst]}{\CC} \ttrans
      \mstate{\PS[\thdid\,{\mapsto} \itst]}{\CC}
      }}
  }}\\
  \ifbool{appendix}{
  \ebpisem{JOIN-ERR1}{{
      \Hypo{\code{
      \itst{=} \tstate{\_}{\_}{\joinfunction{\thdid_1}{\thdid_2}}          
      }}
      \Hypo{\code{
      \PS=\PS'[\thdid_1{\mapsto} \tstate{\_}{\_}{\expr}]
      [\thdid_2{\mapsto} \error ]
      }}
      \Infer2{\code{
      \mstate{\PS[\thdid\,{\mapsto} \itst]}{\CC} \ttrans
      \error
      }}
  }} \\ }{}
  \ifbool{appendix}{
  \ebpisem{JOIN-ERR2}{{
      \Hypo{\code{
      \itst{=} \tstate{\_}{\_}{\joinfunction{\thdid_1}{\thdid_2}}          
      }}
      \Hypo{\code{
      \PS=\PS'[\thdid_1{\mapsto} \error]
      [\thdid_2{\mapsto} \tstate{\_}{\_}{\expr}]
      }}
      \Infer2{\code{
      \mstate{\PS[\thdid\,{\mapsto} \itst]}{\CC} \ttrans
      \error
      }}
  }}  \\ }{}
\end{array}
$
}}}


\newcommand{\iopsemthread}[1]{
\booltrue{#1}
\ifbool{appendix}{\captionsetup[subfigure]{labelformat=empty}}{}
\subfloat[][\ifbool{paper}{Per-Thread Reduction}{}]{
\label{fig.semantics.thread}
\tabcolsep=0pt
$
\begin{array}{c}
  \ebpisem{SEND}{{
     \Hypo{\code{
     (\CS,\prot) {=}\CC \quad
     \chanvar,\rolevar^*=\CS(\echanvar)
     \quad
     \rolevar \in \rolevar^*
     \quad
     \SAFE = \safesnd{\project{\chanvar}{\prot}}{\rolevar}{}{} 
     }}
     \Infer[no rule]1{\code{
     (\SAFE,{v{\cdot}\fmsg},\prot') = \fid(\prot,\chanvar,\rolevar) \quad
     \CC' := (\CS,\prot')
     \quad
     \tst {=}\tst' {\uplus} \tst_x 
     \quad
     \Satisf{\tst_x}{[x/v]\St}
     }}
     \Infer1{\code{
     {\langle} \tstate{\tst}{\rolevar}{send(\echanvar,x)},\CC{\rangle} 
     \ttrans
     {\langle} \tstate{\tst'}{\rolevar}{\ccskip},\CC'{\rangle}
     }}
  }}\\
  \ifbool{appendix}{
  \ebpisem{SEND-ERACE}{{
     \Hypo{\code{
     (\CS,\prot) {=}\CC \quad
     \chanvar,\rolevar^* =\CS(\echanvar)
     \quad
     \rolevar \in \rolevar^*
     \quad
     \notiferr{\FAIL = \safesnd{\project{\chanvar}{\prot}}{\rolevar}{}{} }
     }}
     \Infer1{\code{
     {\langle} \tstate{\tst}{\rolevar}{send(\echanvar,x)},\CC{\rangle} 
     \ttrans
     \RACE
     }}
  }} \\  }{}
  \ifbool{appendix}{
  \ebpisem{SEND-EPROT}{{
     \Hypo{\code{
     (\CS,\prot) {=}\CC
     \quad
     \chanvar,\rolevar^*=\CS(\echanvar)
     \quad
     \rolevar \in \rolevar^*
     \quad
     \notiferr{(\FAIL,\_,\_) = \fid(\prot,\chanvar,\rolevar) }
     }}
     \Infer1{\code{
     {\langle} \tstate{\tst}{\rolevar}{send(\echanvar,x)},\CC{\rangle} 
     \ttrans
     \UNEXPPROT
     }}
  }}\\ } {}
  \ifbool{appendix}{\vspace{+5mm}}{}
  \ifbool{appendix}{
  \ebpisem{SEND-ERES}{{
     \Hypo{\code{
     (\CS,\prot) {=}\CC
     \quad
     \chanvar,\rolevar^*=\CS(\echanvar)
     \quad
     \rolevar \in \rolevar^*
     \quad
     \SAFE = \safesnd{\project{\chanvar}{\prot}}{\rolevar}{}{} 
     }}
     \Infer[no rule]1{\code{
     (\SAFE,{v{\cdot}\fmsg},\_) = \fid(\prot,\chanvar,\rolevar)
     \quad
     \color{red}{\neg(\exists \tst_x,\tst' \cdot
     (\tst {=}\tst' {\uplus} \tst_x)  
     \wedge
     (\Satisf{\tst_x}{[x/v]\St}))}
     }}
     \Infer1{\code{
     {\langle} \tstate{\tst}{\rolevar}{send(\echanvar,x)},\CC {\rangle}
     \ttrans
     \RESERR
     }}
  }} \\ } {}
  \ebpisem{RECV}{{
     \Hypo{\code{
     (\CS,\prot) {=}\CC
     \quad
     \chanvar,\rolevar^*=\CS(\echanvar)
     \quad
     \rolevar \in \rolevar^*
     \quad
     \SAFE = \safercv{\project{\chanvar}{\prot}}{\rolevar}
     }}
     \Infer[no rule]1{\code{
     (\SAFE,{v{\cdot}\fmsg},\prot') = \fid(\prot,\chanvar,\rolevar) \quad
     \CC' := (\CS,\prot') \quad 
     \tst' {:=} \tst {\uplus} \tst_{\res} \quad 
     \Satisf{\tst_{\res}}{[\res/v]\St}
     }}
     \Infer1{\code{
     {\langle} \tstate{\tst}{\rolevar}{\trecv{\echanvar}},\CC{\rangle} 
     \ttrans
     {\langle} \tstate{\tst'}{\rolevar}{\ccskip},\CC'{\rangle}
     }}
  }}\\
  \ifbool{appendix}{
  \ebpisem{RECV-ERACE}{{
     \Hypo{\code{
     (\CS,\prot) {=}\CC
     \quad
     \chanvar,\rolevar^* =\CS(\echanvar)
     \quad
     \rolevar \in \rolevar^*
     \quad
     \notiferr{\FAIL = \safercv{\project{\chanvar}{\prot}}{\rolevar}}
     }}
     \Infer1{\code{
     {\langle} \tstate{\tst}{\rolevar}{\trecv{\echanvar}},\CC{\rangle} 
     \ttrans
     \RACE
     }}
  }}\\ } {} 
  \ifbool{appendix}{
  \ebpisem{RECV-EPROT}{{
     \Hypo{\code{
     (\CS,\prot) {=}\CC
     \quad
     \chanvar,\rolevar^* = \CS(\echanvar)
     \quad
     \rolevar \in \rolevar^*
     \quad
     \notiferr{(\FAIL,\_,\_) = \fid(\prot,\chanvar,\rolevar)} \quad
     }}
     \Infer1{\code{
     {\langle} \tstate{\tst}{\rolevar}{\trecv{\echanvar}},\CC{\rangle} 
     \ttrans
     \UNEXPPROT
     }}
  }}\\ } {} 
  \ebpisem{RECV-BLOCK}{{
  \Hypo{\code{
     (\CS,\prot) {=}\CC
     \quad
     \chanvar,\rolevar^* =  \CS(\echanvar)
     \quad
     \rolevar \in \rolevar^*
     \quad
     \BLOCK = \safercv{\project{\chanvar}{\prot}}{P} \quad
  }}
  \Infer1{\code{
  {\langle} \tstate{\tst}{\rolevar}{\trecv{\echanvar}},\CC {\rangle} 
  \ttrans
  {\langle} \tstate{\tst}{\rolevar}{\trecv{\echanvar}},\CC {\rangle}
  }}
  }}\\
  \hide{
  \hspace{-5mm}
  \ebpisem{NOTIFYALL}{{
  \Hypo{\code{
  (\prot,\_) {=}\sstore \quad
  \btt{HasHappen}(\ev,\prot) }}
  \Infer[no rule]1{\code{ \itst{=}{\langle}\tst,\rolevar, notifyAll(\ev){\rangle}
  }}
  \Infer1{\code{
  {\langle} \itst,\CS,\sstore{\rangle} 
  \ttrans  {\langle}{\langle}\tst,\rolevar, \ccskip{\rangle},\CS,\sstore{\rangle}
  }}
  }}\quad
  \ebpisem{WAIT-BLOCK}{{
  \Hypo{\code{
  (\prot,\_) {=}\sstore \quad
  \neg \btt{HasHappen}(\ev,\prot) 
  }}
  \Infer[no rule]1{\code{ \itst{=}  {\langle}\tst,P, wait(\ev){\rangle}
  }}
  \Infer1{\code{
  {\langle}\itst,\CS,\sstore{\rangle} 
  \ttrans
  {\langle}\itst,\CS,\sstore{\rangle}
  }}
  }}\quad
  \ebpisem{WAIT}{{
  \Hypo{\code{
  (\prot,\_) {=}\sstore \quad
  \btt{HasHappen}(\ev,\prot) 
  }}
  \Infer[no rule]1{\code{ \itst{=}  {\langle}\tst,P, wait(\ev);e{\rangle}
  }}
  \Infer1{\code{
  {\langle}\itst,\CS,\sstore{\rangle} 
  \ttrans {\langle}{\langle}\tst,P, e{\rangle},\CS,\sstore{\rangle}
  }}
  }}\	}
\end{array}
$     
}}


\newcommand{\iopsemopenclose}[1]{
\booltrue{#1}
\ifbool{appendix}{\captionsetup[subfigure]{labelformat=empty}}{}
\subfloat[][\ifbool{paper}{Ghost Transition}{}]{
\label{fig.semantics.ghost}
\tabcolsep=0pt
$
\begin{array}{c}
  \ebpisem{OPEN}{{
    \Hypo{\code{\CC {=} {\langle}\CS,\prot {\rangle}}}
    \Hypo{\code{\CC' {:=} {\langle}\CS[\res{\mapsto}(\lchanvar,\rolevar^*)],\prot {\rangle}}}
    \Infer2{\code{
    {\langle} {\langle}\tst,\_, open()~with~ (\lchanvar,\rolevar^*){\rangle},\CC{\rangle}
    \ttrans 
    {\langle}{\langle}\tst,\_, \ccskip {\rangle},\CC' {\rangle}
    }}
  }}\\
  \ebpisem{CLOSE}{{
    \Hypo{\code{
    \CC{=}{\langle}\CS'[\echanvar{\mapsto}(\lchanvar,\rolevar^*)],\prot {\rangle}
    }}
    \Hypo{\code{
    \CC'{:=}{\langle}\CS',\prot {\rangle}
    }}
    \Hypo{\code{
    \btt{\safeempty}(\project{\chanvar}{\prot},\rolevar^*)
    }}
    \Infer3{\code{
    {\langle} {\langle}\tst,\_,close(\echanvar){\rangle},\CC{\rangle} 
    \ttrans  {\langle}{\langle}\tst',\_, \ccskip {\rangle},\CC'{\rangle}
    }}
  }}\\
  \ifbool{appendix}{
  \ebpisem{CLOSE-ELEAK}{{
    \Hypo{\code{
    \CC{=}{\langle}\CS'[\echanvar{\mapsto}(\lchanvar,\rolevar^*)],\prot {\rangle}
    }}
    \Hypo{\code{
    \color{red}{\neg(\btt{EMPTY}(\project{\chanvar}{\prot},\rolevar^*))}
    }}
    \Infer2{\code{
    {\langle} \tstate{\tst}{\_}{close(\echanvar)},\CC{\rangle} 
    \ttrans
    \LEAKERR
    }}
  }}\\ }{}
\end{array}
$     
}}

\newcommand{\iopsemghost}[1]{
\booltrue{#1}
\ifbool{appendix}{\captionsetup[subfigure]{labelformat=empty}}{}
\subfloat[][\ifbool{paper}{Ghost Transition}{}]{
\label{fig.semantics.ghost}
\tabcolsep=0pt
$
\begin{array}{c}
  \ebpisem{ASSERT-PEER}{{
    \Hypo{}
    \Infer1{\code{
    {\langle} {\langle}\tst,\_, assert~Peer(\rolevar){\rangle},\CC {\rangle} 
    \ttrans
    {\langle}{\langle}\tst,\rolevar, \ccskip {\rangle},\CC {\rangle}
    }}
  }}\\
  \ebpisem{ASSERT-PROT}{{
    \Hypo{\code{(\CS,\_)=\CC}}
    \Hypo{\code{\CC' := (\CS,\prot)}}
    \Infer2{\code{
    {\langle} {\langle}\tst,\_, assert~{\prot(\{\rolevar_1
  .. \rolevar_n\} ,\lchanvar^*)}
  {\rangle},\CC {\rangle} 
    \ttrans
    {\langle}{\langle}\tst,\rolevar, \ccskip {\rangle},\CC' {\rangle}
    }}
  }}\\
\end{array}
$     
}} 


\noindent{\bf \emph{Semantic Model.}}
Following the traditional storage model for heap manipulating
programs, the program state is defined as the pair:
\[
\code{State \defeq Stack \times Heap}
\]
\noindent where a stack \code{s \in Stack} is a total mapping
from local and logical variables \code{Var} to primitive values
\code{Val} or memory locations \code{Loc};
a heap \code{h \in  Heap} is a finite partial mapping
from memory locations to data structures stored
in the heap, \code{DVal}:
\[\arraycolsep=10pt
\begin{array}{cc}
  \code{Stack \defeq Var \rightarrow  Val \cup Loc} &
  \code{Heap \defeq Loc \rightharpoonup_{fin}   DVal}
\end{array}
\]

\figref{fig.gen_semantic_model} models the machine for the proposed
programming language whose configuration comprises a pair of
a \emph{program state} and of \emph{channels states}.
The set of channels used by the program are
described as a map from a program channel identifier
to a \emph{FIFO} list of messages. Moreover, a program is a set of
threads identified by a unique id and described by
their local state.
At any point a thread's execution influences its local
state \code{\tst},
the channels' state \code{\CH}, and
advances the program \code{\Prog}.






\begin{figure}
\begin{center}
\captionsetup[subfigure]{labelformat=empty,farskip=-5pt,captionskip=-15pt}  
\subfloat[][]
{$
  \begin{array}{llll}
    \textit{Machine~Config.} & \code{MC} &\!\!\!\code{::=}&
                                                             \code{\pstate ~ \times ~ \CH} \\
 \textit{Channels} & \code{\CH} &\!\!\!\code{::=}&
                       \code{\bringson{\echantyp}{(Val \cup Loc
                                                   )^*}}\\
 \textit{Program~State} &\code{\pstate} &\!\!\!\code{::=}&
                                                         \code{\bringson{\text{\idt}}{\LS}{}{}} \\
 \textit{Thread~State} &\code{ 
                                \LS} &\!\!\!\code{::=}& \code{\btt{State} \times \btt{\Prog}} \\
 \textit{Local~State} &\code{ 
                          {\btt{State}}}
                       &\!\!\!\code{::=}& \code{\btt{Stack} \times
                                          \btt{Heap} } \\
 \textit{Thread~Config.} &\code{\tconf} &\!\!\!\code{::=}& \code{\LS
                                                       ~\times~ \CH} \\
  \end{array}
$}\\
\subfloat[][]
{$
\begin{array}{lllll}
    \\
    \code{\PS{\in}\pstate} &
    \code{\TC{\in}\tconf} &
    \code{{\tst}{\in}{State}} &
    \code{{\ltst}{\in}{\LS}} &
    \code{\channel{\in}\CH} 
\end{array}
$}
\vspace*{3mm}
\hrule
\end{center}
\caption{A Semantic Model of the Core Language }\label{fig.gen_semantic_model}
\end{figure}


\noindent{\bf \emph{Operational Semantics.}}
The operational semantics is given as a set of reduction
rules between machine or 
thread configurations. Each reduction step is indicated
by \code{\ttrans}, where \code{\ttrans^*} denotes the
reflexive and transitive closure of \code{\ttrans}.
Moreover, \code{\ttrans^*} between 
thread configurations in the machine step,
indicates that the proposed semantic is not
constrained to a specific scheduler, quantifying
over all permissible executions.

Similar to the semantics of \cite{Brookes:2007:SCS:1235896.1236120},
a thread reduction could either lead to another thread state and
interfere in the states of program channels,
\code{{\langle} \ltstate{\tst}{\expr},\CH{\rangle}
       \ttrans
     {\langle} \ltstate{\tst'}{\expr'},\CH'{\rangle}
   }, or it could signal an
   \emph{error}
\code{{\langle} \ltstate{\tst}{\expr},\CH{\rangle}
  \ttrans
  \error
}. However, a program which is proved correct
should never reach an error state. A program terminates
in an error-free state if the final configuration reaches the
\code{\ccskip}
expression.

For convenience, only the semantic reduction rules which
do not fault are listed down in the subsequent.
If any of the premises in these rules do not hold,
then the considered reduction leads to \code{\error}. Moreover,
any state which is not captured by
the given reduction rules is forced to fault.

\begin{figure*}
\begin{center}
{
\renewcommand{\arraystretch}{4}
\opsemmachine{appendix}
}
\hrule
\end{center}
\caption{Semantic Rules: Machine Reduction} \label{fig.appendix_semantics_machine}
\end{figure*}

\begin{figure*}
\begin{center}
{
\renewcommand{\arraystretch}{4}
\opsemghost{appendix}
\vspace{-10mm}
\opsemthread{appendix}
}
\hrule
\end{center}
\vspace*{-3mm}
\caption{Semantic Rules: Per-Thread Reduction} \label{fig.appendix_semantics_thread}
\vspace*{-3mm}
\end{figure*}


\noindent{\bf \emph{Instrumented Semantic Model.}}
As depicted in \figref{fig.semantic_model},
the semantic model of the program state
is instrumented to capture the
communication specification as well as the usual heap and stack.
The machine configuration is a pair
of program state and channel configuration,
where a channel configuration captures
both a map from program channels to logical
channels, as well as a global protocol
describing the expected communication.
Besides the local thread state and the
current program, an instrumented state \code{\itst \in \TS} also
accounts for the role played by the instrumented
thread. A thread configuration \code{\TC\in\tconf} is thus defined as:\\
\centerline{
  \code{\TC \eqdef \tconfig{\itst}{\CC}}}\\
\centerline{~where~
  \code{\itst \eqdef \tstate{\tst}{\rolevar}{\expr}} and
  \code{\CC \eqdef (\CS,\prot)}.
}

For brevity
we implicitly assume the existence of a set of
function definitions in the program's environment.







\begin{figure}
\begin{center}
\captionsetup[subfigure]{labelformat=empty,farskip=-5pt,captionskip=-15pt}  
\subfloat[][]
{$
  \begin{array}{llll}
    \textit{Machine~Config.} & \code{MC} &\!\!\!\code{::=}&
                                                             \code{\pstate ~ \times ~ \cconf} \\
 \textit{Channels~Config} & \code{\cconf} &\!\!\!\code{::=}&
                            \code{\cstore ~ \times ~ \prot } \\
  \textit{Channel~Store} & \code{\cstore} &\!\!\!\code{::=}& \code{\bringson{\echantyp}{\lchantyp\times\roletyp^*}} \\

 \textit{Program~State} &\code{\pstate} &\!\!\!\code{::=}&
                                                         \code{\bringson{\text{\idt}}{\TS}{}{}} \\
 \textit{Instr.~Thread~State} &\code{ 
                                \TS} &\!\!\!\code{::=}& \code{\btt{State} \times {\roletyp} \times \btt{\Prog}} \\
 \textit{Local~State} &\code{ 
                          {\btt{State}}}
                       &\!\!\!\code{::=}& \code{\btt{Stack} \times
                                          \btt{Heap} \times \morders} \\
 \textit{Thread~Config.} &\code{\tconf} &\!\!\!\code{::=}& \code{\TS
                                                       ~\times~ \cconf} \\
  \end{array}
$}\\
\subfloat[][]
{$
\begin{array}{lll}
    \\
    \code{\PS{\in}\pstate} &
    \code{\TC{\in}\tconf} &
    \code{{\tst}{\in}{State}} \\
    \code{{\itst}{\in}{\TS}} &
    \code{\CS{\in}\cstore} &
    \code{\CC{\in}\cconf} 
\end{array}
$}
\vspace*{3mm}
\hrule
\end{center}
\caption{An Instrumented Semantic Model of the Core Language }\label{fig.semantic_model}
\end{figure}

\def\fidelity{
\begin{figure*}
\begin{flushleft}
\captionsetup[subfigure]{labelformat=empty,farskip=-4pt,captionskip=0pt}  
\subfloat[][]
{$
  \begin{array}{ll}
    \code{\safefid{\lchanvar}{\rolevar}{(\atransmit{S}{R}{v{\cdot}\fmsg}{\lchanvar_0}{\atid}{\seq}\prottail)}} 
    &\code{\defeq~} 
      \code{(\SAFE,~{v{\cdot}\fmsg},~\atransmit{\HOLE}{R}{v{\cdot}\fmsg}{\lchanvar_0}{\atid}{\seq}\prottail)} 
      \text{~when~} 
      \code{\lchanvar_0 {=} \lchanvar ~\wedge~S {=} \rolevar}\\
    
    \code{\safefid{\lchanvar}{\rolevar}{(\atransmit{S}{R}{v{\cdot}\fmsg}{\lchanvar_0}{\atid}{\seq}\prottail)}} 
    &\code{\defeq~} 
      \code{(\SAFE,~{v{\cdot}\fmsg},~\atransmit{S}{\HOLE}{v{\cdot}\fmsg}{\lchanvar_0}{\atid}{\seq}\prottail)} 
      \text{~when~} 
      \code{\lchanvar_0 {=} \lchanvar ~\wedge~R{=} \rolevar }\\

    \code{\safefid{\lchanvar}{\rolevar}{(\atransmit{S}{R}{v{\cdot}\fmsg}{\lchanvar_0}{\atid}{\seq}\prottail)}}
    &\code{\defeq~}
      \code{(\ANS,~\msgvar,~\atransmit{S}{R}{v{\cdot}\fmsg}{\lchanvar_0}{\atid}\seq\protres)}
      \text{~when~} 
      \code{\rolevar {\notin} \{S,~R\} ~\wedge~ \lchanvar{\neq}\lchanvar_0
      }
    {\code{~\wedge~ (\ANS,~\msgvar,~\protres) {=} \safefid{\lchanvar}{\rolevar}{\prottail} }}\\

    \code{\safefid{\lchanvar}{\rolevar}{(\prot_1 \vee \prot_2){\seq}\prottail}}
    &\code{\defeq~}
    \code{(\ANS,~\msgvar,~(\protres \vee \prot_2) \seq\prottail)}
    \text{~when~} 
    \code{(\ANS,~\msgvar,~\protres) {=}
      \safefid{\lchanvar}{\rolevar}{\prot_1}} 
    { \code{~\wedge~ (\FAIL,~{\_},~{\_}) {=} \safefid{\lchanvar}{\rolevar}{\prot_2} }}\\
&   \text{~or~}
    \code{(\ANS,~\msgvar,~(\prot_1 \vee \protres) \seq\prottail)}
    \text{~when~} 
    \code{(\FAIL,~{\_},~{\_}) {=}
  \safefid{\lchanvar}{\rolevar}{\prot_1} }
    {\code{~\wedge~
    (\ANS,~\msgvar,~\protres) {=}
    \safefid{\lchanvar}{\rolevar}{\prot_2}}}\\
    
    \code{\safefid{\lchanvar}{\rolevar}{(\prot_1 \useq \prot_2){\seq}\prottail}}
    &\code{\defeq~}
      \code{(\ANS,~\msgvar,~(\protres \useq \prot_2) \seq\prottail)}
      \text{~when~} 
      \code{(\ANS,~\msgvar,~\protres) {=}
      \safefid{\lchanvar}{\rolevar}{\prot_1}}
    {\code{~\wedge~ (\FAIL,~{\_},~{\_}) {=} \safefid{\lchanvar}{\rolevar}{\prot_2} }}\\
    &   \text{~or~}
      \code{(\ANS,~\msgvar,~(\prot_1 \useq \protres) \seq\prottail)}
      \text{~when~} 
      \code{  (\FAIL,~{\_},~{\_}) {=}
      \safefid{\lchanvar}{\rolevar}{\prot_1}}
      {\code{~\wedge~ (\ANS,~\msgvar,~\protres) {=} \safefid{\lchanvar}{\rolevar}{\prot_2}}}
  \end{array}
  $} \hfill \\
\end{flushleft}
\vspace*{-3mm}
\hrule
\vspace*{3mm}
\caption{Safety Check: Protocol Conformance}\label{fig.semantics.fidelity}
\end{figure*}
}

\def\racefreecheck{
\begin{figure*}
  \begin{flushleft}
\captionsetup[subfigure]{labelformat=empty,farskip=-4pt,captionskip=0pt}  
\subfloat[][]
{$
  \begin{array}{ll}
    \code{\safesnd{(\atransmit{S}{R}{v{\cdot}\fmsg}{}{\atid}{\seq}\cprojtail)}{\rolevar}{\msgvar}{\sigma}} &\code{\defeq~} 
    \code{\SAFE } 
    \text{~when~}  \code{\rolevar {=} S
}\\
    \code{\safesnd{(\atransmit{\HOLE}{R}{v{\cdot}\fmsg}{}{\atid}{\seq}\cprojtail)}{\rolevar}{\msgvar}{\sigma}}
    &\code{\defeq~}
     \code{\safesnd{\cprojtail}{\rolevar}{\msgvar}{\sigma}}
     \text{~when~} 
     \code{\rolevar {\neq} R    }\\
    \code{\safesnd{((\cproj_1 {\vee} \cproj_2){\seq}\cprojtail)}{\rolevar}{{\msgvar}}{\sigma}}
    &\code{\defeq~}
      \code{ \safesnd{\cproj_1}{\rolevar}{\msgvar}{\sigma} }
      \text{~when~} 
      \code{\FAIL {=} \safesnd{\cproj_2}{\rolevar}{\msgvar}{\sigma}}\\
    & \text{\quad ~or~~  }
      \code{\safesnd{\cproj_2}{\rolevar}{\msgvar}{\sigma}}
      \text{~when~} 
      \code{\FAIL {=} \safesnd{\cproj_1}{\rolevar}{\msgvar}{\sigma}}    
  \end{array}
$}
\hfill 
\subfloat[][]
{$
  \begin{array}{ll}
    \code{\safercv{(\atransmit{\HOLE}{R}{v{\cdot}\fmsg}{}{\atid}{\seq}\cprojtail)}{\rolevar}} &\code{\defeq~} 
    \code{\SAFE} 
    \text{~when~} 
    \code{\rolevar {=} R}\\
    \code{\safercv{(\atransmit{S}{R}{v{\cdot}\fmsg}{}{\atid}{\seq}\cprojtail)}{\rolevar}} &\code{\defeq~} 
    \code{\BLOCK}
    \text{~when~} 
    \code{\rolevar {=} R}\\
    \code{\safercv{(\atransmit{\HOLE}{\HOLE}{\_}{}{\atid}{\seq}\cprojtail)}{\rolevar}}
    &\code{\defeq~}
    \code{\safercv{\cprojtail}{\rolevar}}\\
    \code{\safercv{((\cproj_1 {\vee} \cproj_2){\seq}\cprojtail)}{\rolevar}}
    &\code{\defeq~}
    \code{\safercv{\cproj_1}{\rolevar} }
      \text{~when~} 
    \code{\FAIL {=} \safercv{\cproj_2}{\rolevar}}\\
    & \text{\quad ~or~~  }
    \code{\safercv{\cproj_2}{\rolevar} }
    \text{~when~} 
    \code{\FAIL {=} \safercv{\cproj_1}{\rolevar} }
    \\
  \end{array}
  $}\hfill
\end{flushleft}
\vspace*{-3mm}
\hrule
\vspace*{3mm}
\caption{Safety Checks: Race-free}\label{fig.safe_transmission}
\end{figure*}
}

\def\figsafeemp{
\begin{figure}
\begin{center}
\captionsetup[subfigure]{labelformat=empty,farskip=-4pt,captionskip=0pt}  
\subfloat[][]
{$
  \begin{array}{ll}
    \code{\safeemp{\rolevar^*}{(\atransmit{S}{R}{\_}{}{\atid})}} 
    &\code{\defeq~}
      \code{\forall \rolevar \in \rolevar^* \implies  \rolevar \notin \{S,R\}}\\
    \code{\safeemp{\rolevar^*}{\cproj_1 \seq \cproj_2}} 
    &\code{\defeq~}
      \code{\safeemp{\rolevar^*}{\cproj_1}  \wedge}
      \code{\safeemp{\rolevar^*}{\cproj_2}}\\
    \code{\safeemp{\rolevar^*}{\cproj_1 \vee \cproj_2}} 
    &\code{\defeq~}
      \code{\safeemp{\rolevar^*}{\cproj_1}  \vee}
      \code{\safeemp{\rolevar^*}{\cproj_2}}
  \end{array}
  $}\hfill
\end{center}
\hrule
\vspace*{3mm}
\caption{Safety Check: Leak-free}\label{fig.safe_empty}
\end{figure}
}

\noindent{\bf \emph{Small-steps Operational Semantics.}}
Since a machine cannot run
such an instrumented semantics, once we prove the soundness
of the verifier, we show how the instrumented semantics
is correlated to the initial semantics. 

The small-step operational semantics are defined by the
semantic rules in \figref{fig.appendix_isemantics_machine},
\figref{fig.appendix_isemantics_thread} and
\figref{fig.appendix_isemantics_ghost}.
These semantic rules are defined using
the transition relation \code{\ttrans}
between machine configurations
\code{{\langle} \PS,\CC {\rangle} \ttrans
  {\langle} \PS',\CC' {\rangle}
}, and between
thread configurations
\code{  {\langle} \itst,\CC{\rangle}
   \ttrans
  {\langle} \itst',\CC' {\rangle}
}, respectively. Similar to \secref{sec.gen_semantics},
we use \code{\ttrans^*}
to denote the transitive closure of the transition relation
\code{\ttrans}.

\figref{fig.appendix_isemantics_machine} describes the machine
reduction rules.

\begin{figure*}
\begin{center}
{
\renewcommand{\arraystretch}{4}
\iopsemmachine{appendix}
}
\hrule
\end{center}
\caption{Instrumented Semantic Rules: Machine Reduction} \label{fig.appendix_isemantics_machine}
\end{figure*}

We next distinguish between the possible
instantiations of \code{\error}, namely those
reduction faults
resulted as a consequence of communication errors:
\begin{description}
\item [\code{\UNEXPPROT}] indicates that the current reduction
  refers to a transmission which is not expected within the
  communication protocol.
\item [\code{\RACE}] is raised when the \code{\safesend}  or
  \code{\saferecv} identify a sending or receiving race condition.
\item [\code{\RESERR}] is raised when the communicating thread
  refers to resources it does not owe.
\item [\code{\LEAKERR}] indicates a possible data leak towards
  unintended recipients.
\end{description}

The expressions which manipulate the channels are given a semantic via
the rules in \figref{fig.appendix_isemantics_openclose}.

\figsafeemp

\begin{figure*}
\begin{center}
{
\renewcommand{\arraystretch}{3.5}
\iopsemopenclose{appendix}
}
\hrule
\end{center}
\vspace*{-3mm}
\caption{Instrumented Semantic Rules: Channel Manipulation} \label{fig.appendix_isemantics_openclose}
\vspace*{-3mm}
\end{figure*}

To test for protocol conformance we define a predicate,
\code{\safefid{\lchanvar}{\rolevar}{\prot} \defeq (safe,\msgvar,\prot')},
which checks whether the current transmission is correctly captured
within the global protocol. This predicate takes as input
a logical channel \code{\lchanvar} which corresponds to the channel
which performs the current transmission,
a logical party \code{\rolevar} corresponding to the thread
which is performing the transmission
and the current state of the
communication protocol \code{\prot}. The predicate then
encapsulates the safety of the transmission, \code{safe
  \in\{\SAFE,\FAIL\}}
as well as the logical description of the transferred message
and an updated communication protocol. To capture that a
transmission has been consumed, the corresponding transmission is replaced by
a \code{\HOLE} in the updated global protocol. 
The \code{\fid}
predicate is inductively defined over the structure of
protocol \code{\prot}, as depicted in
\figref{fig.semantics.fidelity}:
the first two cases correspond to a send and
receive, respectively, while the third case
permits a transmission to overtake the head of the protocol
should the current transmission be consumed on a different
sender/receiver and channel than expected. The cases for
choice and parallel communication emphasize the
non-deterministic character of this approach where
the current transmission can safely be consumed 
only by strictly one branch.
\fidelity

To force the machine to fault when it encounters
a communication race, we define the predicates
\code{\safesend} and \code{\saferecv}
to check for send or receive race, respectively.
The predicates expect a logical peer - corresponding to the
role played by the current thread -
and a channel specification 
as arguments. The channel specification is obtained
by dynamically projecting the global protocol
onto a logical channel as per the projection rules described in
\figref{fig.projection.party}. The holes of the global protocol
are ignored
during the projection. The predicates are then recursively
defined on the structure of the channel specification.
Since sending is asynchronous, \code{\safesend}
also ensures that the program order and
the communication protocol agree on the order
of transmissions (\code{\rolevar {\neq} R} in the second case).
\code{\safesend} is defined in terms of \code{safe
  \in\{\SAFE,\FAIL\}}, while \code{\saferecv} also
captures the blocking behavior of the receive,
\code{safe
  \in\{\SAFE,\FAIL,\BLOCK\}}.

Any case which is not explicitly captured by 
\code{\fid}, \code{\safesend} and \code{\saferecv}
is deemed to fail.

\racefreecheck

The communication related expressions are given a semantic via
the rules in \figref{fig.appendix_isemantics_thread}.

\begin{figure*}
\begin{center}
{
\renewcommand{\arraystretch}{4}
\iopsemthread{appendix}
}
\hrule
\end{center}
\vspace*{-3mm}
\caption{Instrumented  Semantic Rules: Per-Thread Reduction} \label{fig.appendix_isemantics_thread}
\vspace*{-3mm}
\end{figure*}

Finally, the \code{{\ioprulen{ASSERT-PEER}}} in
\figref{fig.appendix_isemantics_ghost}
updates the configuration
of a thread to indicate the role it plays.

\begin{figure*}
\begin{center}
{
\renewcommand{\arraystretch}{3.5}
\iopsemghost{appendix}
}
\hrule
\end{center}
\vspace*{-3mm}
\caption{Instrumented Semantic Rules: Ghost Transition} \label{fig.appendix_isemantics_ghost}
\vspace*{-3mm}
\end{figure*}

\section{Soundness}
This section aims to prove the soundness of the proposed
multiparty solution
with respect to the given small-step operational semantics
by proving progress and preservation. But before proving
soundness, it is worth discussing the 
interference and locality issues addressed
by the concurrency formalism approaches in general, and by
separation logic based approaches in particular.

\noindent{\bf \emph{Interference.}}
We stress on the fact that, for clarity,
the current thesis
highlights the effects explicit
synchronization has strictly over the communication.
The effects that explicit
synchronization has over the local heap are orthogonal issue
tackled by works such as  
\cite{Brookes:2007:SCS:1235896.1236120,
  Reddy:2012:SCI:2103656.2103695,
  Feng:2007:RCS:1762174.1762193}.
This explains the semantic choice for the programming
model of \figref{fig.semantic_model} which 
separates
the resources owned by a thread
from those owned by a communication channel,
and where the
environment interference
only affects the state of communication and not the local state.

\noindent{\bf \emph{Locality.}}
Assuming that each message is characterized by 
a precise formula, each time a thread performs
a read it acquires the resource ownership of exactly
that heap portion needed
to satisfy the formula corresponding to the received message.
Similarly, 
on sending a message the thread releases a resource,
i.e. it transfers the
ownership of exactly that heap portion determined by the message formula.
Since the formulae describing the messages are
precise, a transmission can only modify the state
of the local heap in one way, releasing or acquiring
the resource which is being transmitted, or in other words
there is only one possible local transmission
which is safe, 
as highlighted 
in \figref{fig.appendix_isemantics_thread}.

Moreover, a parallel computation
\code{\expr_1 \cpar \expr_2}
can be decomposed into the local computations
\code{\expr_1} and \code{\expr_2}
which are interference-free, 
except for the communication
related
interactions  carefully guided by
a global protocol. 

\begin{defn}[Compatible instrumented states]
  Two instrumented states \code{\itst^1,\itst^2 \in \TS} are
  compatible, written
  \code{\comp{\itst^1}{\itst^2}},
  if and only if there exists
  \code{ \rolevar\in\roletyp,~
    \expr_1,\expr_2 \in \Prog}, and \code{\tst_1,\tst_2 \in \LS
  } such that
  \code{\itst^1 {=} \tstate{\tst_1}{\rolevar}{\expr_1}}
  and
  \code{\itst^2 {=} \tstate{\tst_2}{\rolevar}{\expr_2}}
  and  \code{\tst_1 \disjoint \tst_2}
  and either \code{\expr_1 = \cskip } or \code{\expr_2 = \cskip }.
\end{defn}

\begin{defn}[Composition of instrumented states]
  The composition of two instrumented states
  \code{\itst^1,\itst^2 \in \TS}, where
  \code{\comp{\itst^1}{\itst^2 }} and
  \code{\itst^1 {=} \tstate{\tst_1}{\rolevar}{\expr}}
  and
  \code{\itst^2 {=} \tstate{\tst_2}{\rolevar}{\cskip}}
  is defined as:\\
  \centerline{\code{\itst^1 \uplus \itst^2 \eqdef \tstate{\tst_1
        \uplus \tst_2}{\rolevar}{\expr} } }\\
  with \code{\itst^1 \uplus \itst^2 \in \TS} .
\end{defn}

\begin{lemma}
  For any two instrumented states
  \code{\itst^1,\itst^2 \in \TS}, where
  \code{\comp{\itst^1}{\itst^2 }} and
  \code{\itst^1 \uplus \itst^2 \in \TS},
  there exists \code{\CC\in\cconf}
  such that:\\
  (1) \code{\tconfig{\itst^1 \uplus \itst^2}{\CC}
    \ttrans^*
    \error 
    ~~\implies~~
    \itst^1 \ttrans \error
  }\\
  (2) \code{\tconfig{\itst^1 \uplus \itst^2}{\CC}
    \ttrans^*
    \tconfig{\itst'}{\CC'}
    ~~\implies~~}\\
    \code{
    \exists {\itst^0 \in \TS \cdot \itst'=\itst^0 \uplus \itst^2 }
    ~and~
    \code{
      \tconfig{\itst^1}{\CC}
      \ttrans^*
      \tconfig{\itst^0}{\CC'}
    }.
  }
\end{lemma}

\begin{proof}
  Proving (1) is straightforward. Proving (2) requires induction
  on the length of the derivation and a case analysis
  on \code{\expr} which is similar to the
  standard locality principle of
  separation logic \cite{Brookes:2007:SCS:1235896.1236120},
  with the extra judgement on \code{\CC} . 
\end{proof}

\begin{defn}[Instrumented Satisfaction]
  An assertion \code{\St} is satisfied in an instrumented thread
  state \code{\itst {=} \tstate{\tst}{\rolevar}{\_}}
  and channel configuration \code{\CC}
  written \code{\ISatisf{\itst}{\CC}{\St}}, if \code{\St} is satisfied in
  the thread's local state:\\
  \centerline{\code{
      \ISatisf{\tstate{\tst}{\rolevar}{\_}}{\CC}{\St}
      {~\Leftrightarrow~}
      \Satisf{\tst}{\St}.
    }}
\end{defn}

\begin{lemma}\label{lem.soundness.sep}
  For any two instrumented states
  \code{\itst^1,\itst^2 \in \TS}, where
  \code{\comp{\itst^1}{\itst^2 }} and
  \code{\itst^1 \uplus \itst^2 \in \TS},
  there exists \code{\CC\in\cconf}, and formulae
  \code{\St_1,\St_2}
  such that:\\
  \centerline{\code{
    \ISatisf{\itst^1}{\CC}{\St_1}
    ~\wedge~
    \ISatisf{\itst^2}{\CC}{\St_2}
    ~~\implies~~
    \ISatisf{\itst^1 \uplus \itst^2}{\CC}{\St_1 \sep \St_2}
  }. }
\end{lemma}
\begin{proof}
  This proof is straightforward
  from the definition of
  instrumented states composition and
  the semantic of \code{\sep},
  provided that \code{\St_1,\St_2} are
  precise.  
\end{proof}

The validity of a triple is inductively defined with respect to the
reduction rules and satisfaction relation as follows:
\begin{defn}[Validity]
  A triple \code{\triple{\St_1}{\expr}{\St_2}} is valid,
written \code{\Valid{\triple{\St_1}{\expr}{\St_2}}}
if:\\
{\code{
  \forall {\itst \in \TS,\CC \in \cconf} ~ \cdot}}\\
{\code{({\ISatisf{\itst}{\CC}{\St_1}}) ~\wedge~
  ({\itst = \tstate{\_}{\_}{\expr}})  ~\wedge~
  ({\tconfig{\itst}{\CC} \ttrans \tconfig{\itst'}{\CC'}}) ~\wedge~}}\\
\code{{  ({\itst' = \tstate{\_}{\_}{\expr'}})
  }}
{\code{
    {~\implies~ \exists \St \cdot (\ISatisf{\itst'}{\CC'}{\St})
      ~\wedge~
      ({\triple{\St}{\expr'}{\St_2}})
    }.
}}
\end{defn}

\begin{thm}[Preservation]
  For expression \code{\expr} and states \code{\St_1} and \code{\St_2}, if \code{\htriple{\St_1}{\expr}{\St_2}}
  then \code{\Valid{\triple{\St_1}{\expr}{\St_2}}}.
\end{thm}

{
\begin{proof}
We first show that each proof rule
is sound: if the premisses are valid, then the conclusion is valid. It
then follows, by structural induction on e, that every provable
formula is valid. We only focus on the communication
related rules, since the rest are standard. 
Assume \code{\tconfig{\itst}{\CC}} as the initial
configuration for each of the case studies below:
\begin{description}
  \item [Parallel Decomposition] ~\\ Suppose that
    \code{{\ISatisf{\itst}{\CC}{\St_1 \sep \St_2}}}. Since we assume
    only precise formulae, it results that there exists
    \code{{\itst^1},{\itst^2} \in \TS} such that
    \code{\itst = \itst^1 \uplus \itst^2},
    and \code{{\ISatisf{\itst^1}{\CC}{\St_1}}}
    and \code{{\ISatisf{\itst^2}{\CC}{\St_2}}}.
    From the dynamic semantics then either
    \code{ \tconfig{\itst^1}{\CC}
      \ttrans^*
      \error
    }, or
    \code{ \tconfig{\itst^2}{\CC}
      \ttrans^*
      \error
    }, in which case the whole program faults as per
    \code{{\ioprulen{JOIN-ERR1}}} and
    \code{{\ioprulen{JOIN-ERR2}}}, respectively,
    or
    there exists 
    \code{{\itst^{1'}},{\itst^{2'}} \in \TS}
    such that 
    \code{ \tconfig{\itst^1}{\CC}
      \ttrans^*
      \tconfig{\itst^{1'}}{\CC_1}
    }, and
    \code{ \tconfig{\itst^2}{\CC}
      \ttrans^*
      \tconfig{\itst^{2'}}{\CC_2}
    }.
    with \code{\itst^{1'}=\tstate{\tst_1'}{_}{\expr_1'}}
    and
    \code{\itst^{2'}=\tstate{\tst_2'}{_}{\expr_2'}}.
    If \code{\neg({\CC_1} {\equiv} {\CC_2})},
    then it must the case that either
    \code{\expr_1' \neq \cskip}
    or
    \code{\expr_2' \neq \cskip}, corresponding thus
    to either
    \code{{\ioprulen{JOIN-SKIP1}}}
    or
    \code{{\ioprulen{JOIN-SKIP2}}}, respectively.
    If there is no race,
    then both 
    threads stabilize according to
    \code{{\ioprulen{JOIN}}},
    meaning that there exists
    \code{{\itst^{1''}},{\itst^{2''}} \in \TS}
    such that
    \code{ \tconfig{\itst^{1'}}{\CC_1'}
      \ttrans^*
      \tconfig{\itst^{1''}}{\CC_1''}
    }, and
    \code{ \tconfig{\itst^{2'}}{\CC_2'}
      \ttrans^*
      \tconfig{\itst^{2''}}{\CC_2''}
    }, with
    \code{{\CC_1''} {\equiv} {\CC_2''}},
    and
    \code{\itst^{1''}=\tstate{\tst_1''}{_}{\cskip}}
    and
    \code{\itst^{2''}=\tstate{\tst_2''}{_}{\cskip}}.
    Therefore, there exists precise formulae \code{\St_1',\St_2'}
    such that
    \code{{\ISatisf{\itst^{1''}}{\CC}{\St_1'}}}
    and \code{{\ISatisf{\itst^{2''}}{\CC}{\St_2'}}}.
    It results from \lemref{lem.soundness.sep}
    that \code{{\ISatisf{\itst^{1''} \uplus
          \itst^{2''}}{\CC}{\St_1'\sep\St_2'}}},
    hence the conclusion.
  \item [Open] If \code{\ISatisf{\itst}{\CC}{\init(\lchanvar)}}
    then opening a channel according to \code{{\ioprulen{OPEN}}}
    involves updating the channel store with a  map from
    \code{\res} to \code{\lchanvar,\rolevar^*}, that is a state which,
    according to \figref{fig.fig_semantics_prim_pred},
    satisfies the \code{{~\opened{\lchanvar}{\rolevar^*}{\res}}}
    predicate, therefore the conclusion holds.
  \item [Close] If
    \code{\ISatisf{\itst}{\CC}{\emptyc{\lchanvar}{\echanvar}}}
    then \code{\safeemp{\rolevar^*}{\project{\lchanvar}{\prot}}}
    holds
    (assuming \code{\HOLE}s are ignored during projection),
    and this thread cannot, therefore, dynamically reach fault according to the
    operational semantics of
    \figref{fig.appendix_isemantics_openclose}.
    The post-state trivially holds in this case.    
  \item [Send]
    If
    \code{\ISatisf{\itst}{\CC}
      {\chani{\lchanvar}{\rolevar}{{
            \send{v}{V(v)}} {\seq} {\proj}} {\sep}
        {V(x)} {\sep}
        \peer{\rolevar} \sep
        \opened{\lchanvar}{\rolevar^*}{\echanvar} \wedge \rolevar
        {\in} \rolevar^*
      } 
    } (1) holds, it then follows immediately that
    the first three premises of \code{{\ioprulen{SEND}}}
    also hold: \code{
      (\CS,\prot) {=}\CC}
    \code{
      \wedge~
      (\chanvar,\rolevar^*=\CS(\echanvar))
      \wedge
      (\rolevar \in \rolevar^*)
    }. We next need to prove that \code{\safesend} also holds.
    Since \code{\chani{\lchanvar}{\rolevar}{{
          \send{v}{V(v)}} {\seq} {\proj}}}
    with the race-free specification
    \code{{\send{v}{V(v)}} {\seq} {\proj}}
    for channel \code{\lchanvar} and party \code{\rolevar}
    is satisfied in the configuration
    \code{\itst,\CC}, it implies
    that indeed the next sending operation
    expected by \code{{\project{\lchanvar}{\prot}}}
    get consumed on \code{\rolevar}, hence
    \code{\SAFE = \safesnd{\project{\chanvar}{\prot}}{\rolevar}{}{} }
    also holds. This ensures that the current
    execution cannot fault with \code{\RACE},
    as depicted by \code{\ioprulen{SEND-ERACE}}.
    We next check whether \code{\fid} holds.
    The fact that \code{\safesend} holds,
    guarantees that the next operation described by \code{\prot}
    \wrt \code{\lchanvar} and \code{\rolevar} is
    a send, we need to check that \code{\rolevar}
    is not expected to perform another transmission
    on a channel different than \code{\lchanvar}.
    That follows directly from the pattern of the communication
    specification: \code{{\send{v}{V(v)}}{\seq}{\proj}}.
    From \propref{prop.proj.fid} and the projection
    rules in \figref{fig.projection.channel}, it results that
    should there be any other transmission \wrt party
    \code{\rolevar} but on a different channel,
    then the specification should have contained
    a guard of the form \code{\guard{\event{\rolevar}{\atid}}}.
    Since the channel specification holds for
    \code{{\send{v}{V(v)}}} as the next transmission (2),
    it implies that there are no other intermediary
    transmissions on \code{\rolevar}, hence \code{(\SAFE,\_,\_)=
      \fid(\prot,\chanvar,\rolevar)
    }, which is a shorthand for
    \code{\exists \fmsg \cdot (\SAFE,{v{\cdot}\fmsg},\_)=
      \fid(\prot,\chanvar,\rolevar) 
    } (3), hence this computation cannot fault with \UNEXPPROT~
    (\code{\ioprulen{SEND-EPROT}}).
    From (2) and (3) we can conclude that
    \code{V(v) = \St}. Since the current state satisfies
    \code{V(x)}, it follows immediately that 
    \code{\exists \tst_x,\tst' \cdot
      (\tst {=}\tst' {\uplus} \tst_x)  
      \wedge
      (\Satisf{\tst_x}{[v/x]\St})} (4),
    or in other words the current state owns
    the message it sends on \code{\echanvar}. This proves that the
    current execution cannot fault with \RESERR
    ~(\code{\ioprulen{SEND-ERES}}).
    We can conclude therefore that only \code{\ioprulen{SEND-EPROT}}
    can be fired in this case, meaning that:\\
    \code{\ISatisf{\tstate{\tst'}{\rolevar}{\_}}{\CC'}
      {{\chani{\lchanvar}{\rolevar}{\proj} 
          \sep
          \peer{\rolevar} \sep
        \opened{\lchanvar}{\rolevar^*}{\echanvar} \wedge \rolevar
        {\in} \rolevar^*}}
  }, which trivially holds by (1) and (4)
  since the only updates to the configuration
  are represented by the thread state \code{\tst'},
  reflecting now the fact that the current thread lost
  the ownership of the transmitted message,
  and \code{\CC'} whose global protocol \code{\prot}
  is updated with a \code{\HOLE} to denote the consumed transmission:
  \code{\project{\prot}{\rolevar,\lchanvar} \equiv \proj}.
  
  \item [Receive]
   If
   \code{\ISatisf{\itst}{\CC}
     {
       \chani{\lchanvar}{\rolevar}{{
           \recv{v}{V(v)}} {\seq} {\proj}} {\sep}
        \peer{\rolevar} \sep
        \opened{\lchanvar}{\rolevar^*}{\echanvar} \wedge \rolevar
        {\in} \rolevar^*
      } 
    } (1) holds, similar to \code{{\ioprulen{SEND}}},
    it follows immediately by \figref{fig.fig_semantics_prim_pred}
    that
    the first three premises of \code{{\ioprulen{RECV}}}
    also hold: \code{
      (\CS,\prot) {=}\CC}
    \code{
      \wedge~
      (\chanvar,\rolevar^*=\CS(\echanvar))
      \wedge
      (\rolevar \in \rolevar^*)
    }. If the corresponding send was not fired yet,
    the thread will stay in the same state, and
    when it will be eventually fired, (1) is still satisfied.
    Since \code{\chani{\lchanvar}{\rolevar}{{
          \recv{v}{V(v)}} {\seq} {\proj}}} (2)
    specifies a race-free communication 
    \wrt channel \code{\lchanvar} and party \code{\rolevar},
    it follows immediately
    that \code{{\project{\lchanvar}{\prot}}}
    expects for the next receiving operation
    over \code{\lchanvar}
    is on \code{\rolevar}, hence
    \code{\SAFE = \safercv{\project{\chanvar}{\prot}}{\rolevar}} (3),
    assuming the state is not blocked.
    This ensures that the current
    execution cannot fault with \code{\RACE},
    as depicted by \code{\ioprulen{RECV-ERACE}}.
    We next check whether \code{\fid} holds.
    (3) implies that the next transmission
    \wrt \code{\lchanvar} and \code{\rolevar}
    is indeed a receive, we need to check that there
    are no intermediate transmissions on different channels.
    From \propref{prop.proj.fid}, the projection
    rules \figref{fig.projection.channel} and (1),
    that is the communication
    specification \code{\chani{\lchanvar}{\rolevar}{{
          \recv{v}{V(v)}} {\seq} {\proj}}}
    expects a receive, it results that there
    are no other prior unconsumed transmissions expected
    over \code{\lchanvar}, since that would have involved
    for the specification to be guarded by
    \code{\guard{\event{\rolevar}{\atid}}} prior to the received
    specified by (2).
    It then follows that
    \code{\exists \fmsg \cdot (\SAFE,{v{\cdot}\fmsg},\_)=
      \fid(\prot,\chanvar,\rolevar) 
    } (4), hence this thread cannot fault with \UNEXPPROT~
    (\code{\ioprulen{RECV-EPROT}}).
    From (2) and (4) we can conclude that
    \code{V(v) = \St}. Executing the transmission
    according to \code{\ioprulen{RECV-EPROT}}
    results in the update of the current thread state
    such that it reflects the message ownership gain:
    \code{\exists \tst_{\res} \cdot \tst' {:=} \tst {\uplus} \tst_{\res}} where
    \code{\Satisf{\tst_{\res}}{[\res/v]\St}}.
    Since \code{V(v) = \St}, then the following also holds:
    \code{\Satisf{\tst_{\res}}{V(res)}} (5).
    With (1), (5) and the update of \code{\prot} to reflect
    the consumed receive operation, where
    \code{\project{\prot}{\rolevar,\lchanvar} \equiv \proj},
    we can conclude that the updated thread state
    satisfies the consequence of \code{\rulen{RECV}}:
    \code{\ISatisf{\tstate{\tst'}{\rolevar}{\_}}{\CC'}
     {
       \chani{\lchanvar}{\rolevar}{{\proj}} {\sep}
        V(res) \sep \peer{\rolevar} \sep
        \opened{\lchanvar}{\rolevar^*}{\echanvar} \wedge \rolevar
        {\in} \rolevar^*
      } 
    }.    
\end{description}
\end{proof}

\anay{Can abstract predicates have a semantic?
  how to describe the fact that init/opened... etc, hold?
}

\begin{thm}[Progress]
  If \code{\htriple{\St_1}{\expr}{\St_2}} and 
  \code{\exists {\itst \in \TS}} \code{{\CC \in \cconf} ~\cdot}
  \code{\ISatisf{\itst}{\CC}{\St_1}} then either
  \code{\expr} is a value or \\
  \code{\exists {\itst' \in \TS,\CC' \in \cconf} ~\cdot}
  \code{{\tconfig{\itst}{\CC} \ttrans \tconfig{\itst'}{\CC'}}}.
\end{thm}
\begin{proof}
  By induction on the length of the execution and by case analysis on the steps
  taken we could show that if \code{\tconfig{\itst}{\CC}} is a non-final,
  fault-free configuration, then \code{\tconfig{\itst}{\CC}} doesn't
  get stuck. The communication related cases are straightforward
  assuming the well-formedness of communication protocols. It may
  appear as if \code{\rulen{RECV}} could cause a process to get
  stuck, however, if the protocol which describes the communication
  is well-formed, as per \defref{gloabl.well-formed}, it is guaranteed
  for a corresponding sender to get fired within a finite number of
  machine steps.  
\end{proof}
}


\end{document}